\documentclass[traditabstract]{aa} 

\usepackage{graphicx}
\usepackage{txfonts}
\usepackage{longtable}
\usepackage{color}
\usepackage{url}

\begin{document}

   \title{A Virgo Environmental Survey Tracing Ionised Gas Emission (VESTIGE).IX. The effects of ram pressure stripping 
   down to the scale of individual HII regions in the dwarf galaxy IC 3476\thanks{Based on observations obtained with
   MegaPrime/MegaCam, a joint project of CFHT and CEA/DAPNIA, at the Canada-French-Hawaii Telescope
   (CFHT) which is operated by the National Research Council (NRC) of Canada, the Institut National
   des Sciences de l'Univers of the Centre National de la Recherche Scientifique (CNRS) of France and
   the University of Hawaii. Based on observations made with ESO Telescopes at the La Silla Paranal Observatory under programme ID 095.D-0172. Based  on  observations  collected  
   at  the  Observatoire de Haute Provence (OHP) (France), operated by the CNRS.}
      }
   \subtitle{}
  \author{A. Boselli\inst{1},  
  	  A. Lupi\inst{2},
	  B. Epinat\inst{1},
	  P. Amram\inst{1},
	  M. Fossati\inst{3,4},
 	  J.P. Anderson\inst{5},
	  S. Boissier\inst{1},  
	  M. Boquien\inst{6},
          G. Consolandi\inst{7},
  	  P. C{\^o}t{\'e}\inst{8},
          J.C. Cuillandre\inst{9},
          L. Ferrarese\inst{8},
	  L. Galbany\inst{10},
          G. Gavazzi\inst{3},
	  J.A. G\'omez-L\'opez\inst{1},
          S. Gwyn\inst{8}, 
          G. Hensler\inst{11},
	  J. Hutchings\inst{8},
	  H. Kuncarayakti\inst{12,13},
	  A. Longobardi\inst{1},
          E.W. Peng\inst{14},
	  H. Plana\inst{15},
 	  J. Postma\inst{16},
          J. Roediger\inst{8},
	  Y. Roehlly\inst{1},
	  C. Schimd\inst{1},
	  G. Trinchieri\inst{7},
	  B. Vollmer\inst{17}
       }

\institute{     
                Aix Marseille Univ, CNRS, CNES, LAM, Marseille, France
                \email{alessandro.boselli@lam.fr}
        \and  
                Scuola Normale Superiore, Piazza dei Cavalieri 7, I-56126 Pisa, Italy
	\and
		Universit\'a di Milano-Bicocca, piazza della scienza 3, 20100 Milano, Italy
	\and	
		Institute for Computational Cosmology and Centre for Extragalactic Astronomy, Department of Physics, Durham University, South Road, Durham DH1 3LE, UK
        \and
                European Southern Observatory, Alonso de C\'ordova 3107, Casilla 19, Santiago, Chile
        \and
                Centro de Astronom\'a (CITEVA), Universidad de Antofagasta, Avenida Angamos 601, Antofagasta, Chile
	\and
	        INAF - Osservatorio Astronomico di Brera, via Brera 28, 20159 Milano, Italy
        \and
                NRC Herzberg Astronomy and Astrophysics, 5071 West Saanich Road, Victoria, BC, V9E 2E7, Canada
        \and        
                AIM, CEA, CNRS, Universit\' Paris-Saclay, Universit\'e Paris Diderot, Sorbonne Paris Cit\'e, Observatoire de Paris, PSL University, F-91191 Gif-sur-Yvette Cedex, France
        \and
                Departamento de F\'isica Te\'orica y del Cosmos, Universidad de Granada, E-18071 Granada, Spain                
        \and
                Department of Astrophysics, University of Vienna, T\"urkenschanzstrasse 17, 1180 Vienna, Austria
	\and
		Tuorla Observatory, Department of Physics and Astronomy, FI-20014 University of Turku, Finland
	\and
		Finnish Centre for Astronomy with ESO (FINCA), FI-20014 University of Turku, Finland
        \and
                Department of Astronomy, Peking University, Beijing 100871, PR China
        \and
                Laborat\'orio de Astrof\'isica Te\'orica e Observacional, Universidade Estadual de Santa Cruz - 45650-000, Ilh\'eus-BA, Brasil
	\and
		University of Calgary, 2500 University Drive NW, Calgary, Alberta, Canada
	\and
		Universit\'e de Strasbourg, CNRS, Observatoire astronomique de Strasbourg, UMR 7550, F-67000 Strasbourg, France
                }

\authorrunning{Boselli et al.}
\titlerunning{VESTIGE}

   \date{}

 
  \abstract  
{ We study the IB(s)m galaxy IC 3476 observed in the context of the Virgo Environmental Survey Tracing Ionised Gas Emission (VESTIGE), a blind narrow-band H$\alpha$+[NII] imaging survey 
of the Virgo cluster carried out with MegaCam at the CFHT. The deep narrow-band image reveals a very pertubed ionised gas distribution, characterised by a prominent
banana-shaped structure in the front of the galaxy formed of giant HII regions crossing the stellar disc, with star forming structures at $\sim$ 8 kpc from the edges of the stellar disc, detected also in a deep FUV ASTROSAT/UVIT image. This particular morphology indicates that the galaxy is undergoing an almost 
edge-on ram pressure stripping event. The same H$\alpha$+[NII] image also shows that the star formation activity is totally quenched in the leading edge of the disc, where the gas has been removed during the interaction 
with the surrounding medium. The SED fitting analysis of the multifrequency data indicates that this quenching episode is very recent ($\sim$ 50 Myr), and roughly corresponds to an increase
of the star formation activity by a factor of $\sim$ 161\%\ in the inner regions with respect to what expected for secular evolution. The analysis of these data, whose angular resolution allows the study of the induced effects of the 
perturbation down to the scale of individual HII regions ($r_\mathrm{eq}$ $\simeq$ 40 pc), also suggests that the increase of the star formation activity
is due to the compression of the gas along the stellar disc of the galaxy, which is able to increase its mean electron density and boost the star formation process producing bright HII regions 
with luminosities up to $L(H\alpha)$ $\simeq$ 10$^{38}$ erg s$^{-1}$. 
The combined analysis of the VESTIGE data with deep IFU spectroscopy gathered with MUSE and with high spectral resolution Fabry Perot data also indicates that 
the hydrodynamic interaction has deeply perturbed the velocity field of the ionised gas component while leaving unaffected that of the stellar disc. The comparison of the data with tuned high-resolution
hydrodynamic simulations accounting for the different gas phases (atomic, molecular, ionised) consistently indicates that the perturbing event is very recent (50-150 Myr), once again confirming that
ram pressure stripping is a violent phenomenon able to perturb on short timescales the evolution of galaxies in rich environments.
 }
   {}
   {}
   {}
   {}
   {}


   \maketitle
%

\section{Introduction}

Galaxies inhabiting rich environments are subject to different kind of perturbations that modify their evolution. In rich clusters, 
which are characterised by a hot ($T$ $\simeq$ 10$^7$-10$^8$ K) and dense ($\rho$ $\simeq$ 10$^{-3}$ cm$^{-3}$) intergalactic
medium (IGM; e.g. Sarazin 1986), the hydrodynamic pressure exerted on the interstellar medium (ISM) of gas-rich members moving at high velocity ($\simeq$ 1000 km s$^{-1}$) 
can remove the gas and quench the activity of star formation of the perturbed galaxies. A lot of observational evidence, such as the presence of
galaxies with long cometary tails in the different gas phases (atomic cold - e.g. Chung et al. 2007; ionised - Yagi et al. 2010; hot - e.g. Sun et al. 2007) without any associated stellar counterpart, or the 
high frequency of spirals with radially truncated gaseous discs (e.g. Cayatte et al. 1994; Koopmann \& Kenney 2004), indicates that this 
process, which is commonly referred to as ram pressure stripping (Gunn \& Gott 1972), dominates the evolution of gas-rich late-type systems in nearby massive clusters 
(e.g. Boselli \& Gavazzi 2006, 2014).

While many representative objects undergoing a ram pressure stripping event have been already studied in detail using multifrequency observations 
(Kenney et al. 2004, 2014, 2015; Crowl et al. 2005; Boselli et al. 2016a; Sun et al. 2006, 2007, 2010; Merluzzi et al.
2013; Abramson \& Kenney 2014; Abramson et al. 2011, 2016; Jachym et al. 2013, 2014, 2017, 219; Fumagalli et al. 2014; Fossati et al. 2016; 
Poggianti et al. 2019; Bellhouse et al. 2019; Deb et al. 2020; Moretti et al. 2020a) and tuned models and 
simulations (Vollmer et al. 1999, 2000, 2004a,b, 2006a, 2008a,b, 2012, 2018; Roediger \& Hensler 2005; Boselli et al. 2006), we are still far from understanding the physics of this perturbing mechanism and the fate of the stripped gas once removed from the galaxy disc.
The behaviour of galaxies undergoing ram pressure stripping might drastically change from object to object in a way still not totally predictable. Indeed, 
in some galaxies gas removal is followed by a drastic reduction of the star formation activity mainly in the outer regions 
(e.g. NGC 4569 - Vollmer et al. 2004a, Boselli et al. 2006, 2016a, or NGC 4330 in Virgo - Abramson et al. 2011; Vollmer et al. 2012, 2020; Fossati et al. 2018). These objects, where the overall
activity is significantly reduced, populate the green valley between blue star forming and red quiescent systems (Boselli et al. 2014a). 
In other objects, the interaction induces a gas compression in the leading side of the perturbed galaxies 
which locally boosts the activity of star formation (e.g. CGCG 097-073 and 097-079 in A1367, Gavazzi et al. 2001), moving them 
above the main sequence (e.g. Vulcani et al. 2018). The overall increase of the star formation activity of these objects, however, could 
last only a few million years, and the continuum stripping process might sooner or later transform them into quiescent systems.
Furthermore it is also unclear what is the fate of the stripped gas, which in some objects is observed in the cool atomic hydrogen phase (Chung et al. 2007), 
in others is mainly ionised (Yagi et al. 2010 ; Boselli et al. 2016a; Gavazzi et al. 2018), or hot, but rarely at the same time in all these three phases.
This different behaviour can be due to the different properties of the perturbed galaxies such as their total mass 
(given their deeper gravitational potential well, massive systems are more resistent
to any external perturbations than dwarfs) and gas content and distribution, to the impact parameters of the infalling systems (infall velocity,
orbital parameters, interaction angle), and to the properties of the IGM (radial variation of the gas density and temperature).  

VESTIGE (A Virgo Environmental Survey Tracing Ionised Gas Emission) is a large program carried out with MegaCam at the CFHT (Boselli et al. 2018a).
This project aims at covering the whole Virgo cluster region (104$\degr$$^2$) through a narrow-band (NB) filter centered on the H$\alpha$ line to trace the 
distribution of the ionised gas at unprecedented sensitivity 
and angular resolution (see Sect. 2). The survey has been designed to study all kinds of environmental perturbations on galaxies in high density regions.
This survey, which is still ongoing, allowed us to detect several objects with perturbed morphologies suggesting the presence of an undergoing perturbation, including
minor merging events (NGC 4424, Boselli et al. 2018b; M87, Boselli et al. 2019), galaxy harassment (NGC 4254, Boselli et al. 2018c), and ram pressure stripping 
in two massive galaxies (NGC 4569, Boselli et al. 2016a; NGC 4330, Fossati et al. 2018, Vollmer et al. 2020, Longobardi et al. 2020; NGC 4522, Longobardi et al. 2020). 
In this work we present a detailed analysis of the 
dwarf galaxy IC 3476 (see Table \ref{gal}) which presents tails of ionised gas suggesting an ongoing ram pressure stripping event. As in our previous
studies, the analysis is based on a unique set of 
multifrequency data many of which still unpublished, including the VESTIGE NB and ASTROSAT/UVIT FUV imaging, and MUSE and Fabry-Perot IFU spectroscopy (Sect. 2).  
The derived physical and kinematical properties of the galaxy (Sect. 3) are then compared to the predictions of tuned simulations (Sect. 4).
The results gathered from this analysis are discussed in the framework of galaxy evolution in a rich environment (Sect. 5).

\begin{table}
\caption{Properties of the galaxy IC 3476 (VCC 1450)}
\label{gal}
{
\[
\begin{tabular}{ccc}
\hline
\noalign{\smallskip}
\hline
Variable                	& Value                 				& Ref.          \\      
\hline
$Type$          		& IB(s)m:                                               & 1       \\
$cz$            		& -170 km s$^{-1}$                                      & 2       \\
$M_\mathrm{star}^a$    		& 1.0$\times$10$^{9}$ M$_{\odot}$                       & 3       \\ 
$M(H_\mathrm{2})^b$        	& 1.4$\times$10$^{8}$ M$_{\odot}$                       & 4       \\ 
$M(HI)$         		& 2.6$\times$10$^{8}$ M$_{\odot}$                       & 4       \\ 
$HI-def$        		& 0.67			                                & 4       \\ 
$R_\mathrm{ISO}(r)$		& 5.16 kpc						& 3       \\		
$Distance$      		& 16.5 Mpc                                              & 5,6,7,8 \\
$Proj.~ distance~from~M87$      & 0.5 Mpc, 0.32 $r/r_\mathrm{vir}$                      & 9       \\
log$f(H\alpha+[NII])^c$      	& -11.90$\pm$0.02  erg s$^{-1}$ cm$^{-2}$    		& 9       \\
$SFR^a$         		& 0.27  M$_{\odot}$yr$^{-1}$                            & 9       \\
\noalign{\smallskip}
\hline
\end{tabular}
\]
References: 1) NED; 2) GOLDMine (Gavazzi et al. 2003), from HI observations; 3) Cortese et al. (2012); 4) Boselli et al. (2014b); 5) Mei et al. (2007); 6) Gavazzi et al. (1999); 7) Blakeslee et al. (2009); 
8) Cantiello et al. (2018); 9) this work.\\
Notes:  all quantities have been scaled to a distance of 16.5 Mpc, the mean distance of the cluster, for a fair comparison with other VESTIGE works. The 
uncertainty on the distance can be assumed to be $\sim$ 1.5 Mpc, the virial radius of the Virgo cluster. 
a) $M_\mathrm{star}$ and $SFR$ are derived assuming a Chabrier (2003) IMF and the Kennicutt (1998) calibration.
b) derived using the luminosity dependent CO-to-H$_\mathrm{2}$ conversion factor of Boselli et al. (2002); c) corrected for Galactic attenuation. }
\end{table}

\section{Observations and data reduction}

\subsection{VESTIGE narrow-band imaging}

H$\alpha$ NB imaging observations of IC 3476 have been carried out with MegaCam at the CFHT during the VESTIGE survey (Boselli et al. 2018a).
The galaxy has been observed during the blind survey of the Virgo cluster using the MP9603 filter ($\lambda_c$ = 6591 \AA; $\Delta\lambda$ = 106 \AA), whose
transmissivity at the redshift of the galaxy is $T$ $\simeq$ 93\%. At this redshift the filter includes the H$\alpha$ line and the [NII] doublet.\footnote{Hereafter
we refer to the H$\alpha$+[NII] band simply as H$\alpha$, unless otherwise stated.}
A detailed description of the observing strategy and of the data reduction procedures is given in Boselli et al. (2018a). Briefly, the Virgo cluster was mapped 
following a sequence of different pointings optimised to minimise the variations in the sky background necessary for the detection of low surface brightness features.
For this purpose, a large dithering of 15 arcmin in RA and 20 arcmin in Dec has been used. The total integration time per pixel was 7200 s in the NB and 720 s in the $r$-band 
necessary for the subtraction of the stellar continuum. The typical sensitivity of the survey is $f(H\alpha)$ $\simeq$ 4$\times$10$^{-17}$ erg s$^{-1}$ cm$^{-2}$ (5 $\sigma$) 
for point sources and $\Sigma(H\alpha)$ $\simeq$ 2 $\times$10$^{-18}$ erg s$^{-1}$ cm$^{-2}$ arcsec$^{-2}$ (1 $\sigma$ after smoothing the data to $\sim$ 3\arcsec resolution)
for extended sources. The data for IC 3476 have been acquired in excellent seeing conditions ($FWHM$ = 0.74\arcsec).

The data were reduced using the Elixir-LSB data reduction pipeline (Ferrarese et al. 2012) expressly designed to detect 
the diffuse emission of extended low surface brightness features formed after the interaction of galaxies with the surrounding environment.
The Elixir-LSB pipeline has been designed to remove any contribution of scattered light in the different images.
This pipeline works perfectly whenever the science frames are background dominated, as indeed is the case for the NB images taken during the VESTIGE survey.
The astrometric and photometric calibration of the images is done by comparing the fluxes and positions of stars in the different frames to those
derived from the SDSS and PanSTARRS surveys using the MegaPipe pipeline (Gwyn 2008). The typical photometric uncertainty in the data is of 0.02-0.03 mag in both bands.

\begin{figure*}
   \centering
   \includegraphics[width=1.0\textwidth]{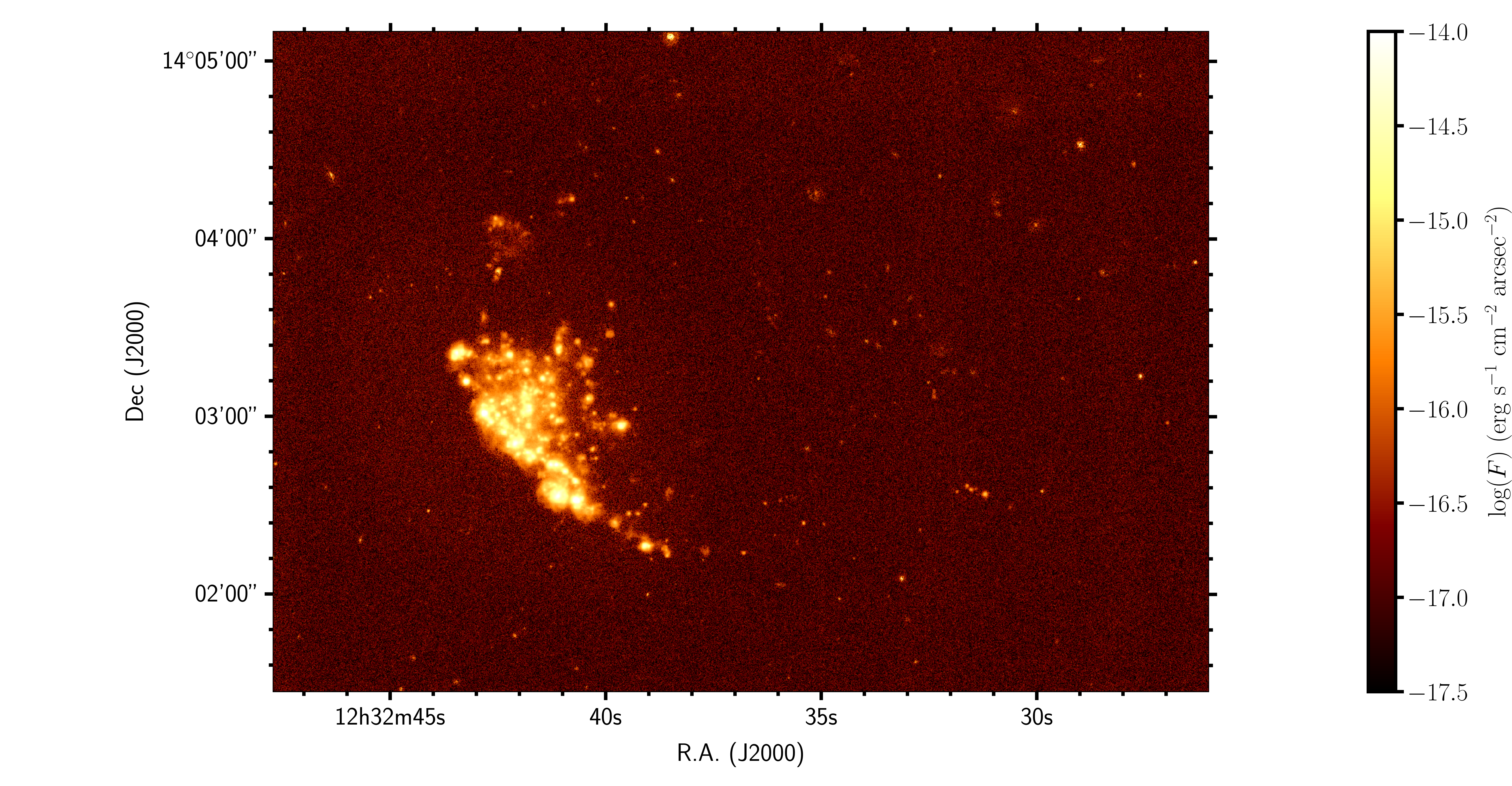}
   \caption{Continuum-subtracted H$\alpha$ image of IC 3476 derived from VESTIGE. With the assumed distance of 16.5 Mpc, 1\arcsec = 80 pc.
 }
   \label{Ha}%
   \end{figure*}

The ionised gas emission has been determined after subtracting the stellar continuum emission present in the NB filter. For this purpose, 
the contribution of the stellar continuum has been derived as in Boselli et al. (2019a) from the $r$-band image combined with the $g$-band frame taken 
from the NGVS survey of the cluster (Ferrarese et al. 2012) to quantify the colour correction necessary to take into account the dependence on 
the spectral shape of the emitting sources (Spector et al. 2012). This continuum subtraction works very well since it does not leave 
any clear residual in the final image of the pure gas emission shown in Fig. \ref{Ha}. 
This is also confirmed by the excellent match with the MUSE spectroscopic data (see Sect. 2.3).

\subsection{FUV ASTROSAT/UVIT imaging}

IC 3476 has been observed with ASTROSAT/UVIT (Agrawal 2006; Tandon et al. 2020) in May 2020 using the far-UV filter BaF2 ($\lambda_c$ = 1541 \AA; $\Delta\lambda$ = 380 \AA)
during a run dedicated to the observation of a representative sample of Virgo cluster galaxies (proposal A08-003: A FUV Survey of Virgo Cluster Galaxies,
PI: J. Hutchings). The observations have been carried out with an integration of 12453 s, reaching a typical surface brightness of 
$\mu(FUV)$ $\simeq$ 26.8 AB mag arcsec$^{-2}$.  The field of view of the instrument has a diameter of $\simeq$ 28\arcmin and 
an angular resolution of $\simeq$ 1.5\arcsec. The data have been reduced following the prescriptions given in Tandon et al. (2020) 
using a zero point of $z_p$ = 17.771 mag and the
astrometry has been checked against the accurate NGVS data (Ferrarese et al. 2012, see below). 

\subsection{MUSE spectroscopy}

The MUSE data were gathered as part of the 095.D-0172 program that studied the host galaxies of core-collapse supernovae
(Kuncarayakti et al.2018). 
Observations were carried out in May 2015 in the Wide Field Mode under excellent seeing conditions ($FWHM$ = 0.74\arcsec).
The data cover the spectral range 4800-9300 \AA\ and have a spectral resolution $R$ $\sim$ 2600 (1.25 \AA\ sampling per pixel), corresponding to 
a limiting velocity dispersion $\sigma$ $\sim$ 50 km s$^{-1}$ at H$\alpha$ (see Appendix A). 
The total integration time was 1800 s, with a second exposure of 200 s taken at 2 arcmin from the galaxy to secure the determination of the sky emission. 
With this integration time, the typical sensitivity of MUSE to low surface brightness emission at H$\alpha$ is $\Sigma(H\alpha)$ $\simeq$ 4 $\times$10$^{-18}$ erg s$^{-1}$ cm$^{-2}$ arcsec$^{-2}$.
The MUSE data were reduced using ad-hoc procedures developed within the team as extensively described in Fossati et al. (2016), Consolandi et al. (2017), and Boselli et al. (2018b).
The astrometry of the MUSE datacube has been finally recalibrated on the VESTIGE data using point sources in the field.
A comparison of the flux of the H$\alpha$+[NII] emission line extracted from MUSE with the one obtained from the NB VESTIGE data gives consistent results
within 0.3\%.

\subsection{Fabry-Perot spectroscopy}

Fabry-Perot 3D spectroscopic observations of the galaxy IC 3476 were gathered using the GHASP instrument on the 1.93 m 
Observatoire de Haute Provence (OHP) telescope in May 2017 during the complete Fabry-Perot survey of the \textit{Herschel} Reference Survey (G\'omez-L\'opez et al. 2019). 
The GHASP focal reducer has a Fabry-Perot 
with a field of view of 5.8 $\times$ 5.8 arcmin$^2$ coupled with a 512 $\times$ 512 Imaging Photon Counting System (IPCS) with a pixel scale of 0.68 $\times$ 0.68 arcsec$^2$ 
(Gach et al. 2002). IC 3476 has been observed in dark time and photometric conditions with a total exposure of 4800 s and a typical seeing of 2.0\arcsec. 
The 378 km s$^{-1}$ free spectral range of the Fabry-Perot was scanned through 32 channels providing a spectral resolution at H$\alpha$ of $R$ $\simeq$ 10000. The galaxy was observed 
through an interference filter of 15 \AA\ using the NeI emission line at 6598.95 \AA\ for wavelength calibrations. The data have been reduced as 
described in Epinat et al. (2008) and G\'omez-L\'opez et al. (2019). To provide the best combination of angular resolution and signal-to-noise at each position, 
we applied an adaptive binning technique based on a 2D Voronoi tassellation on the H$\alpha$ data cube producing radial velocity and velocity dispersion maps.
This has been done aiming at a fixed $S/N$ = 7.

\subsection{Multifrequency data}

A large set of multifrequency data useful for the following analysis is available for IC 3476.
UV data in the far-UV ($\lambda_c$ = 1539 \AA; integration time = 107 s) and near-UV ($\lambda_c$ = 2316 \AA; integration time = 1720 s) 
bands were obtained with GALEX during the GUViCS survey of the cluster (Boselli et al. 2011). 
Deep optical images in the $u$, $g$, $i$, and $z$-bands are available from the NGVS survey of the Virgo cluster (Ferrarese et al. 2012).
The sensitivity of this survey is $g$ $\simeq$ 25.9 AB mag for point sources (10 $\sigma$) and $\mu(g)$ 29 AB mag arcsec$^{-2}$ for extended sources (2$\sigma$), respectively.

  \begin{figure*}
   \centering
   \includegraphics[width=0.49\textwidth]{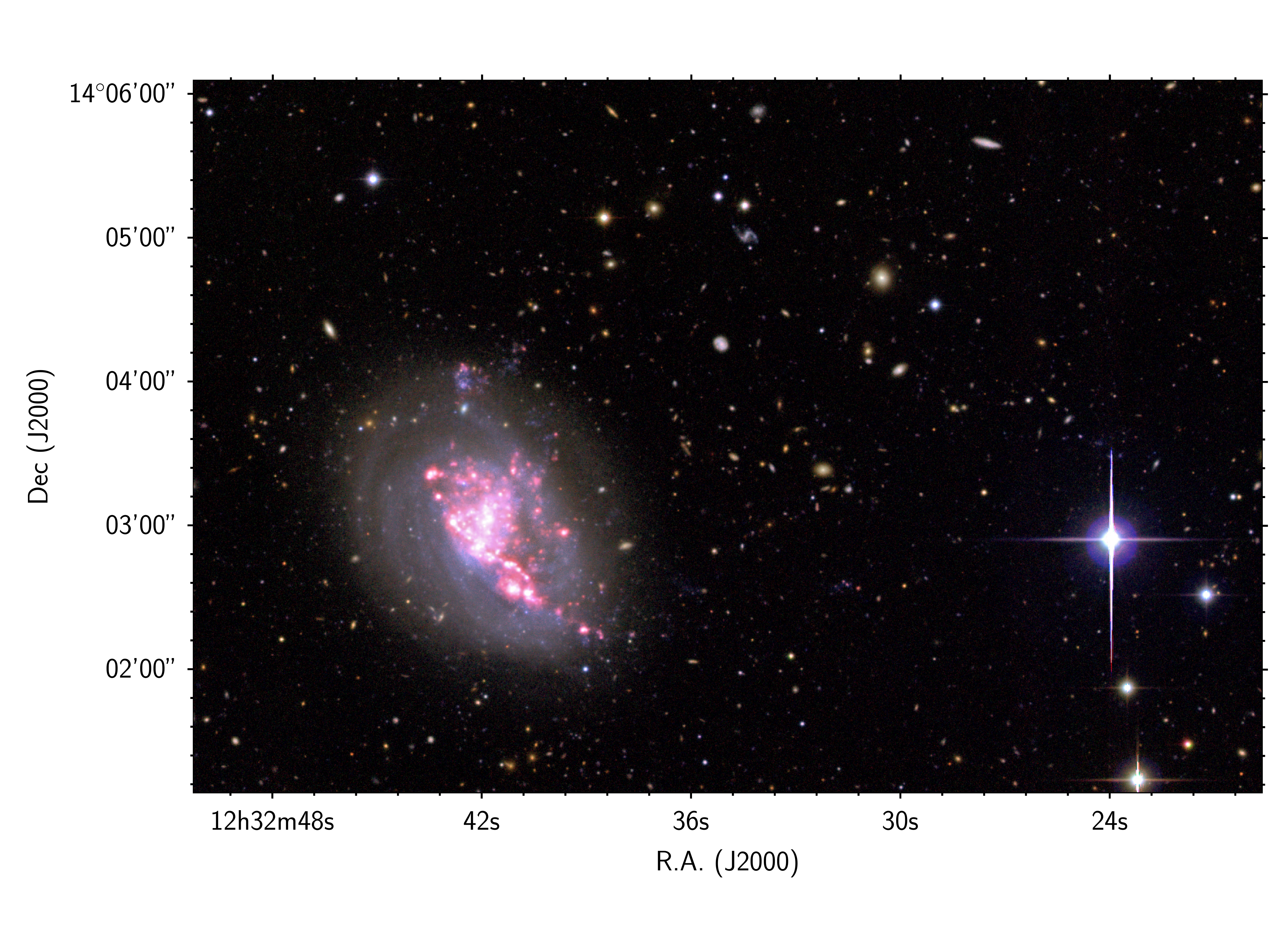}
   \includegraphics[width=0.49\textwidth]{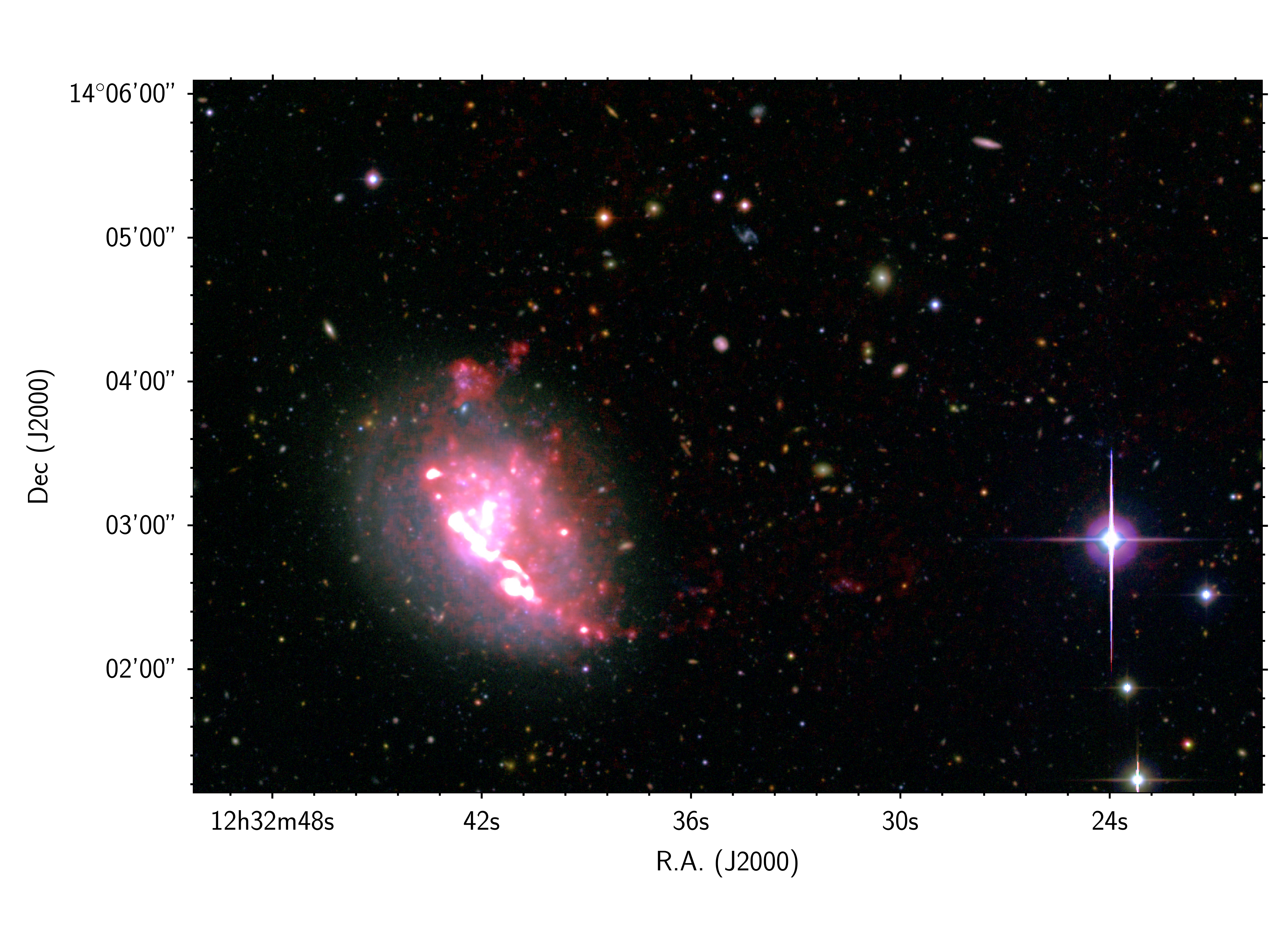}
   \caption{Left: Pseudo-colour image of IC 3476 obtained combining the NGVS (Ferrarese et al. 2012) optical $u$ and $g$
   in the blue channel, the $r$ and NB in the green, and the $i$ and the continuum-subtracted H$\alpha$ in the red. 
   Right: same figure produced using a H$\alpha$ continuum-subtracted frame smoothed to an angular resolution of 2.8\arcsec to highlight low surface brightness features. 
   The diffuse ionised gas is shown by the coherent emission seen in the red channel on the western side of the galaxy, while the granularity in the eastern side is noise.
 }
   \label{NGVS}%
   \end{figure*}

Far-IR data include \textit{Herschel} SPIRE (Ciesla et al. 2012) and PACS (Cortese et al. 2014)
at 100, 160, 250, 350, and 500 $\mu$m gathered during the \textit{Herschel} Reference Survey (HRS, Boselli et al. 2010)
and the \textit{Herschel} Virgo Cluster Survey (HeViCS, Davies et al. 2010), and \textit{Spitzer} data in the IRAC and MIPS bands from the programs: The \textit{Spitzer} IRAC Star Formation
Reference Survey (P.I. G. Fazio), The \textit{Spitzer} Survey of Stellar Structures in Galaxies (S4G, Sheth et al. 2010), 
and VIRGOFIR: a far-IR shallow survey of the Virgo cluster (P.I. D. Fadda).  
HI data from single dish observations gathered with the Arecibo radio telescope are also available (see Table \ref{gal}).

\section{Analysis}

\subsection{Broad-band imaging}

\begin{figure}
   \centering
   \includegraphics[width=0.49\textwidth]{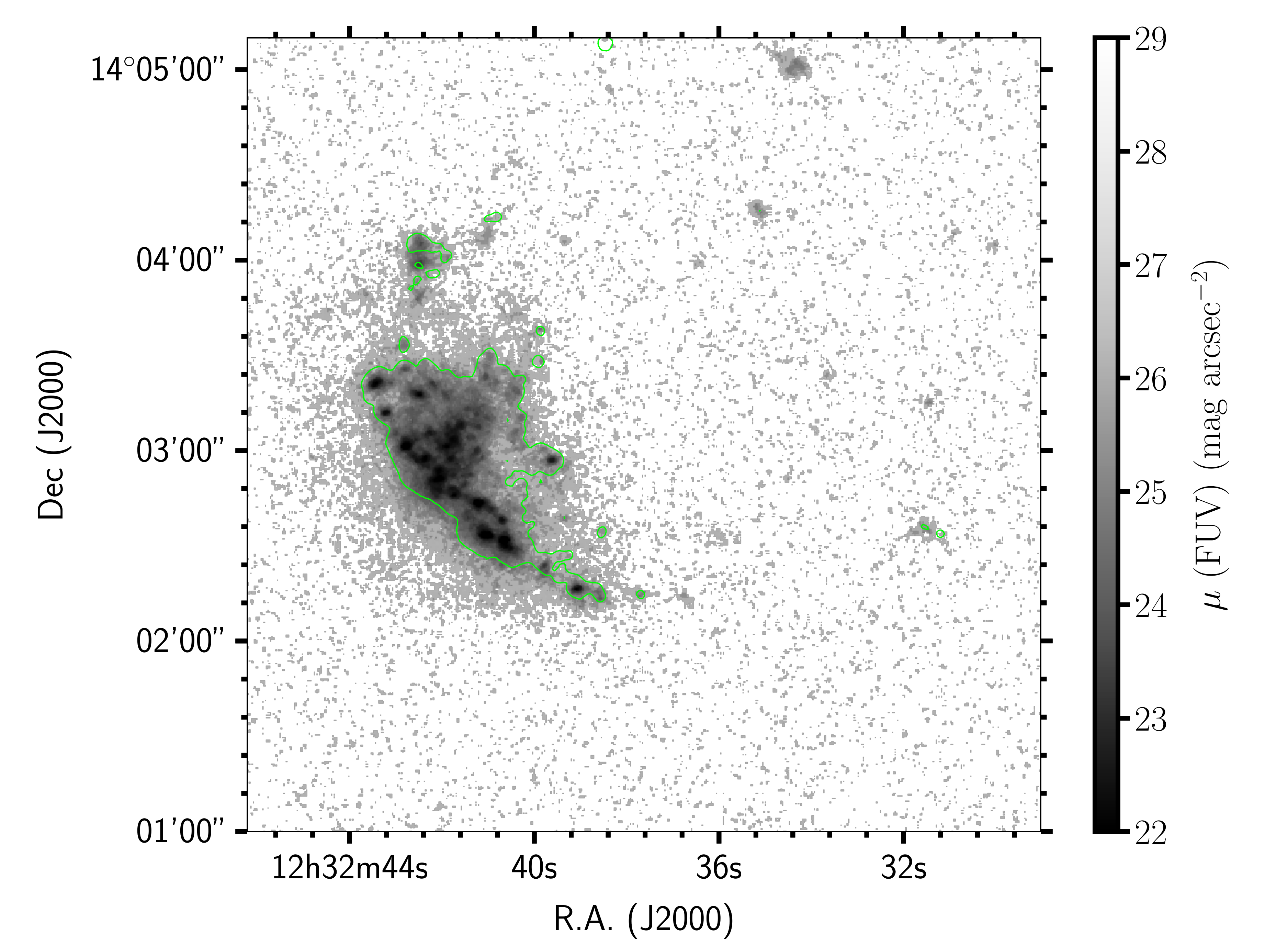}
   \caption{The FUV image of IC 3476 taken with ASTROSAT/UVIT in the BaF2 filter ($\lambda_c$ = 1541 \AA; $\Delta\lambda$ = 380 \AA). Contours show
   the VESTIGE H$\alpha$ emission at a surface brightness level of $\Sigma(H\alpha)$ = 3 $\times$ 10$^{-17}$ erg s$^{-1}$ cm$^{-2}$ arcsec$^{-2}$.
 }
   \label{UVIT}%
   \end{figure}

The colour image of IC 3476 constructed using a combination of NGVS and VESTIGE data (Fig. \ref{NGVS}) shows a star forming galaxy with an asymmetric morphology, 
with prominent blue compact regions located along a banana-shaped structure crossing the old stellar disc at its south-eastern extension 
from the south-west to the north-east. Similar blue regions, showing the dominance of a young stellar population,
are also present at the edge of this elongated structure. The exquisite quality of the CFHT data also shows a dust filament almost parallel to the elongated structure of young stars,
suggesting that the south-eastern edge of the galaxy is the one closer to the observer. 
The dominance of a young stellar population is also visible in the new ASTROSAT/UVIT FUV image (Fig. \ref{UVIT}), which shows a clumpy and elongated structure extending up to the outer edges of the stellar disc. The deep NGVS images do not show any 
low surface brightness extended and diffuse feature outside the stellar disc down to a surface brightness limit of $\mu(g)$ $\simeq$ 29 mag arcsec$^{-2}$,
ruling out any possible ongoing tidal interaction.

\subsection{Narrow-band imaging}

\begin{figure}
   \centering
   \includegraphics[width=0.5\textwidth]{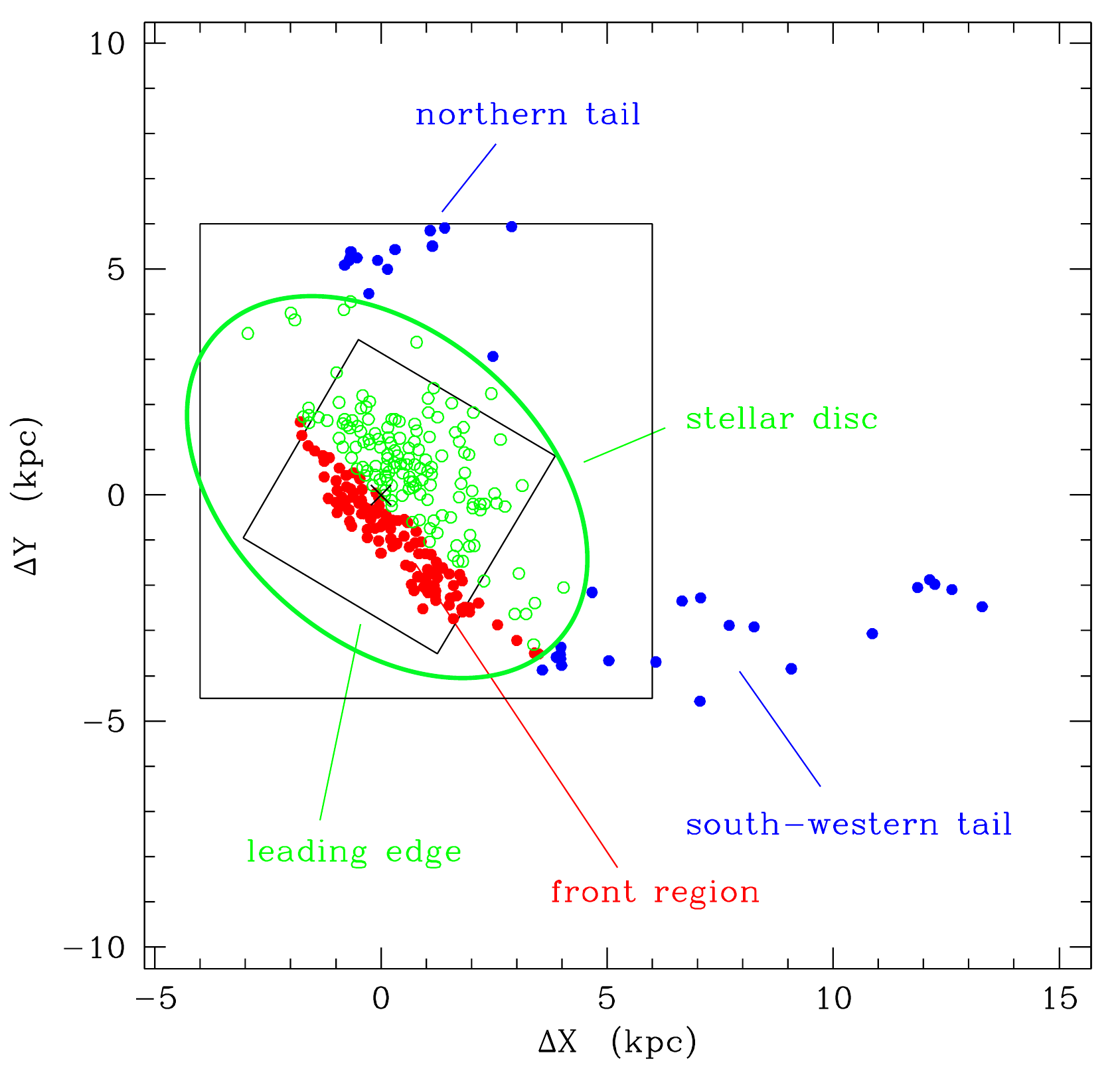}
   \caption{Distribution of the HII regions of IC 3476 with luminosity $L(H\alpha)$ $\geq$ 10$^{36}$ erg s$^{-1}$ and $S/N$ $>$5. Red filled circles are for HII regions in the front structure (101 objects),
   green empty circles for the other HII regions within the stellar disc (142), and
   blue filled symbols for HII regions located outside the stellar disc of the galaxy (35), whose extension is measured within the
   $r$-band isophotal radius at 24 mag arcsec$^{-2}$ (from Cortese et al. 2012, see Table 1). The black cross shows the position of the photometric nucleus.
   The large black box shows the region of the Fabry-Perot datacube analysed in this work, while the small black box the field-of view of MUSE IFU spectrograph, respectively.
 }
   \label{dist}%
   \end{figure}

The continuum-subtracted H$\alpha$ image is sensitive to the distribution of HII regions made by newly formed ($\lesssim$ 10 Myr) and massive ($M_\mathrm{star}$ $\geq$ 10 M$_{\odot}$) 
stars (Kennicutt 1998, Boselli et al. 2009). In IC 3476 the most luminous HII regions are located along the same banana-shaped structure in the south-eastern part of the disc,
crossing it from the south-west to the north-east up to its edges (Fig. \ref{Ha}). Other HII regions of lower luminosity are also 
present on the western side of the galaxy up to the edge of the stellar disc, while they are totally lacking in the eastern direction. This galaxy is also characterised by 
the presence of several HII regions located outside the stellar disc at the two edges of the elongated banana-shaped structure, at the north and at the south-west of the galaxy.
In the south-west they seem to follow a chain starting from the edge of the banana-shaped structure and extending up to $\sim$ 12 kpc in projected distance from the galaxy nucleus
($\sim$ 8 kpc from the edge of the stellar disc), where a complex of several HII regions is present. A few other regions are possibly detected at $\sim$ 17 kpc, 
as depicted in Fig. \ref{dist}.
All these regions are also present in the ASTROSAT/UVIT FUV image (Fig. \ref{UVIT}).

\subsection{Spectroscopy}

\subsubsection{Gas kinematics}

   \begin{figure*}
   \centering
   \includegraphics[width=1\textwidth]{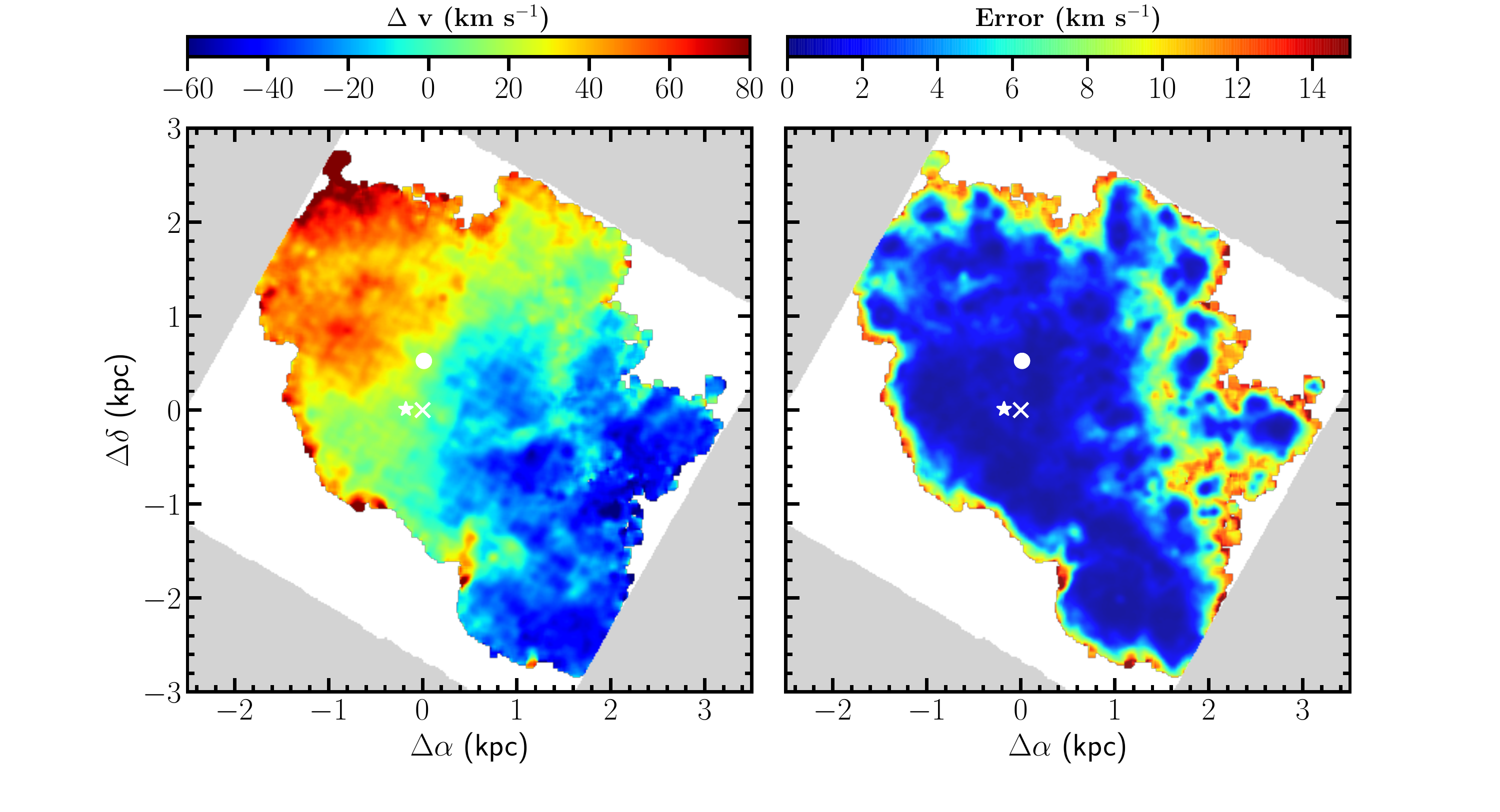}
   \caption{Velocity map (left) and associated error (right) of the ionised gas derived from MUSE for emission lines with a $S/N$ $>$ 5. The velocity of the gas is given relative to the systemic velocity of
   the galaxy of -170 km s$^{-1}$ derived from HI observations. The white cross, star, and filled circle show the position of the photometric, stellar and gas kinematical centres. 
   }
   \label{vel}%
   \end{figure*}


   \begin{figure*}
   \centering
   \includegraphics[width=1\textwidth]{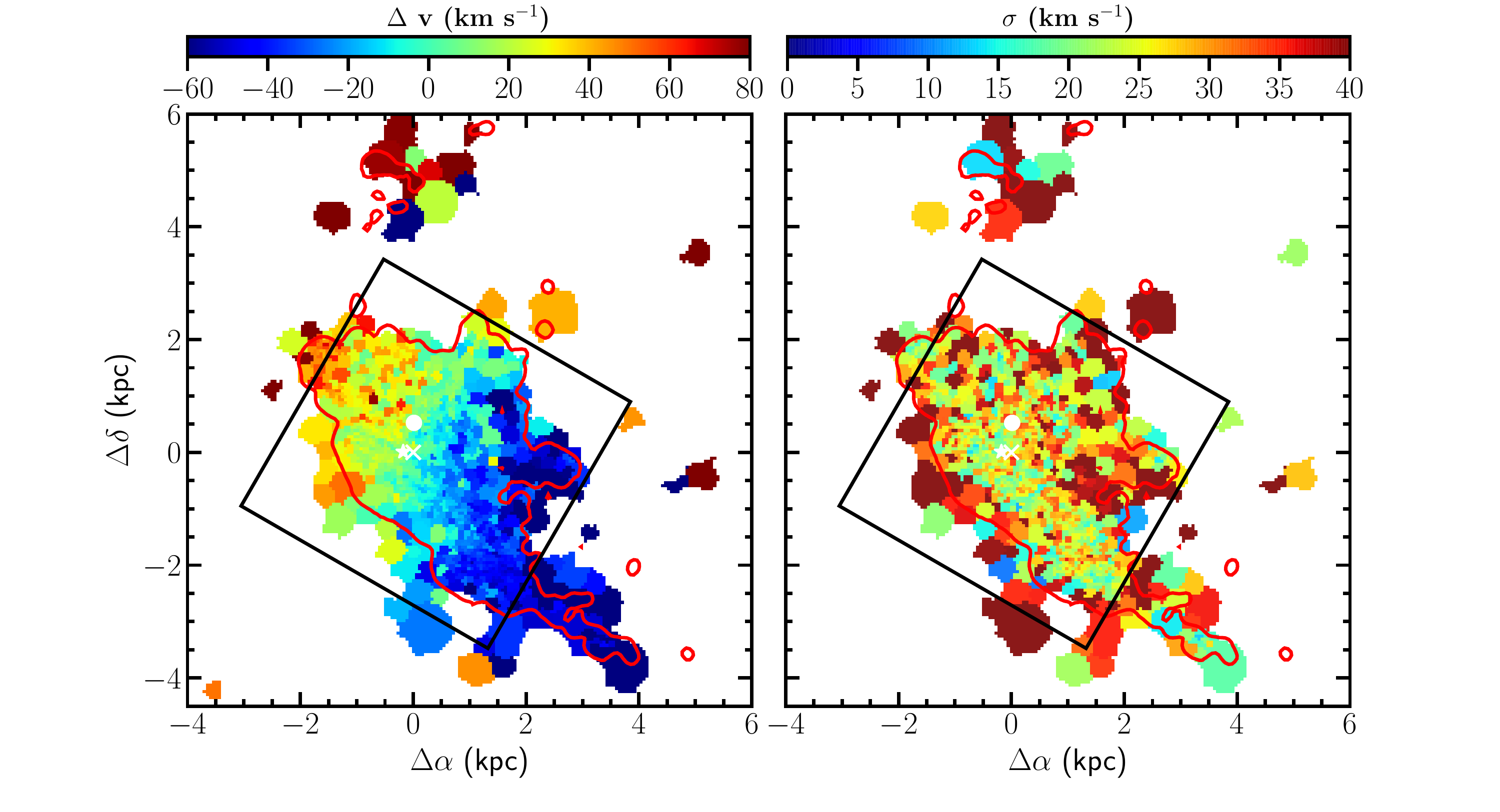}
   \caption{Velocity (left) and dispersion (right) maps of the ionised gas derived from the Fabry-Perot data.  The black box indicates the region covered by the MUSE exposure. 
   Red contours show the VESTIGE H$\alpha$ surface brightness at 3 $\times$ 10$^{-17}$ erg s$^{-1}$ cm$^{-2}$ arcsec$^{-2}$.
   The white cross, star, and filled circle show the position of the photometric, stellar and gas kinematical centres, respectively.
   As for the previous figures, the velocity of the gas is given relative to the systemic velocity of
   the galaxy of -170 km s$^{-1}$. 
 }
   \label{FP}%
   \end{figure*}
   
The 2D velocity field and the velocity dispersion maps of the galaxy derived for 
the gaseous component using the MUSE and Fabry-Perot data are shown in Fig. \ref{vel}, and \ref{FP}, respectively.
In order to determine the ionised gas kinematical centre, we have adjusted a rotating disc model to the MUSE data as in Epinat et al. (2008)\footnote{For this exercise we use the MUSE data
since we want to compare the kinematical properties of the gas to those of the stellar disc, derived using the same set of data.}. 
This model has a fixed centre, systemic velocity, position angle and inclination over the disc and it is built using a 
Courteau (1997) rotation curve (with $\beta=0$). The method used in this work\footnote{
\url{https://gitlab.lam.fr/bepinat/MocKinG}} has been improved with respect to the one of Epinat et al. (2008) 
so that uncertainties on velocity, flux distribution, and spatial resolution are taken into account. 
The kinematic centre, the position angle of the major axis, and the inclination of the disc model are left free to vary within a box of $24\times24$ square 
arcseconds centred on the coordinates of the morphological centre, between 20\degr\ and 60\degr\, and between 50\degr\ and 70\degr, respectively. 
Two $\chi^2$ minimisations were used: the Levenberg-Marquardt algorithm and the multinest bayesian approach (Feroz et al. 2009). 
While the first method can converge to local minima close the initial guesses, the second one is more robust in finding the absolute minimum. Both approaches led to similar results.
The model identified the kinematic centre of the gas at 6.6 arcseconds from the morphological centre (corresponding to $\sim$ 500 pc), as depicted in Fig. \ref{vel}, with a
position angle $P.A.$=54\degr, an inclination of $i$=52\degr, and a systemic velocity of $V_\mathrm{s}(gas)$=-159 km~s$^{-1}$.
Despite the ionised gas kinematics looking distorted, the axisymmetric model fits fairly well the velocity field 
with no strong residuals.

The mean kinematical properties of the ionised gas can also be compared to those of the HI gas as derived from the integrated spectrum given in Haynes et al. (2011). 
The width of the HI line measured at 50\%\ of the peak ($WHI_\mathrm{50}$ = 118 km s$^{-1}$) is comparable to the dynamic range observed in Fig. \ref{vel}. The HI spectrum
is not a two horns profile as expected for a galaxy of similar mass and inclination, but it is highly asymmetric, suggesting a lack of HI gas in the 
North-Eastern part of the disc (the galaxy has an HI-deficiency of $HI-def$ = 0.67, Boselli et al. 2014b). As noted by Haynes et al. (2011), however, 
the HI profile could be partly affected by the Galactic emission.

\subsubsection{Stellar kinematics}

   \begin{figure*}
   \centering
   \includegraphics[width=1\textwidth]{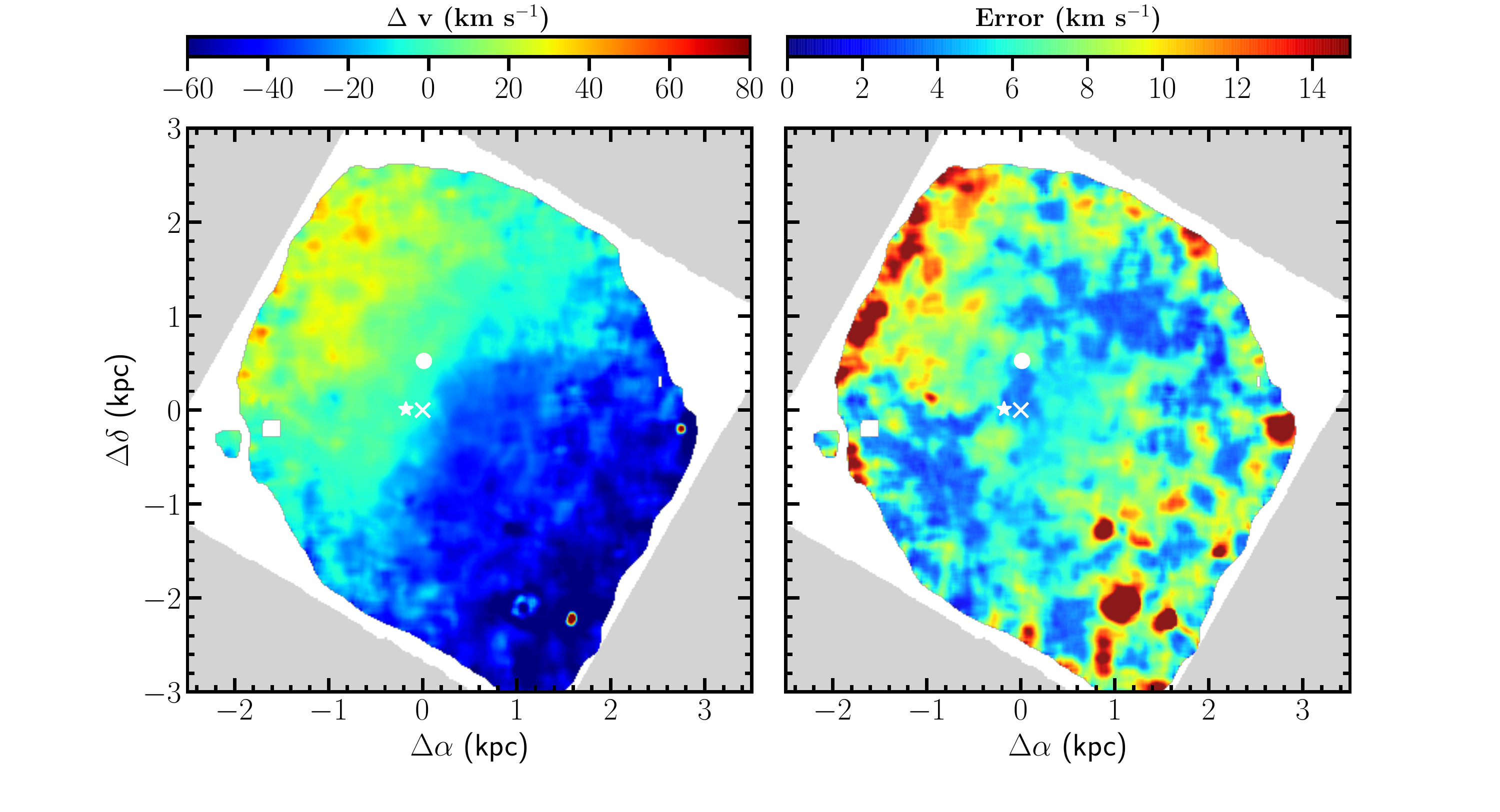}
   \caption{Velocity map (left) and associated error (right) of the stellar disc derived from MUSE for absorption lines with a $S/N$ $>$ 15. The velocity of the stars is given relative to the systemic velocity of
   the galaxy of -170 km s$^{-1}$ derived from HI observations. The white cross, star, and filled circle show the position of the photometric, stellar and gas kinematical centres,
   respectively.
   }
   \label{vel_star}%
   \end{figure*}

As for the ionised gas, we derived the stellar kinematical centre by fitting the same model to the stellar velocity field using the same constraints.
The stellar kinematical centre is located at 2.2 arcseconds ($\sim$ 180 pc) from the morphological centre, as shown in Fig. \ref{vel_star}, and has a position angle of $P.A.$=47\degr, an inclination of $i$=62\degr, 
and a systemic velocity of $V_\mathrm{s}(stars)$=-176 km~s$^{-1}$.
Both the position angles and inclinations of the gas and the stars are in quite good agreement, while their centres are separated by $\sim$ 600 pc. 
Whereas a different centre could induce a different systemic velocity, it seems clear that the observed difference in systemic velocity of the gas and the stars is not 
due to the different centres since an offset in velocity is observed in any place.

\subsubsection{Emission line properties}

\begin{figure}
   \centering
   \includegraphics[width=0.5\textwidth]{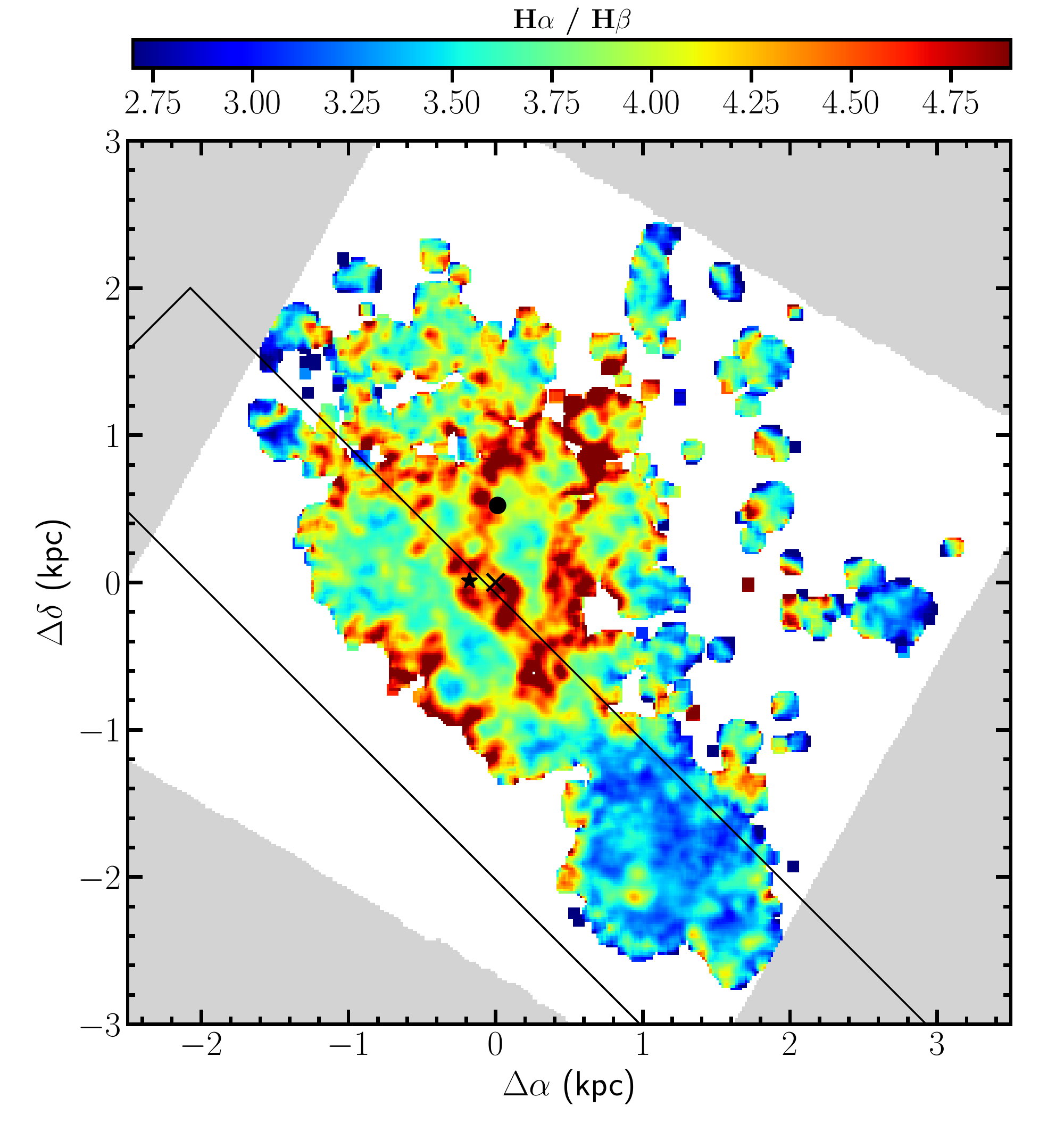}
   \caption{Distribution of the Balmer decrement derived using the H$\beta$ and H$\alpha$ lines with a $S/N$ $>$5. 
   The black cross, star, and filled circle show the position of the photometric, stellar and gas kinematical centres, respectively.
   The black rectangle indicates the front region of the galaxy (see Fig. \ref{dist}).
   }
   \label{BMD}%
   \end{figure}

The excellent quality of the MUSE data allows us to derive several important physical parameters of the ionised gas component of IC 3476.
First of all, they can be used to derive the dust attenuation within the gas using the Balmer decrement (Fig. \ref{BMD}). The MUSE data indicate that
the typical Balmer decrement of IC 3476 is H$\alpha$/H$\beta$ = 3.72$\pm$0.26 as measured in different regions over the disc of the galaxy,
corresponding to $A(H\alpha)$ = 0.68 mag adopting a Cardelli et al. (1989) extinction law, a value comparable to the one observed in galaxies of similar morphological type 
and stellar mass (Boselli et al. 2009, 2013). It also shows that the highest dust attenuation is observed along the dust features located in the central region
and in the NW disc visible 
in the optical image (Fig. \ref{NGVS}), as expected for an efficient screen located in between the emitting regions and the observer. It is interesting 
to note that the mean dust attenuation in the front region, where most of the bright HII regions are located, is $A(H\alpha)$ = 0.57 mag, 
and is thus lower than the mean attenuation of the galaxy. Part of the dust generally found in the most active star forming regions might have been 
displaced in the north-west direction during the interaction.

The [SII]$\lambda$6716/6731 \AA\ line ratio can be used to derive the 2D-distribution of the gas density. The typical value observed in IC 3476 is
[SII]$\lambda$6716/6731 $\simeq$ 1.4-1.5 all over the disc of the galaxy, which indicates that the electron density is $n_\mathrm{e}$ $\lesssim$ 30 cm$^{-3}$ 
(Osterbrock \& Ferland 2006, Proxauf et al. 2014). This upper limit is consistent with the
value derived from the H$\alpha$ emission of individual HII regions (see Sect. 3.5). 

\begin{figure}
   \centering
   \includegraphics[width=0.5\textwidth]{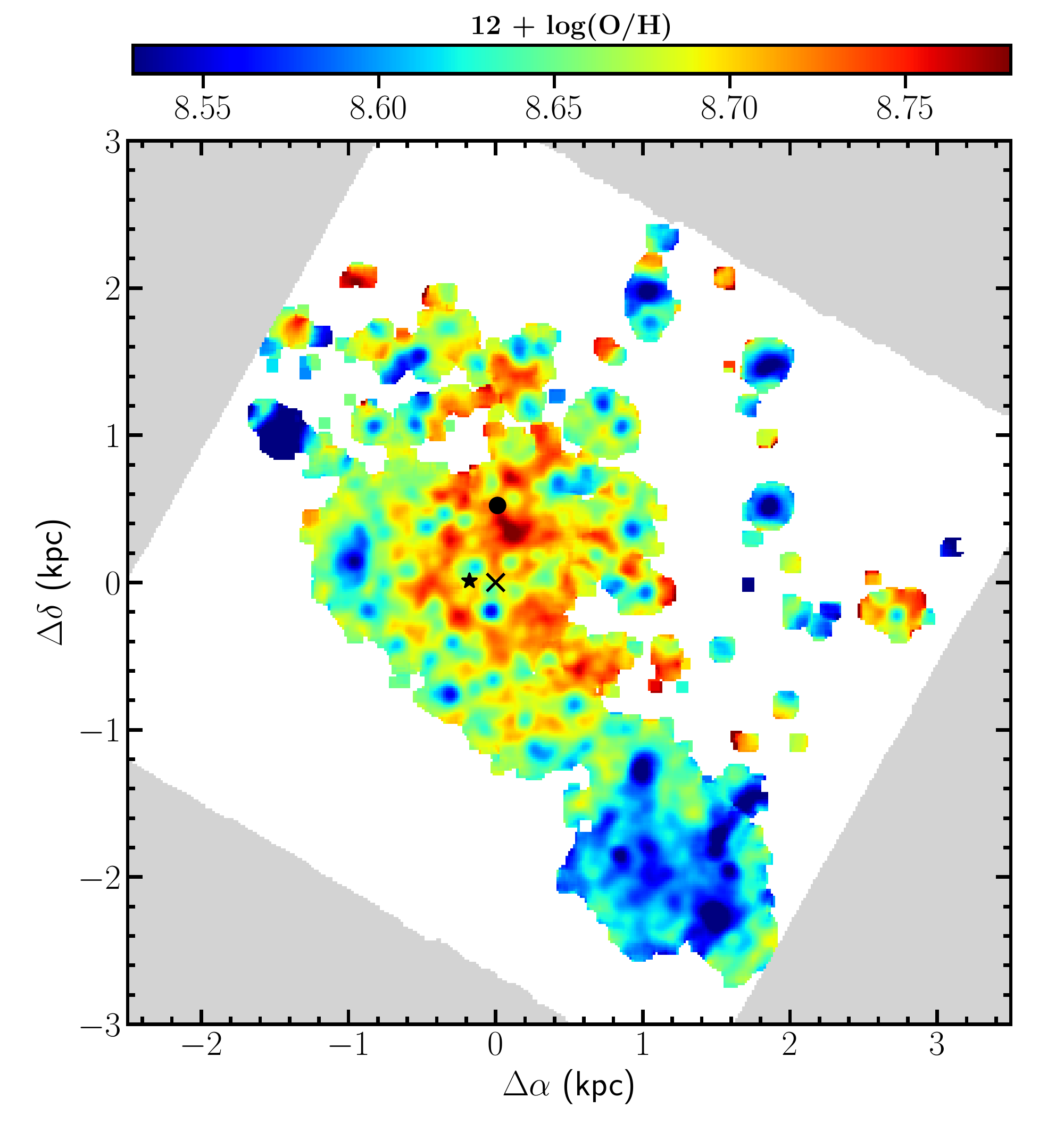}
   \caption{Distribution of the metallicity derived using the calibration of Curti et al. (2017) using the H$\beta$, [OIII], [NII], and H$\alpha$ lines with a $S/N$ $>$5. 
   The black cross, star, and filled circle show the position of the photometric, stellar and gas kinematical centres, respectively.
   }
   \label{metal}%
   \end{figure}

We also measure the metallicity of the gas using the calibration of Curti et al. (2017) based on the [OIII], H$\beta$, [NII], and H$\alpha$ lines and we obtain 
typical metallicities ranging from 12+log O/H $\simeq$ 8.70 in the inner disc down to 12+log O/H $\simeq$ 8.50 in the outer regions, consistent with the mean value 
derived by Hughes et al. (2013) using integrated spectroscopy (12+log O/H = 8.60$\pm$0.12, see Fig. \ref{metal}). Figure \ref{metal} also shows that the highest metallicity
is found close to the gas kinematical centre of the galaxy rather than at its morphological centre, consistently with the idea that all the different components of the ISM (gas, dust, metals) have been displaced
during the interaction of the galaxy with the surrounding environment.

\begin{figure*}
   \centering
   \includegraphics[width=0.8\textwidth]{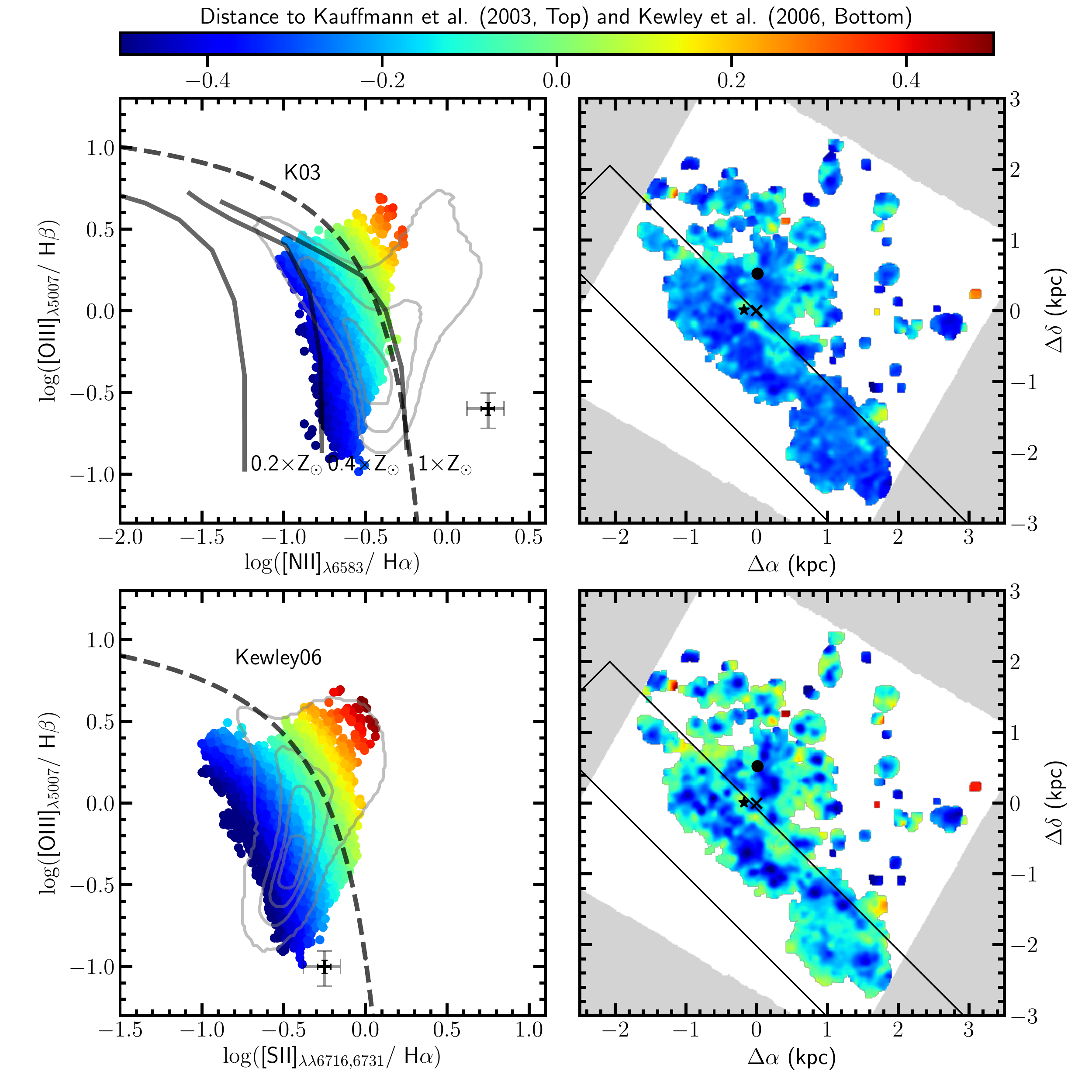}
   \caption{Left: [OIII]/H$\beta$ vs. [NII]/H$\alpha$ (top), and [OIII]/H$\beta$ vs. [SII]/H$\alpha$ (bottom) BPT diagrams for emission lines with a $S/N$ $>$ 5. 
   The dashed curves separate AGN from HII regions (Kauffmann et al. 2003; Kewley et al. 2001; Kewley et al. 2006). 
   Data are colour coded according to their minimum distance 
   from these curves. The black and grey crosses indicate the typical error on the data for lines with S/N $\simeq$ 15 and S/N $\simeq$ 5, respectively.
   The grey contours show the distribution of a random sample of nuclear spectra of SDSS galaxies in the redshift range 0.01 - 0.1 and 
   stellar masses 10$^9$ $\leq$ $M_\mathrm{star}$ $\leq$ 10$^{11}$ M$_{\odot}$. 
   The thick solid lines in the upper left panel show three different photo-ionisation models at different metallicities (0.2, 0.4, 1 Z$_{\odot}$; Kewley et al. 2001).
   Right: Map of the spaxel distribution colour-coded according to their position in the BPT diagram. The cross, star, and filled circle show the position of the photometric, stellar and gas kinematical centres,
   respectively. The black rectangle indicates the front region of the galaxy (see Fig. \ref{dist}).
   }
   \label{BPT}%
   \end{figure*}

The same set of spectroscopic data can be used to derive diagnostic diagrams such as the one proposed by Baldwin et al. (1981, BPT) useful to identify the dominant ionising source of the gas
over the disc of the galaxy. Figure \ref{BPT} shows the BPT diagrams done using the main emission lines detected in the spectrum (H$\beta$, [OIII]$\lambda$5007, 
[NII]$\lambda$6583, H$\alpha$, and [SII]$\lambda$6716,6731, with a $S/N$ $>$ 5). Despite the strong emission of the galaxy and the sensitivity of MUSE, the detection of the [OI]$\lambda$6300 
line is hampered by the contamination of a prominent atmospheric emission line. Figure \ref{BPT} shows that the gas is photo-ionised by young stars all over the disc,
including in the nucleus (both photometric and kinematical) where we do not see the presence of any hard ionising source (AGN). The [OIII]$\lambda$5007/H$\beta$
vs. [SII]$\lambda$6716,6731/H$\alpha$ BPT diagram (lower panels in Fig. \ref{BPT}) indicates that the contribution of stellar photoionisation is dominant on the peaks of the H$\alpha$ emission,
where the HII regions are located, while decreases in the diffuse gas, where the contribution of shocks becomes more important.

\subsection{SED fitting}

IC 3476 presents physical properties similar to those observed in other ram pressure stripped galaxies, i.e. a reduced star formation activity in the outer disc 
with starburst signatures in the inner regions. These properties can be used to quantitatively measure the typical timescales of the quenching phenomenon and of the starburst activity,
important parameters for the comparison of the observations with simulations. For this purpose we used the same methodology applied to NGC 4424 (Boselli et al. 2018b). We 
defined two representative regions within the disc of the galaxy, one located on the leading edge in the eastern part of the disc dominated by old stellar populations 
where star formation is totally lacking (quenched region), and one in the inner region where most of the brightest HII regions are located (starburst region on the front structure). 
We then extracted fluxes in 13 different photometric bands (ASTROSAT $FUV$, GALEX $NUV, u, g, r, i, z$, IRAC 3.6, 4.5, 5.7, 8.0 $\mu$m, 
MIPS 24$\mu$m, and PACS 100 $\mu$m) and in two pseudo-filters (H$\alpha$ in emission and H$\beta$ in absorption, 
as described in Boselli et al. 2016b) using the flux extraction procedure presented in Fossati et al. (2018). The two regions where fluxes are extracted have a typical size of $\sim$ 400
arcsec$^2$ and are thus fully resolved in all these photometric bands. The two selected regions are located within the MUSE field, we thus used the MUSE data to measure the H$\alpha$ flux in
emission uncontaminated by the [NII] lines and corrected for dust attenuation using the mean Balmer decrement within each region ($A(H\alpha)$ = 0.0 mag in the quenched region, 
$A(H\alpha)$ = 0.57 in the starburst region) and the age-sensitive H$\beta$ absorption line (Poggianti \& Barbaro 1997). The H$\beta$ absorption line in the pseudo-filter is measured using
the observed light-weight mean spectrum in the quenched region, while the one corrected for the line emission using the GANDALF code in the starburst region. 
We then fit the observed spectral energy distributions (SED) of these two regions using the CIGALE
fitting code (Noll et al. 2009, Boquien et al. 2019) coupled with the Bruzual \& Charlot (2003) stellar population models derived with a Chabrier IMF for the stellar continuum and the Draine \& Li
(2007) models for the dust emission, assuming a Calzetti attenuation law with 0 $\leq$ $E(B-V)$ $\leq$ 0.8 and a solar metallicity. To quantify the time elapsed since the beginning of the starburst phenomenon in the inner regions or that of the truncation of the star formation
activity in the outer disc we use a parametrised abruptly truncated star formation law characterised by the following free parameters: the rotational velocity of the galaxy (which 
mainly describes the secular evolution of an unperturbed rotating disc, see Boselli et al. 2016b), the quenching age $QA$ (time elapsed since the beginning of 
the quenching of the star formation activity
or of the starburst phase), and the quenching factor $QF$ ($QF$ = 0 for unperturbed SFR, $QF$ = 1 for totally quenched SFR, $QF$ = -1 for a 
SFR increased by a factor of 100\%).

  \begin{figure}
   \centering
   \includegraphics[width=0.5\textwidth]{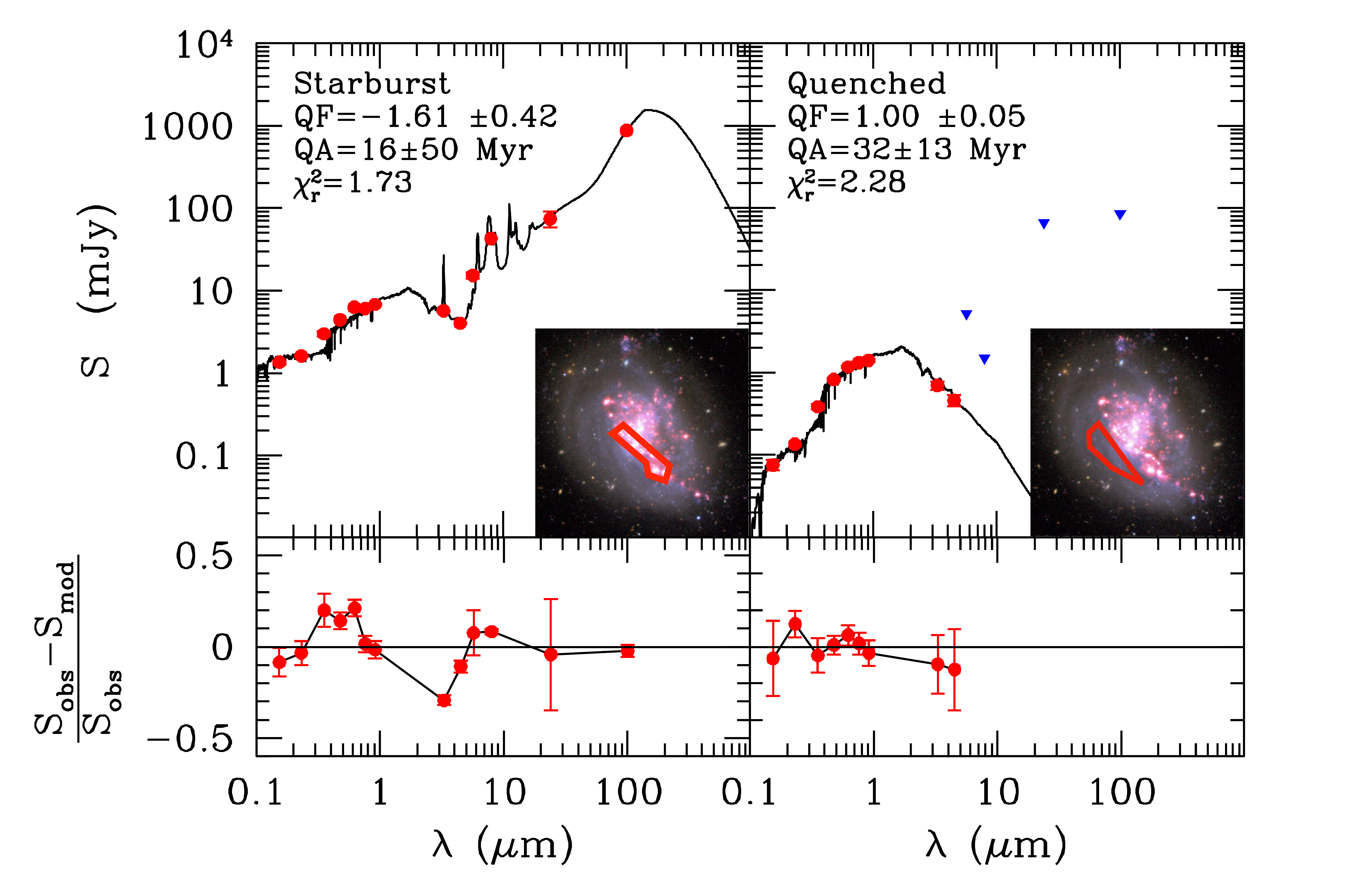}
   \caption{Far-UV to far-IR SED of the two selected regions within IC 3476. In each panel the observational data and their error bars
   are indicated with red filled dots, upper limits with blue triangles. The solid black line shows the best fit model derived with CIGALE. 
   The inset shows the colour image of the galaxy with overlayed in red the region analysed. 
 }
   \label{SED}%
   \end{figure}

  \begin{figure}
   \centering
   \includegraphics[width=0.5\textwidth]{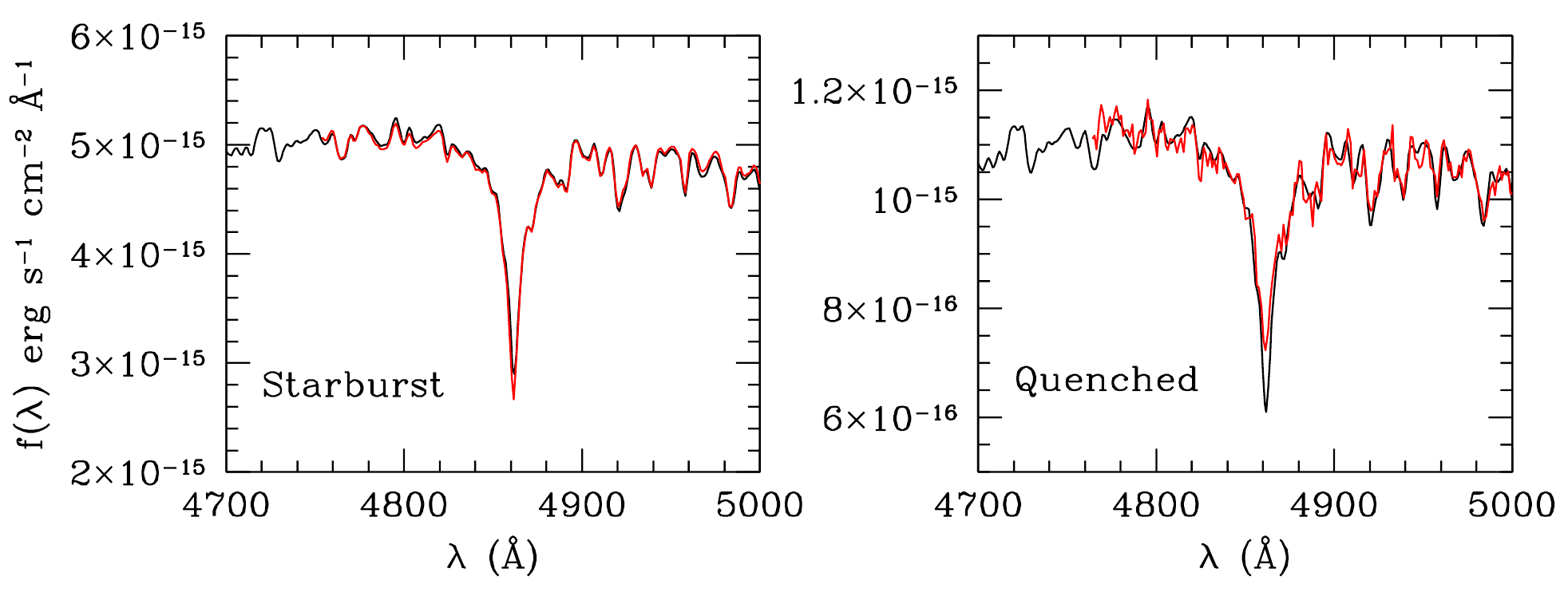}
   \caption{Best model fit of the stellar continuum obtained with CIGALE (black) is compared to the MUSE spectrum (red) in the H$\beta$ line
   after the emission is removed using GANDALF for the starburst (left) and quenched (right) regions. The best model has been scaled on the Y-axis to match the observed spectrum.
 }
   \label{spettro}%
   \end{figure}

The results of the fit are shown in Fig. \ref{SED} and Fig. \ref{spettro}. The quality of the fit is excellent in both regions and
indicates that both the quenching and the starburst episodes are very recent ($QA$ = 32 $\pm$ 12 Myr for the quenched region,
$QA$ = 16 $\pm$ 50 Myr for the starburst region)\footnote{A recent episode of star formation is also consistent with the age of the supernova 1970A (R.A.(J2000): 12:32:40.64, Dec:
+14:02:38.2; Age = 7.18 Myr; Lennarz et al. 2012; Kuncarayakti et al. 2018) which is located within the front starburst region.}. The quenching episode totally stopped the star formation activity in the outer disc
at the front edge of the galaxy ($QF$ = 1) while the star formation activity increased by $\sim$ 161 \%\ in the inner regions. Similar $QF$ and slightly longer
timescales ($QA$ = 103 $\pm$ 30 Myr for the quenched region, $QA$ = 40 $\pm$ 92 Myr for the starburst region) are 
obtained using the smooth quenching episode described in Boselli et al. (2016b). We further checked the robustness of these results by applying the
SED fitting procedure described in Fossati et al. (2018) based on a Monte Carlo Spectro-Photometric technique, which combines both the broad-band photometric points and the 
whole spectrum obtained with MUSE in the fit. Using a similar parametrisation for the description of the secular evolution of the star formation history of the galaxy, and applying an exponentially 
declining star formation model to describe the quenching episode, the fit gives $Q_{Age}$ = 55 $\pm$ 10 Myr for the time elapsed since the beginning of the perturbation and $\tau_Q$ = 5 $\pm$ 1 Myr
for the typical timescale for the decrease of the activity, two values very close to those obtained with CIGALE for an abrupt truncation. We can thus confidently claim that 
the leading edge of the galaxy rapidly quenched its star formation activity $\lesssim$ 100 Myr ago. 
This quenching episode was followed by a starburst activity in the inner regions.

\subsection{Properties of the HII regions}

We use the \textsc{HIIphot} data reduction pipeline (Thilker et al. 2000) to derive the physical properties of individual HII regions inside the
disc of the perturbed galaxy and in the tails of stripped material. This code, which uses a recognition technique based on an iterative 
growing procedure to identify single HII regions over the emission of a varying stellar continuum, is optimised to extract the properties of these 
compact structures from NB imaging data. We refer the reader to Thilker et al. (2000), Scoville et al. (2001),
Helmboldt et al. (2005), Azimlu et al. (2011), Lee et al. (2011), and Liu et al. (2013) for an accurate description and for the utilisation of this code
for this purpose. As described in Boselli et al. (2020), we run this code jointly on the H$\alpha$ continuum-subtracted, the H$\alpha$ NB, and 
the stellar continuum image, the last derived as explained in Sect. 2.1. 

Thanks to the excellent quality of the VESTIGE data in terms of sensitivity and angular resolution, the \textsc{HIIphot} code detects HII 
regions down to luminosities $L(H\alpha)$ $\simeq$ 10$^{36}$ erg s$^{-1}$ and equivalent radii $r_{eq}(H\alpha)$ $\simeq$ 40 pc, defined as in Helmboldt et al. (2005), i.e. the radii 
of the circles of surface equivalent to the area of the detected HII region down to a surface brightness limit of $\Sigma(H\alpha)$ = 3 $\times$ 10$^{-17}$ erg s$^{-1}$ cm$^{-2}$ arcsec$^{-2}$. Equivalent 
radii and diameters are corrected for the effects of the point-spread function (PSF) following Helmboldt et al. (2005). At these low emission levels where crowdness becomes important 
the code might suffer for incompleteness (e.g. Pleuss et al. 2000; Bradley et al. 2006). We thus limit the present analysis to a comparative analysis of the physical 
properties of HII regions located inside and outside the stellar disc, in the stripped material, and detected with a signal-to-noise $S/N$ $>$ 5, avoiding any 
statistical analysis which would require a complete sample. We expect that this comparative analysis, which is based on data extracted from the same images, 
does not suffer from strong systematic biases. We recall, however, that the most crowded regions are probably more frequent within the disc of IC 3476, 
where also the stellar continuum might be dominant, while totally lacking in the stripped tail. We also notice that the two BPT diagrams show a few HII regions 
with Seyfert-like spectra (see Fig. \ref{BPT}). These, however, are only $<$ 0.2 \%\ and $<$ 0.5 \%\ of the spaxels, their contamination in the analysis
presented in the next sections is thus negligible.

   \begin{figure*}
   \centering
   \includegraphics[width=0.99\textwidth]{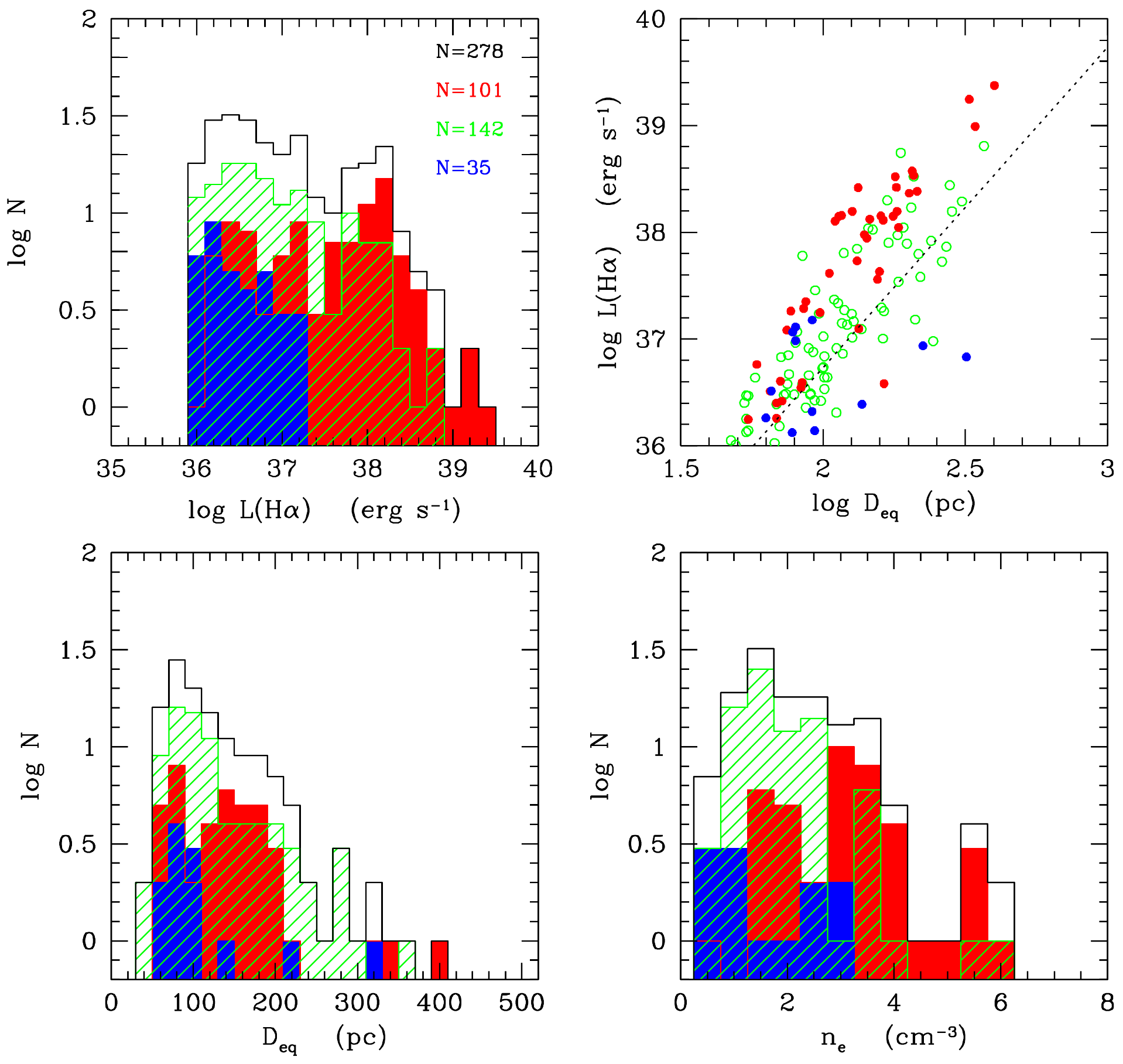}
   \caption{Properties of the HII regions of IC 3476 with luminosity $L(H\alpha)$ $\geq$ 10$^{36}$ erg s$^{-1}$ and $S/N$ $>$5. 
   Upper left: H$\alpha$ luminosity function. Upper right: relationship between the H$\alpha$ luminosity and the equivalent diameter.
   Lower left: distribution of the equivalent diameter. Lower right: distribution of the electron density. Black histograms are for all HII regions,
   red and blue filled histograms and symbols are for HII regions located on the front region within the disc and for HII regions in the tails
   outside the disc, respectively; green hased histograms and empty symbols for HII regions within the stellar disc but outside the front structure. 
   All H$\alpha$ luminosities are corrected for [NII] contamination using the mean [NII]/H$\alpha$ 
   ratio derived within the MUSE field, [NII]/H$\alpha$ = 0.29. Electron densities are also corrected for Balmer decrement ($A(H\alpha)$ = 0.68 mag) 
   for comparison with other works. The dotted line in the upper right panel shows the $n_\mathrm{e}$ = 1 cm$^{-3}$ relation.
   Equivalent diameters and electron densities are plotted only for those regions where the correction for the effect of the PSF is less
   than 50\%.
 }
   \label{HII}%
   \end{figure*}

\subsubsection{Physical properties}

We first identify the galaxy HII regions as those located within the $r$-band isophotal radius measured at 24.0 mag arcsec$^{-2}$ by Cortese et al. (2012; $r_\mathrm{ISO}(r)$ = 64.5\arcsec), 
and assuming $e$ = 0.29, and $P.A.$ = 45 deg. (from North, counterclockwise, where $e$ and $P.A.$ are the ellipticity and the position angle of the corresponding elliptical profile)\footnote{Ellipticity and position angle 
slightly differ from those given in Cortese et al. (2012) since are measured on the deep $r$-band CFHT image which reveals a low surface brightness extended disc undetected in the shallow SDSS image. We stress, however, that
the analysis presented in this section only barely depends on these parameters which are taken here only to geometrically identify different regions within and outside the galaxy.}. 
Those oustide the stellar disc are mainly located along the tails of stripped gas
present in the continuum-subtracted H$\alpha$ image (Fig. \ref{Ha}). \textsc{HIIphot} detects 278 HII regions with $L(H\alpha)$ $\geq$ 10$^{36}$ erg s$^{-1}$ and $S/N$ $>$5 associated to the galaxy, and 35 
outside the stellar disc, as depicted in Fig. \ref{dist} (blue filled symbols). The HII regions located within the stellar disc (green ellipse in Fig. \ref{dist}) can be further 
divided into those in the front region (red filled symbols, located at the west of a line crossing the galaxy at 0.67\arcsec south from its nucleus with a $P.A.$ = 45 deg. measured counterclockwise from the North, 101 objects) 
and those in the back of the disc (green empty symbols, 142 objects).
For these HII regions we derive the H$\alpha$ luminosity function (where 
for comparison with other works $L(H\alpha)$ is corrected only for Galactic extinction and [NII] contamination assuming the mean value
derived from the MUSE data, [NII]/H$\alpha$ = 0.29), the $L(H\alpha)$ vs. size relation (where $D_\mathrm{eq}$ is the equivalent diameter), the size distribution and the 
mean electron density $n_\mathrm{e}$ distribution (Fig. \ref{HII}). The mean electron density $n_\mathrm{e}$ is derived following Scoville et al. (2001) with the relation (case B recombination, from Osterbrock \& Ferland 2006):

\begin{equation}
n_\mathrm{e} = 43\bigg[\frac{(L_{cor}(H\alpha)/10^{37} ~\rm{erg~s^{-1}})(T/10^4 ~\rm{K})^{0.91}}{(D_{eq}/10 ~\rm{pc})^3}\bigg]^{1/2} ~~[\rm{cm^{-3}}]
\end{equation} 

\noindent
where $L_{cor}(H\alpha)$ is the H$\alpha$ luminosity of the individual HII regions corrected for [NII] contamination (see above) and dust attenuation assuming the mean value derived 
from the MUSE IFU data ($A(H\alpha)$ = 0.68 mag), and $T$ the gas temperature (here assumed to be $T$ = 10000 K). Equivalent diameters and electron densities 
are plotted only for those regions where the correction for the effect of the PSF is less than 50\%. 
We notice that the typical values of the electron densities 
derived using Eq. 1 ($n_\mathrm{e}$ $\lesssim$ 8 cm$^{-3}$) are consistent with the upper limit derived using the [SII] doublet ratio using the MUSE data ($n_\mathrm{e}$ $\lesssim$ 30 cm$^{-3}$).

The analysis of Figures \ref{dist} and \ref{HII} indicates that:

\noindent
1) The most luminous HII regions ($L(H\alpha)$ $\geq$ 10$^{38}$ erg s$^{-1}$) are present only within the disc of the galaxy and most of them are located
along the front region crossing the disc, suggesting that they have been probably formed during the interaction of the galaxy ISM with the surrounding intracluster medium (ICM) through ram pressure.

\noindent
2) The HII regions located along the tails of stripped material have typical luminosities $L(H\alpha)$ $\lesssim$ 10$^{37}$ erg s$^{-1}$, most of them sizes $D_\mathrm{eq}$ $\lesssim$ 100 pc, and 
electron densities $n_\mathrm{e}$ $\lesssim$ 3 cm$^{-3}$, corresponding to the faint end distribution of the values found in the HII regions
located within the stellar disc of the galaxy. 
We caution, however, that the parameters derived for the HII regions with the lowest luminosities per unit size might be biased since \textsc{HIIphot} is optimised to identify and measure the emission of HII regions 
located over a strong stellar continuum emission, which is mainly lacking in the tails of stripped material.

\noindent

3) The HII regions located on the front of the galaxy, on the contrary, are on average more luminous per unit size than those located on the back of the disc. 
This effect is mainly evident in the most luminous ($L(H\alpha)$ $\geq$ 10$^{37.5}$ erg s$^{-1}$) and extended ($D_\mathrm{eq}$ $\geq$ 250 pc) HII regions, where $L(H\alpha)$ is
a factor of $\simeq$ 2.5 higher than measured in objects of similar size elsewhere in the disc. This results in a systematic difference in the mean electron density of the HII regions in the front
($n_\mathrm{e}$ = 3.1 $\pm$ 1.3 cm$^{-3}$), in the back of the galaxy ($n_\mathrm{e}$ = 1.9 $\pm$ 0.9 cm$^{-3}$), and in the tail ($n_\mathrm{e}$ = 1.6 $\pm$ 1.0 cm$^{-3}$, where $\pm$ gives 1-$\sigma$ of the distribution).
A Kolmogorov-Smirnov test indicates that the probability that the three distributions are drawn from the same parent population is $P$ = 1.9\%\ ($n_\mathrm{e}(tail)$ vs. $n_\mathrm{e}(front~region)$), 
$P$ = 5.2\%\ ($n_\mathrm{e}(disc)$ vs. $n_\mathrm{e}(front~region)$), and $P$ = 4.6\%\ ($n_\mathrm{e}(disc)$ vs. $n_\mathrm{e}(tail)$). Comparable differences
are obtained if more stringent criteria on the PSF correction are adopted. It must be noticed that an increase of the electron density might artificially result from an underestimate of the size of the HII regions in the crowded regions.
This effect, however, should be negligible because the distribution in size within the crowded, front HII regions is not biased vs. small objects compared to that of the other HII regions in the disc probably because
of the adopted high surface brightness level for the identification of the equivalent isophotal diameters. This is also evident for the HII regions in the tail,
which have the smallest sizes despite 
their very low aggregation. This result is also robust vs. the adopted correction for [NII] contamination and dust attenuation since similar differences are seen when the VESTIGE NB imaging data are corrected 
using a 2D map derived from the fully resolved MUSE data where these are available (thus only on the stellar disc of the galaxy). 

\begin{figure}
   \centering
   \includegraphics[width=0.5\textwidth]{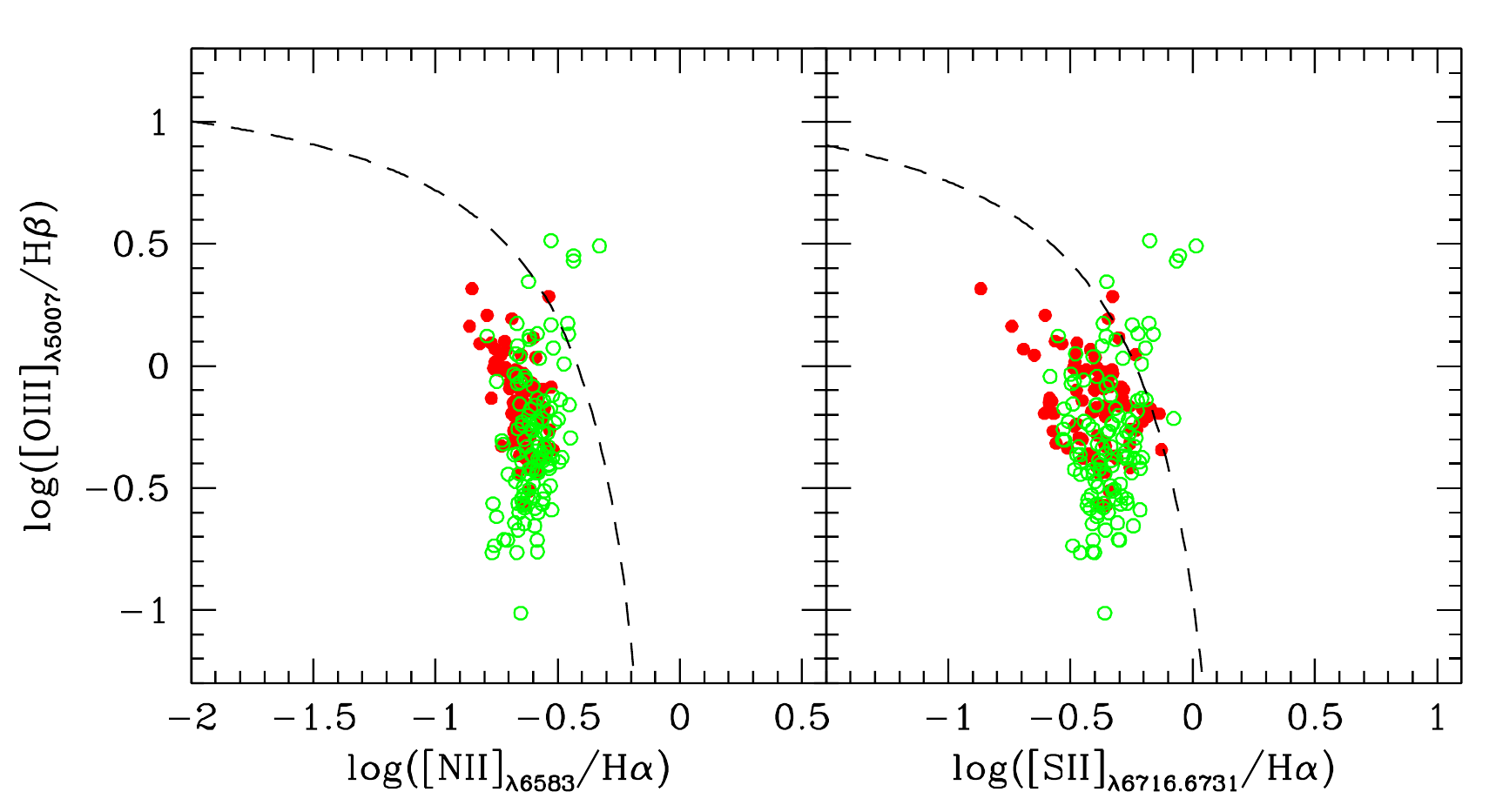}
   \caption{[OIII]/H$\beta$ vs. [NII]/H$\alpha$ (left), and [OIII]/H$\beta$ vs. [SII]/H$\alpha$ (right) BPT diagrams for the HII regions located in the front region (red filled dots)
   and within the galaxy disc (green empty circles) as depicted in Fig. \ref{dist}. 
   The dashed curves separate AGN from HII regions (Kauffmann et al. 2003; Kewley et al. 2001; Kewley et al. 2006).
 }
   \label{BPTHII}%
   \end{figure}

The MUSE data also indicate that the mean 
[OIII]$\lambda$5007/H$\beta$ measured within the front HII regions ([OIII]$\lambda$5007/H$\beta$ = 0.77$\pm$0.35) is slightly higher than that measured on the other HII regions in the disc 
([OIII]$\lambda$5007/H$\beta$ = 0.64$\pm$0.56) (Fig. \ref{BPTHII})\footnote{A Kolmogorov-Sminrov test indicates that the probability that the two distributions 
are driven by the same parent distribution is $P$ $=$ 0.1\%.}. Since the ionisation energy of [OIII] is 54.9 eV, this suggests 
that in the front HII regions the radiation is harder and the ionising stars are possibly slightly younger than in the other regions (Citro et al. 2017), although part of this difference
could also be due to a difference in the mean metallicity of the two regions.
A younger age of these HII regions is also suggested by their smaller size per unit luminosity than the other disc HII regions (e.g. Ambrocio-Cruz et al. 2016, Boselli et al. 2018c).

The typical age of the star forming complexes in the tail of the galaxy can be measured using their H$\alpha$ and FUV emission extracted within given apertures as done in the tail of NGC 4254 
by Boselli et al. (2018c). We refer the reader to that work for a detailed description of the methodology. For this purpose, we chose apertures optimised to match the angular resolution of the UVIT image
($>$ 1.5\arcsec diameter) and extracted fluxes within the FUV and continuum-subtracted H$\alpha$ images using the same procedure described in Sect. 3.4. We then corrected H$\alpha$  
for [NII] contamination assuming the mean value measured within the galaxy disc ([NII]/H$\alpha$ = 0.29) and converted it into Lyman continuum fluxes using the relation (Boselli et al. 2016b):

\begin{equation}
{LyC~\mathrm{[mJy]} = \frac{1.07\times10^{-37}~[L(H\alpha) / \mathrm{erg\,s}^{-1}]}{[D / \mathrm{Mpc}]^2}}
\end{equation}

\noindent
and then into a FUV-H$\alpha$ colour index (where $FUV$ is in AB mag):

\begin{equation}
{H\alpha-FUV = -2.5 \log(LyC) + 20 - FUV}
\end{equation}

\begin{figure}
   \centering
   \includegraphics[width=0.5\textwidth]{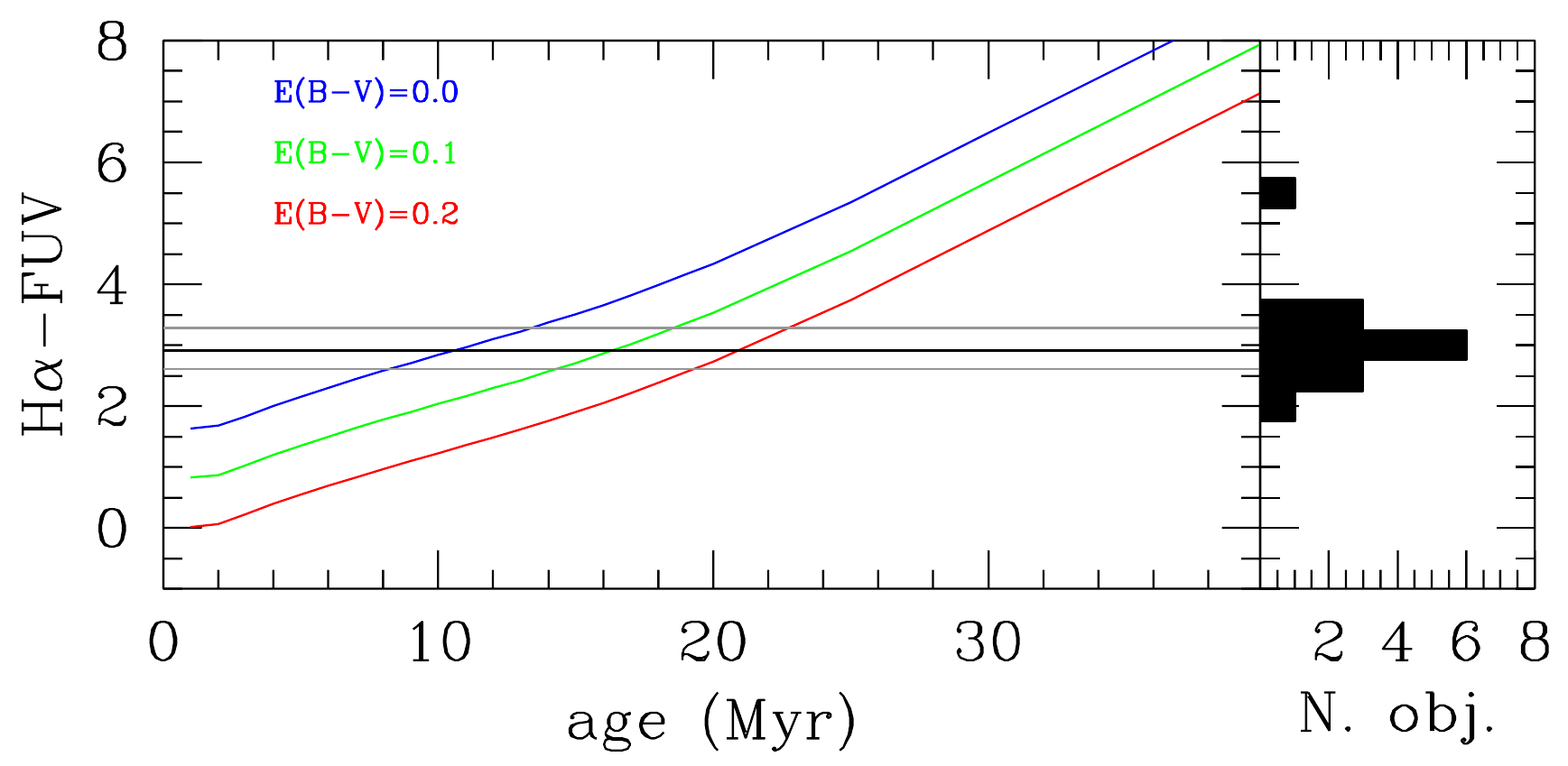}
   \caption{Left panel: variations of the synthetic $H\alpha$-$FUV$ age-sensitive colour index as a function of time derived for the star formation history given in Eq. 4.
   Different colours are used for three different dust attenuations: $E(B-V)$ = 0.0 (blue), 0.1 (green), and 0.2 (red). 
   The black solid line shows the median of the observed colour distribution of the star complexes in the tail of IC 3476, the 
   grey lines the 16\%\ and the 84\%\ quartiles of their distribution.
   Right panel: observed distribution of the $H\alpha$-$FUV$ colour index.
 }
   \label{age}%
   \end{figure}

\noindent
We then compared these observed colours to the synthetic colours of HII regions derived assuming a star formation history defined as:

\begin{equation}
{\mathrm{SFR}(t) = \epsilon\,t\,\mathrm{e}^{-t/\tau}  ~~~~~\rm{M_{\odot} yr^{-1}}}
\end{equation}

\noindent
where the e-folding time $\tau$ is set to 3 Myr, consistent with the typical age of giant 
($L(H\alpha)$ $\simeq$ 10$^{37}$ erg s$^{-1}$) HII regions in the Milky Way
or in other nearby galaxies (Copetti et al. 1985; Tremblin et al. 2014). The synthetic colours of HII regions have been
derived using the CIGALE SED fitting code, assuming Bruzual \& Charlot (2003) population synthesis models with a Salpeter IMF, and adopting three different mean attenuations ($E(B-V)$ = 0.0, 0.1, 0.2).
Attenuations lower than those measured within the stellar disc are expected since these external HII regions have been formed within the metal-poor gas mainly stripped from the outer disc,
as indeed observed in the tail of jellyfish galaxies (e.g. Fossati et al. 2016, Poggianti et al. 2019). 
The comparison between synthetic and observed colours given in Fig. \ref{age} clearly shows that the typical age of the star forming complexes in the tail is $\simeq$ 20 Myr, and reaches 30 Myr
only in one region located close to the galaxy\footnote{This result is robust vs. the assumptions on dust attenuation: as shown in Fig. \ref{age}, an increase of $E(B-V)$ by a factor of $\sim$ 0.1 implies 
an increase of the mean age of the stellar population by $\sim$ 5 Myr.}.
These ages are slightly younger than those measured in the tail of NGC 4254 ($\lesssim$ 100 Myr), where the star forming complexes 
have been selected in the NUV band. The age of the HII regions in the tails is comparable to that of the inner starburst. It is also fairly comparable 
to the time necessary for the galaxy to travel $\sim$ 15 kpc on the plane of the sky ($\sim$ 10 Myr),
the projected length of the tails, at $\sim$ 1500 km s$^{-1}$ = $\sqrt{2}\times \sigma_{Virgo}$, where $\sigma_{Virgo}$ is the typical line of sight velocity dispersion 
of star forming systems within the Virgo cluster ($\sigma_{Virgo}$ $\sim$ 1000 km s$^{-1}$; Boselli et al. 2014a). It is thus
conceivable
that these external HII regions have been formed at the turbulent edges of the galaxy during its interaction with the surrounding medium and stripped away as collapsing gas in giant molecular clouds.

\subsubsection{Kinematical properties}

We compare the physical properties of the HII regions derived in the previous section to their kinematical properties determined using the spectroscopic data. For this purpose 
we use the Fabry-Perot data which have the required spectral resolution ($R$ $\simeq$ 10000) to measure velocity dispersions down to $\sim$ 13 km s$^{-1}$. As indicated in the Appendix, the 
spectral resolution of MUSE (R $\simeq$ 2600, corresponding to a limit in the velocity dispersion of $\sigma$ $\sim$ 50 km s$^{-1}$), is
not sufficient for this purpose. 
To minimise the possible confusion due to the lower angular resolution of the Fabry-Perot data with respect to the imaging data, we limit this analysis to the regions 
where the $S/N$ is sufficiently high and the velocity dispersion is uncontaminated ($\leq$ 10\%) by the adjacent HII regions or the diffuse background emission as indicated in the Appendix. 

   \begin{figure*}
   \centering
   \includegraphics[width=0.99\textwidth]{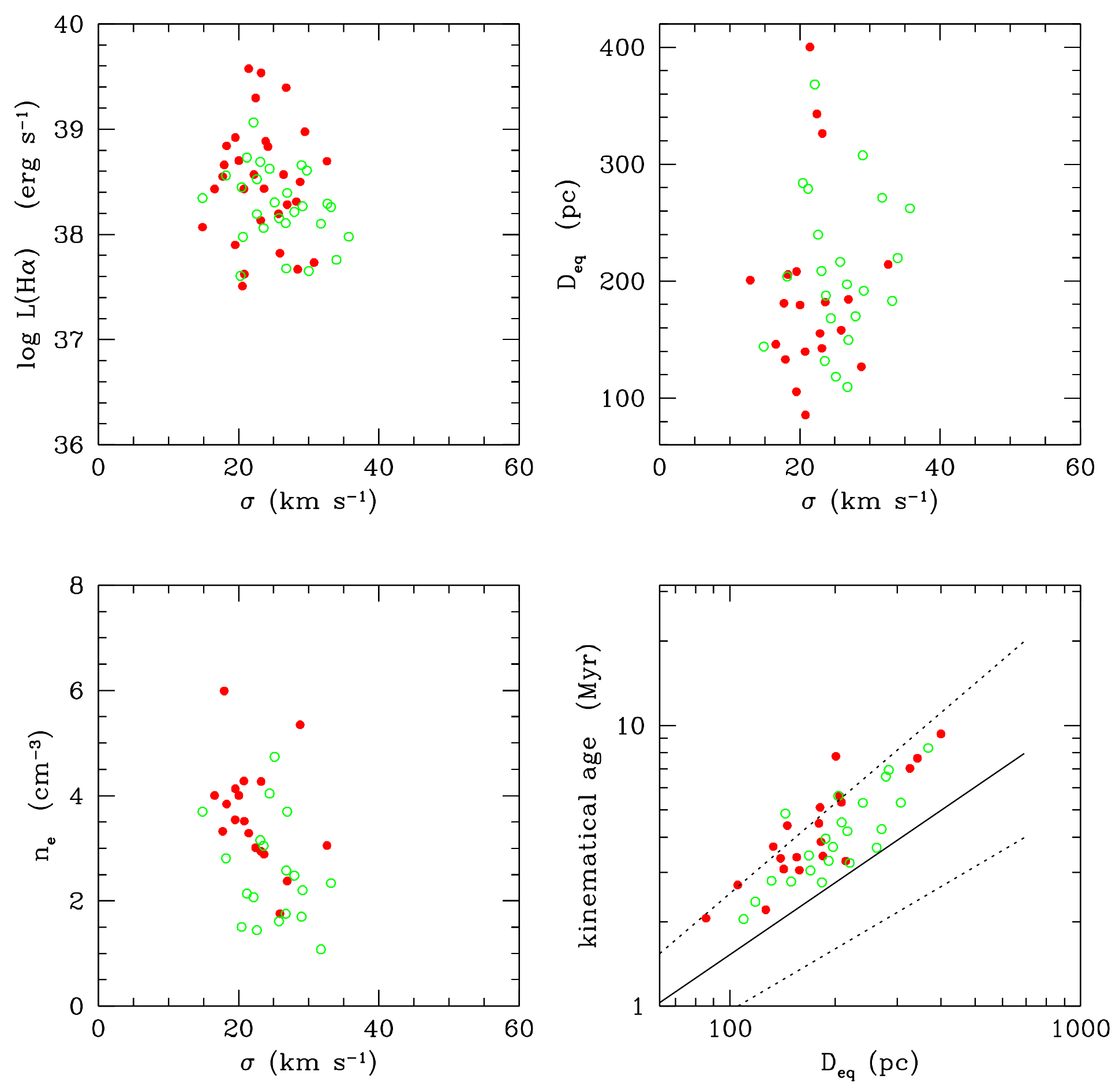}
   \caption{Kinematical properties of the HII regions of IC 3476 with luminosity $L(H\alpha)$ $\geq$ 10$^{36}$ erg s$^{-1}$ and $S/N$ $>$5,
   where the instrumental limit in the velocity dispersion is $\sigma$ $\simeq$ 13 km s$^{-1}$. 
   Upper left: H$\alpha$ luminosity vs. velocity dispersion. Upper right: equivalent diameter vs. velocity dispersion.
   Lower left: electron density vs. velocity dispersion. Lower right: kinematical age vs. equivalent diameter. Red filled symbols are for HII regions
   located on the front structure, green empty symbols for the other HII regions within the stellar disc, 
   respectively. All H$\alpha$ luminosities are corrected for [NII] contamination and for Balmer decrement using the [NII]/H$\alpha$ and H$\alpha$/H$\beta$ 
   ratios measured using the MUSE data within each individual region. Equivalent diameters and electron densities, defined as described in the text,
   are plotted only for HII regions where the PSF correction is less than 50\%. The black solid line and the dotted lines show the mean relation and the envelope of points
   derived in the LMC by Ambrocio-Cruz et al. (2016).
 }
   \label{dispersion}%
   \end{figure*}

Figure \ref{dispersion} shows the relationship between the H$\alpha$ luminosity $L(H\alpha)$, the equivalent diameter $D_\mathrm{eq}$, and the electron density $n_\mathrm{e}$ 
with the velocity dispersion $\sigma$ of the ionised gas within resolved HII regions located on the disc of the galaxy and in the starbursting 
structure. It also shows the relationship between
the kinematical age of the individual HII regions, defined as $k_\mathrm{age}$ = $D_\mathrm{eq}$/$\sigma$, with the equivalent diameter. 
Since all the points plotted in these panels are included within the MUSE field, where the H$\alpha$ emission is sufficiently high to allow an accurate estimate of the velocity
dispersion in the Farby-Perot cube, the H$\alpha$ data are here corrected for [NII] contamination and dust attenuation using their local estimate in the MUSE data. 
The velocity dispersion of the HII regions ranges between 15 $\lesssim$ $\sigma$ $\lesssim$ 40 km s$^{-1}$, and is thus comparable to
that observed in the giant HII regions of other nearby galaxies (e.g. Smith \& Weedman 1970, Chu \& Kennicutt 1994, Ambrocio-Cruz et al. 2016, Bresolin et al. 2020).
The first three plots do not show any correlation,
but suggest that the HII regions located on the front region formed after the dynamical interaction with the IGM, have, on average, 
a higher H$\alpha$ luminosity and electron density per unit velocity dispersion than their counterparts on the stellar disc, with a similar distribution of their size. 
The lower right panel shows that the HII regions formed in the front region and those
located elsewhere in the disc share the same kinematical age vs. equivalent diameter relation\footnote{Overall the HII regions of IC 3476 are located on the upper 
side of the envelope of points tracing the same kinematical age vs. equivalent diameter relation observed in the LMC by Ambrocio-Cruz et al. (2016).}.
Given the relationship between star formation rate, gas column density, and velocity dispersion (e.g. Elmegreen 2015):

\begin{equation}
\Sigma_\mathrm{SFR} = \epsilon_\mathrm{eff}\left(\frac{4}{\sqrt{3}}\right)G\frac{\Sigma_\mathrm{gas}^2}{\sigma}
\end{equation}

\noindent
where $\epsilon_\mathrm{eff}$, the star formation efficiency per unit free fall time, is a fairly constant parameter in different environments (Federrath \& Klessen 2012), 
it is thus conceivable that the observed increase of star formation activity along 
the front region is mainly due to an increase of the gas volume density occured after the displacement of the external gas along the stellar
disc pushed by the dynamical pressure exerted by the surrounding IGM during the mostly edge-on interaction (Elmegreen \& Efremov 1997). The increase of the external pressure favors 
the formation of molecular gas (Blitz \& Rosolowsky 2006) and increases the star formation activity (Bieri et al. 2016). 

\section{Simulations}

\subsection{Methodology}

To understand the origin of the peculiar features observed in the data we perform idealised hydrodynamical 
simulations of a galaxy with properties similar to IC 3476 subject to ram-pressure stripping. We stress that the purpose of this exercise is not that of identifying the
simulation which best reproduces
all the observed properties of IC 3476, which would require to run several low resolution simulations on a large grid of input parameters, but rather to understand, using a high resolution simulation,
the nature of the various physical processes acting on the different galaxy components and its effects on the stripped gas. 
Our setup consists of an isolated galaxy 
at rest at the centre of a 600~kpc cubic periodic box. The galaxy is made of a $2.5\times 10^{11}\rm\, M_\odot$ dark matter halo, a $1.2\times 10^9\rm\, M_\odot$ 
stellar disc, and a $2\times 10^9\rm\, M_\odot$ gaseous disc (see Table \ref{gal} for a comparison with IC 3476). We generated the initial conditions using MakeDisk (Springel et al. 2005), 
which assumes a global equilibrium of the system, and obtained a scale radius of the disc $r_0\sim 2.6$~kpc. We notice that this scalelength is a factor of 1.7 larger than the one measured in IC 3476.
This might moderately overestimate the effects of stripping. We conservatively assume a scale height $h_0=0.1r_0$. These parameters lead to a rotational velocity of the galaxy 
in the range -60 $\lesssim$ $v_\mathrm{rot}$ $\lesssim$ 60 km s $^{-1}$ at 2 kpc, -80 $\lesssim$ $v_\mathrm{rot}$ $\lesssim$ 80 km s $^{-1}$ at 3 kpc, reaching $\sim$ 100 km s$^{-1}$ at $\sim$ 10
kpc.
The box is filled with a hot, low-density gas ($n_\mathrm{e}$ = 2 $\times$ 10$^{-4}$ cm$^{-3}$), aimed at representing the IGM, that can be set in relative motion with the galaxy to mimic the system 
motion throughout the cluster. We note that, although a more accurate description of the IGM would require variable density and velocity of 
the system relative to the IGM (Tonnesen 2019), for the purpose of the current study we assume for simplicity a constant density/velocity gas.

In order to accurately assess the effect of ram-pressure stripping on the system, we performed two different runs, where the ICM is moving along the $z$-direction 
at 1200 and 2000~$\rm km\, s^{-1}$, respectively. These two velocities roughly correspond to the observed stellar line of sight velocity with respect to the mean velocity of the cluster
(1131 km s$^{-1}$), which can be considered as a lower limit, and the same velocity $\times$ $\sqrt{3}$ (1959 km s$^{-1}$) which takes into account the velocity of the galaxy 
on the plane of the sky. The resulting pressure for a velocity of 2000 km s$^{-1}$ is $P$ = 1.5 $\times$ 10$^{-11}$ dyn cm$^{-2}$.
We also performed a reference run where we assume the ICM at rest, aimed at representing the evolution of the system 
in the case of no ram-pressure. To be consistent with the observations, the galaxy is inclined by 70$\degr$ relative to the $z$-axis.

We performed our runs with \textsc{gizmo} (Hopkins 2015), descendant of Gadget 2 (Springel 2005) and Gadget 3 (Springel et al. 2008),
employing the sub-grid models for star formation and stellar feedback (supernovae, wind, and stellar radiation) described in Lupi \& Bovino (2020) and Lupi et al. (2020). 
Gas cooling and chemistry are modelled via \textsc{krome} (Grassi et al. 2014), using the metal network described in Bovino et al. (2016) and already employed 
in several works (Capelo et al. 2018, Lupi \& Bovino 2020, Lupi et al. 2020).
The mass resolution for the galaxy is set to $2.5\times 10^5\rm\,M_\odot$ for dark matter, and $2000\rm\,M_\odot$ for gas and stars. 
Although a higher resolution would allow for a better sampling of small star clusters in the tails, our choice is mainly dictated by the 
fact that a proper sampling of the IMF requires at least ~1000 Msun/particle, and by the computational cost of the runs.
The spatial resolution is defined by the Plummer-equivalent gravitational softening, that is kept fixed for dark matter and stars at 80 and 10 pc, respectively. 
For gas, we employ fully adaptive gravitational softening, with the kernel around each cell defined to encompass an effective number of 32 
neighbours, and the maximum resolution set at 1 pc. The wind is initially sampled 
with a 100 times lower resolution, but it is refined on-the-fly in the simulation via particle splitting, when at least 75 per cent of the kernel of the wind cell is 
covered by higher-resolution gas. This approach allows us to keep a large enough box to avoid boundary effects on the galaxy and at the same time to save computational 
time evolving the low-density ICM in the regions where no interaction with the galaxy occurs.
Because of the initial configuration of the system, where turbulence due to stellar feedback is not present, we expect a star formation burst to occur at the beginning 
of the simulation. To reduce this numerical effect we let the stellar particles already present in the system to explode as supernovae stochastically, assuming a uniform 
age distribution with average 3.5~Gyr. In addition, instead of starting all the runs at $t=0$, we start the two ram-pressure simulations taking the results of the reference 
run at $t=100$~Myr, i.e. at the end of the initial burst.
The different gas phases, neutral atomic, molecular, and ionised, are a direct output of the simulations. They results from the consistent non-equilibrium chemical
evolution of the gas (Lupi et al. 2018, Capelo et al. 2018). For the following analyses, we select the particles that initially belong to the 
galaxy, and exclude the gas belonging to the ICM, independently of its thermodynamic/chemical state.

\subsection{Results}

\subsubsection{2D distribution}

\begin{figure*}
   \centering
   \includegraphics[width=0.99\textwidth]{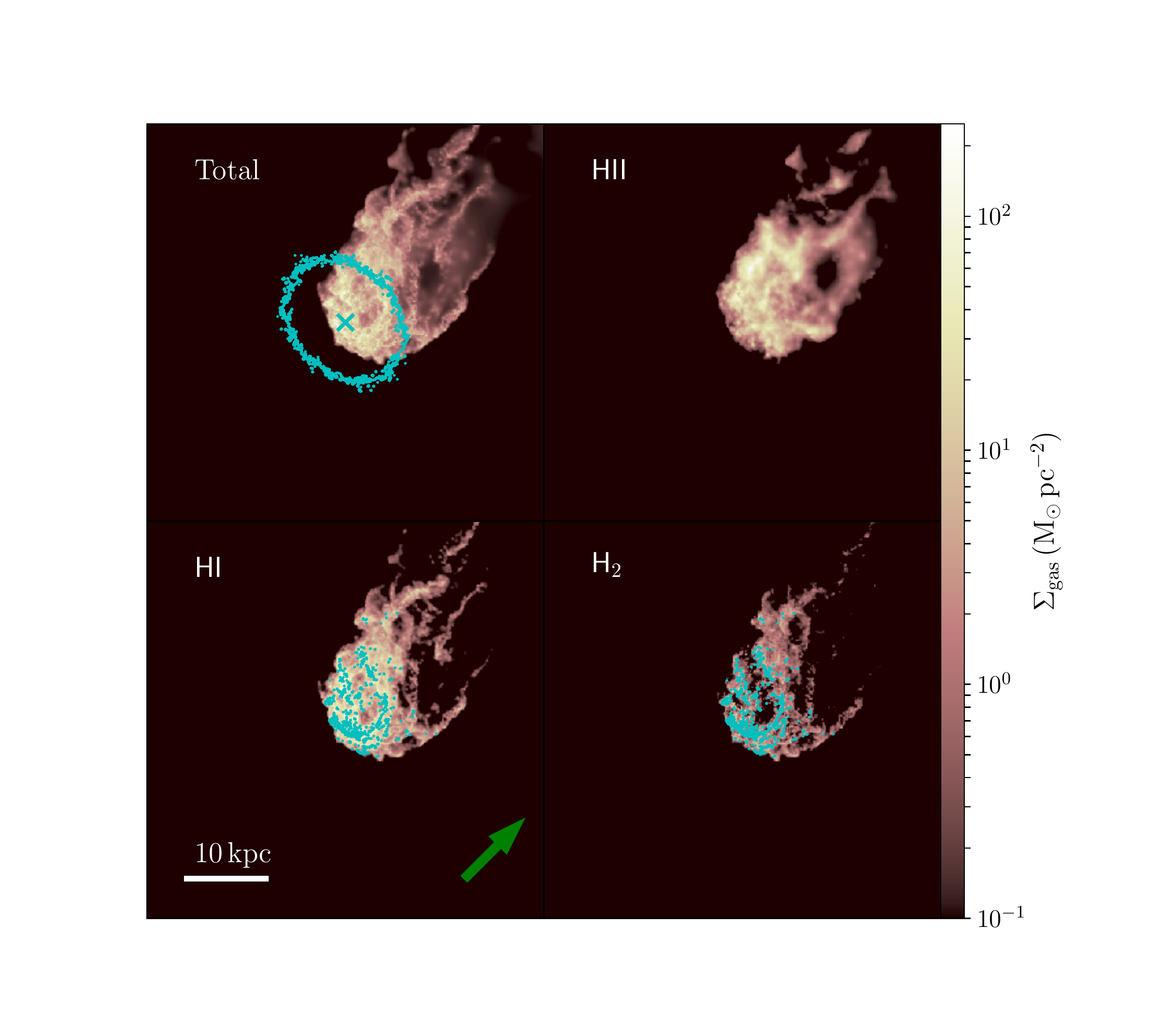}
   \caption{2D distribution of the total (HI, H$_\mathrm{2}$, HII, He, and metals; upper left panel), ionised (upper right), atomic (lower left), and molecular (lower right) gas components of the simulated galaxy undergoing 
   a wind of 2000 km s$^{-1}$, 150 Myr after the beginning of the interaction. The model galaxy is oriented as IC 3476 on the plane of the sky. The cyan cross 
   in the upper left panel shows the stellar centre, the contours the stellar disc, while those in the lower panels the distribution of the star forming regions. The green arrow indicates the direction of the wind.
   }
   \label{sim_gas}%
   \end{figure*}

\begin{figure}
   \centering
   \includegraphics[width=0.5\textwidth]{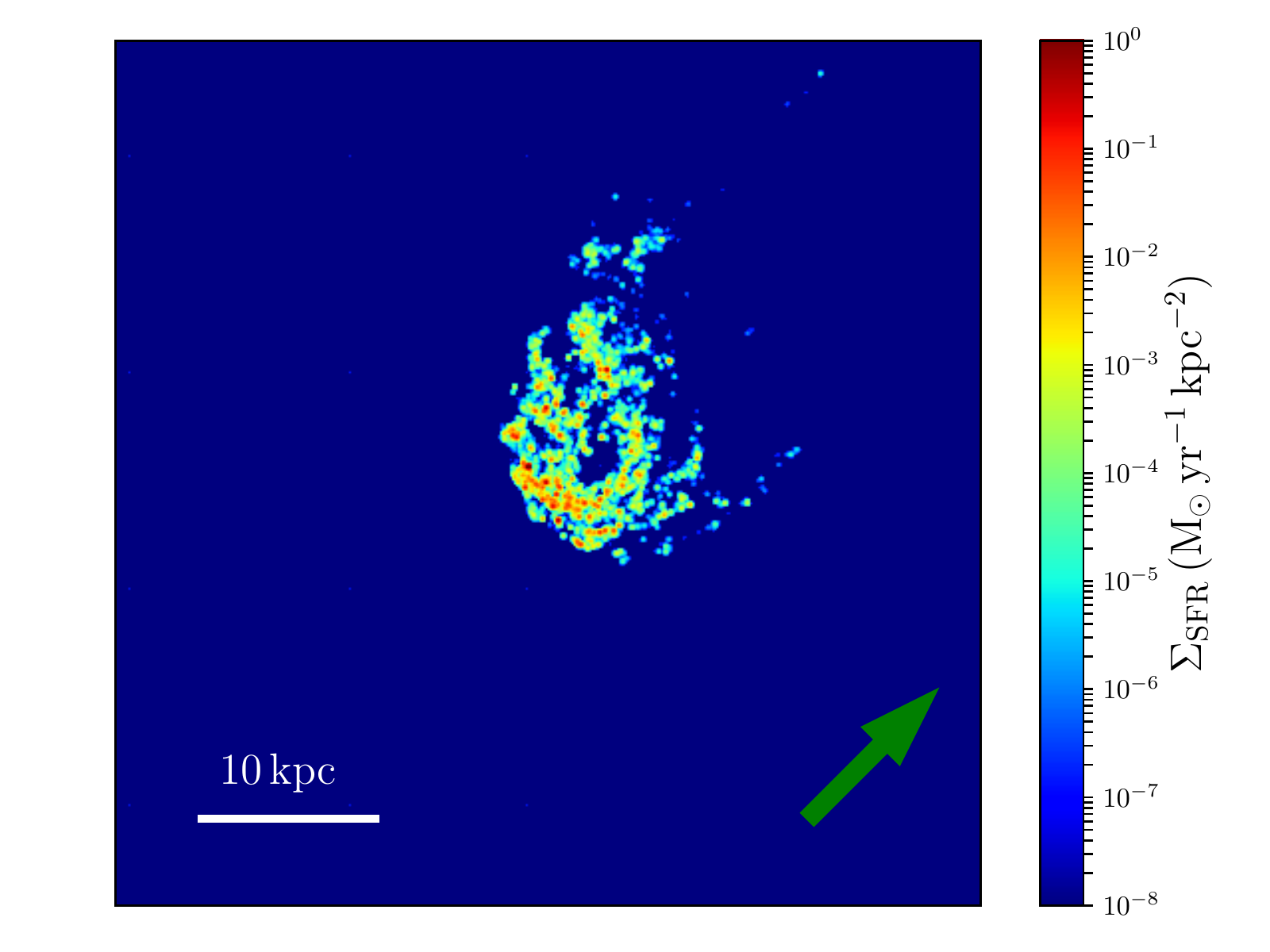}
   \caption{2D distribution of the star formation activity of the simulated galaxy undergoing a wind of 2000 km s$^{-1}$ 150 Myr after the beginning of the interaction. The green arrow indicates the direction of the wind.
   }
   \label{sim_sfr}%
   \end{figure}

The results of the simulations are shown in Figures \ref{sim_gas}  to \ref{vel_sim}. For a comparison with the observations
we plot here the output of the simulations at those epochs which better match the observed data. To ease the comparison with IC 3476,
the simulated galaxy is first projected at an inclination of 45\degr, then rotated on the plane of the sky by 45\degr. For this projection, this implies that
the line-of-sight velocity of the wind is $V_\mathrm{los}(wind)$ = $\cos(25\degr) \times v(wind)$, while those on the plane of the sky $V_\mathrm{X,Y}(wind)$ = $\sqrt{2} \times \cos(65\degr) \times v(wind)$,
which, for a wind of 2000 km s$^{-1}$, gives respectively $V_\mathrm{los}(wind)$ $\simeq$ 1800 km s$^{-1}$ and $V_\mathrm{X,Y}(wind)$ $\simeq$ 600 km s$^{-1}$. 
Given that the line-of-sight velocity of IC 3476 with respect to the cluster is 1131 km s$^{-1}$, the velocity of the galaxy on the plane of the sky should be $\sim$ 1650 km s$^{-1}$, thus 
the projected extension of the gas and of the stellar tails
are underestimated in Fig. \ref{sim_gas} and \ref{sim_sfr}. We focus on the simulations done with a wind of 2000 km s$^{-1}$
which are more representative of IC 3476. The results obtained with a wind of 1200 km s$^{-1}$ are similar, although less pronounced.

Figures \ref{sim_gas} and \ref{sim_sfr} show the 2D distribution of the different gas components (HI, H$_\mathrm{2}$, HII, and total, defined as HI+H$_\mathrm{2}$+HII+He+metals) and of the star formation activity  
of the simulated galaxy undergoing an external wind of 2000 km s$^{-1}$, 150 Myr after the beginning of the interaction. 
The gas distribution is highly asymmetric, with most of the gas (in all its phases) located on one side of the galaxy
and in prominent tails extending up to $\sim$ three times the stellar disc. The leading side of the galaxy is completely deprived of gas. The dominant gas phase in the tail is the ionised
one\footnote{We recall that the simulations shows the distribution of the ionised gas, and not that of H$\alpha$, which is just one of the lines emitted after the recombination
of HII into HI.}, 
which has a diffuse distribution as the atomic gas, while the molecular phase is only located in thin filamentary structures. Because of the lack of gas, the star formation activity is totally quenched in the leading
side of the stellar disc, while boosted in the front edge of the gas at the interface between the IGM and the ISM ($\Sigma_\mathrm{SFR}$ $\simeq$ 0.1-0.5 M$_{\odot}$ yr$^{-1}$ kpc$^{-2}$) where its density is maximal
($\Sigma_\mathrm{gas}$ $\simeq$ 20-50 M$_{\odot}$ pc$^{-2}$). Star formation is also present far from the stellar disc in the condensed regions located along the filamentary structures seen in H$_\mathrm{2}$.
Figure \ref{sim_sfr} also shows that the brightest star forming regions have been formed at the leading edge located at the the South-East of the galaxy, at the boundary region between the hot ICM and the cold ISM (front region),
consistently with the observations. 

\subsubsection{Gas deficiency}

\begin{figure}
   \centering
   \includegraphics[width=0.5\textwidth]{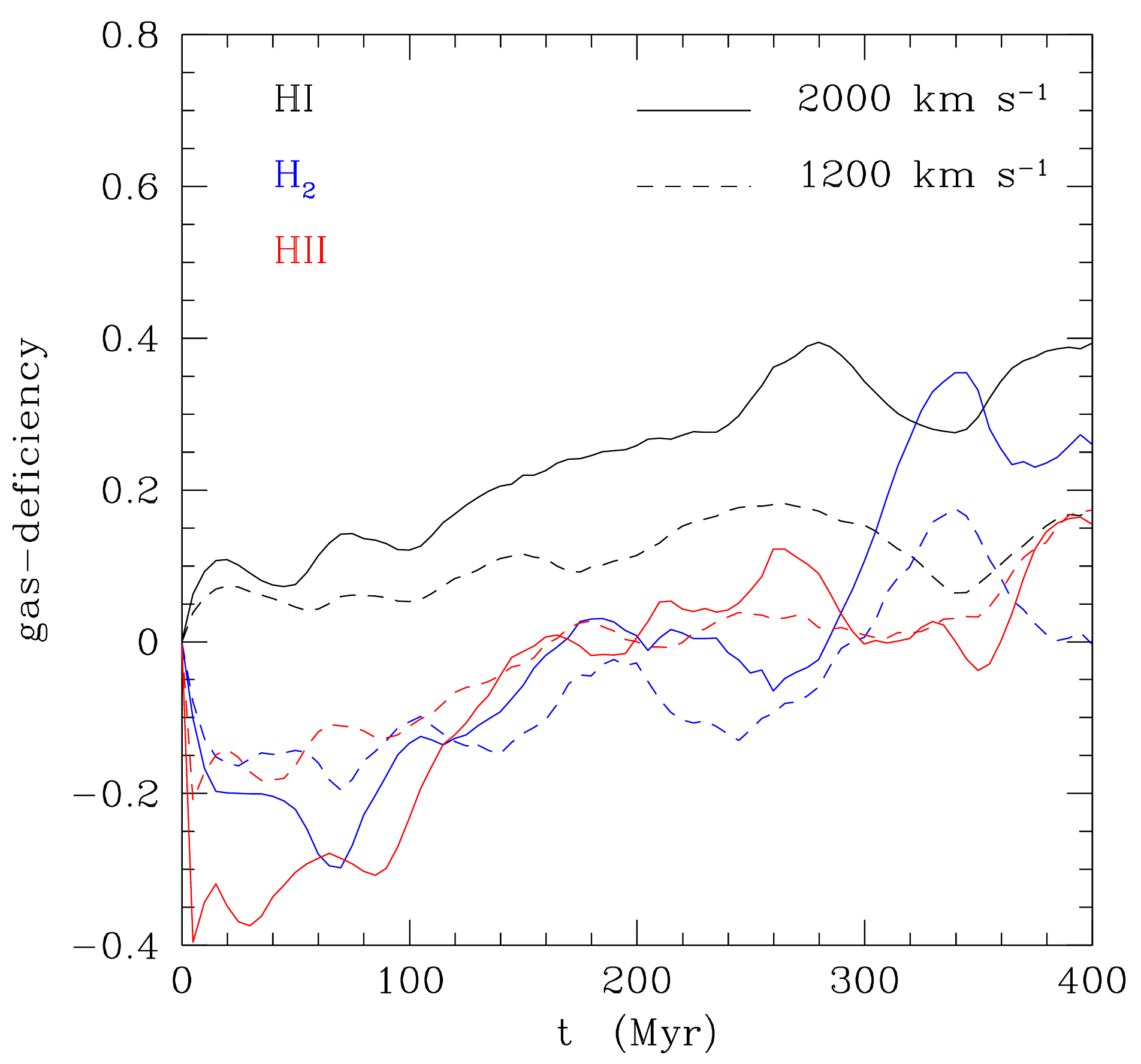}
   \caption{Variation of the gas deficiency parameter in the different phases (black: HI; blue: H$_\mathrm{2}$; red: ionised) as a function of time for the simulated galaxy
   subject to an external wind of 2000 km s$^{-1}$ (solid) and 1200 km s$^{-1}$ (dashed). 
 }
   \label{HIdef}%
   \end{figure}

Figure \ref{HIdef} shows the time variation of the gas deficiency parameter, defined as
the logarithmic difference between the total mass of the atomic, molecular, and ionised gas mass of the unperturbed galaxy and that of the simulated perturbed systems measured at the
same epoch. All entities are geometrically measured within a circular aperture (projected as an ellipse according to the inclination of the galaxy)
with radius 3.4 times the stellar disc scalelength of the simulated galaxy, corresponding roughly to 9 kpc, thus at $\sim$ 2 optical isophotal radii.
While the aperture is centered at each time step on the centre of mass of the stellar component of the galaxy, the radius is kept fixed in time since the new stellar mass formed is not 
large enough to significantly change the disc size over the short timescales of the simulations. Figure \ref{HIdef} can be thus directly compared with the observations.
Figure \ref{HIdef} shows that the HI-deficiency parameter monotonically increases with time reaching $HI-def$ $\simeq$ 0.4 at $t$ $\simeq$ 280 Myr
for an external wind of 2000 km s$^{-1}$, a value slightly lower than the one observed in IC 3476 ($HI-def$ = 0.67; Boselli et al. 2014b)\footnote{The typical
dispersion in the scaling relation of isolated galaxies used to calibrate the HI-deficiency parameter is $\simeq$ 0.3-0.4 dex, which should be considered as the typical uncertainty on $HI-def$.}. 
On the contrary, the H$_\mathrm{2}$ and HII-deficiency parameters
first abruptly decrease to $\simeq$ -0.2 and -0.4 and later increase to $H_\mathrm{2}-def$ $\simeq$ 0.3 after 350 Myr and $HII-def$ $\simeq$ 0.4 after 400 Myr.
The variation with time of the HI-deficiency parameter can be easily explained by a continuous gas stripping due to the wind flow. The fact that 
the simulation does not reach the observed value of $HI-def$ = 0.67 just indicates that the simulated stripping process is not sufficiently efficient, and this can be due to 
the assumed impact parameters (the inclination of the galaxy disc with respect to the wind flow might be lower than 70$\degr$), by an underestimate of
the assumed infall velocity or density profile, or by the limited resolution of the simulations. On the other hand, the model galaxy is more extended that IC 3476, thus stripping should be more efficient
in the model. The almost edge-on stripping process pushes on short timescales ($t$ $\lesssim$ 100 Myr) the gas located on the 
leading side of the galaxy over the stellar disc. The increase of the 
external pressure facilitates the formation of molecular clouds on the disc which leads to a decrease of the molecular gas deficiency and to
an increase of the HI deficiency. On longer timescales ($t$ $\gtrsim$ 200 Myr)
the molecular gas condensed in high density regions is transformed into stars while the diffuse component, which is located within the inner disc, is reached by the external pressure
and removed along with the HI component. The observed abrupt decrease of the ionised gas deficiency just after the beginning of the interaction could be due
to a combined effect of an increase of the star formation activity boosted by the supply of fresh gas on the disc with a change of phase of the gas located on the leading side,
shock-ionised during the interaction on the interface between the cold ISM and the hot IGM.
 
\subsubsection{Gas phases in the tail}

\begin{figure}
   \centering
   \includegraphics[width=0.5\textwidth]{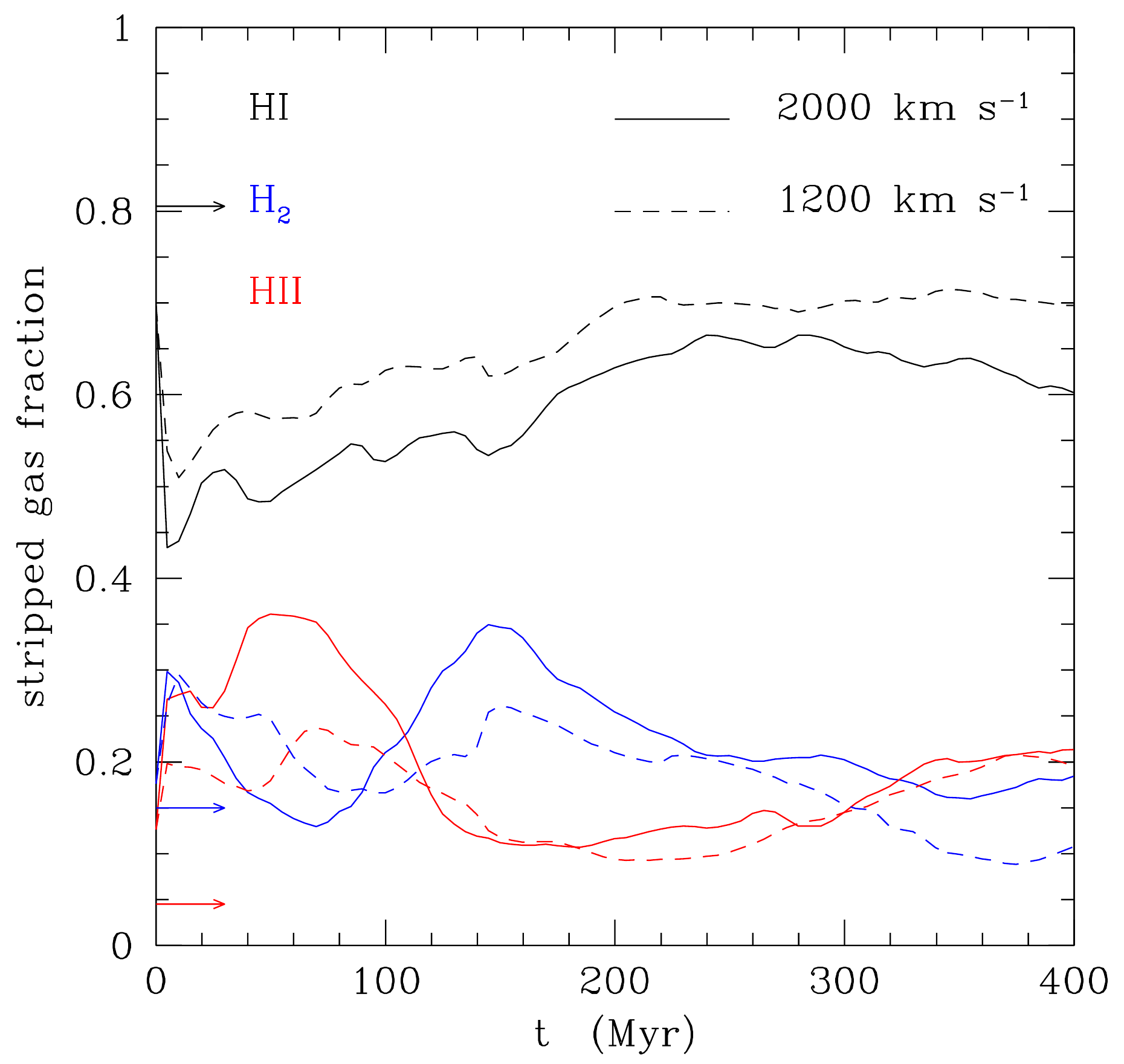}
   \caption{Variation of the stripped gas fraction within the tail in its different phases (black: HI; blue: H$_\mathrm{2}$; red: ionised) as a function of time for the simulated galaxy
   subject to an external wind of 2000 km s$^{-1}$ (solid) and 1200 km s$^{-1}$ (dashed). The black, blue, and red arrows on the Y-axis indicate for reference the mean HI, H$_\mathrm{2}$, and HII gas fractions
   on the leading side of the unperturbed galaxy disc.
 }
   \label{phases}%
   \end{figure}

Figure \ref{phases} shows the time variation of the fraction of the different gas phases-to-total gas measured outside the disc of the simulated galaxy at a radius $r$ $>$ 9 kpc
from the nucleus. The interpretation of this figure is complicated because both the numerator (HI, H$_\mathrm{2}$, HII) and the denominators (HI + H$_\mathrm{2}$ + HII) of 
the stripped gas fraction change with time. At the beginning of the stripping process the gas in the tail is at $\sim$ 40\%\ HI, and 30\%\ H$_\mathrm{2}$ and HII, respectively.
The HI fraction increases to $\sim$ 60\%\ after $\sim$ 250 Myr, and slightly decreases afterwards. The molecular gas fraction first drops to $\sim$ 20\%\ after 50 Myr, increases up to
$\sim$ 35\%\ at 150 Myr, and slowly decreases to $\sim$ 20\%\ at 400 Myr after the beginning of the interaction. The ionised gas component has an opposite trend, with an increase in the
first 50 Myr, a rapid decrease after 150 Myr, followed by a mild increase up to $\sim$ 20\%\ after 400 Myr. These trends can be explained if most of the gas is stripped in 
the HI phase, which is the dominant gas phase in the outer regions of the disc ($\sim$ 80\%), thus the component less bound to the gravitational potential well. A fraction of this gas is shocked 
at the interface of the ISM-IGM front and rapidly removed from the galaxy as ionised gas. Indeed, the fraction of ionised gas on the leading edge of the unperturbed galaxy disc
from where most of the stripped gas comes from is only 4.5\%\ while it reaches $\sim$ 35\%\ in the tail in less than 100 Myr.
The mild increase of the ionised gas fraction at $t$ $\gtrsim$ 200 Myr, instead, could be due to 
a change of phase of the cold gas in the tail heated by the surrounding IGM (heat conduction). The initial decrease of the molecular gas fraction could be due 
to a rapid change of phase of the molecular component once in contact with the surrounding medium. Its increase at $t$ $\sim$ 150 Myr can be due either to the supply of extra molecular
gas mainly located within the inner disc of the galaxy reached by the outside-in stripping of the gas, or to a fraction of cold atomic gas whose density is sufficiently high to allow self shielding and 
cooling within the tail. This would be the gas component responsible for the birth of new stars in the tail. 
The following decrease of the molecular gas fraction is probably due to the heating of the gas after its mixing with the hot IGM, 
hence a less efficient cooling in the stripped material. This gas would become ionised only if sufficiently heated by the surrounding medium.

\subsubsection{Star formation history}

\begin{figure*}
   \centering
   \includegraphics[width=0.99\textwidth]{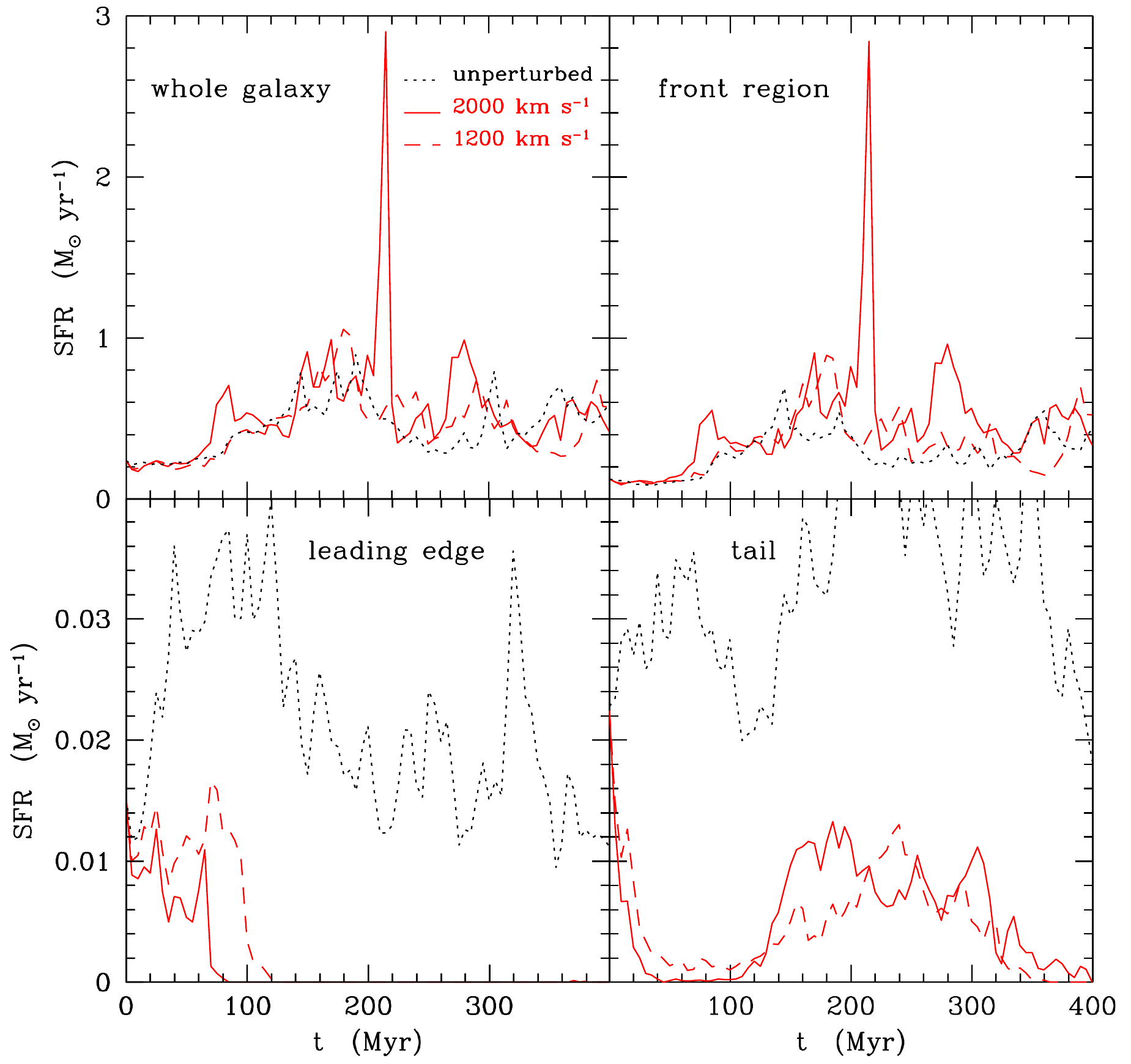}
   \caption{Variation of the star formation rate within the whole galaxy (upper left panel), the front region (upper right), the leading side (lower left), and the 
   tail of stripped gas (lower right)  as a function of time for the simulated galaxy
   subject to an external wind of 2000 km s$^{-1}$ (solid red) and 1200 km s$^{-1}$ (dashed red) compared to an unperturbed system (dotted black). 
 }
   \label{sfr}%
   \end{figure*}

Figure \ref{sfr} shows the time variation of the star formation activity of the perturbed galaxy as a whole, in a front region formed by the compressed gas 
after the interaction (similar to the one observed in IC 3476 and described in Sect. 3.2), in the leading side of the galaxy, and within the tail (in the model defined as the regions at more than 
9 kpc from the galaxy nucleus). In all panels the star formation rates are compared to those measured within the same regions of the unperturbed simulated galaxy. Figure \ref{sfr} shows that the 
overall activity of star formation of the galaxy can be significantly boosted during the interaction, probably because of the supply of the gas originally located in the
outer disc to the inner regions. Because of the external pressure, the gas increases density and collapses to form giant molecular clouds, where star formation takes place.
The induced activity is bursty, with peaks lasting 20-40 Myr at maximum. If the external wind is sufficiently high (2000 km s$^{-1}$), the star formation activity can increase by a
factor of up to $\sim$ 4. Figure \ref{sfr} suggests that most of the increase of the activity is due to the HII regions in the inner front structure formed after the interaction
as the one observed in Fig. \ref{Ha} (Steyrleithner et al. 2020).
On the contrary, in the leading side of the disc the gas is totally removed on short timescales, totally quenching the activity after $\sim$ 50-100 Myr, as indeed estimated from the SED
fitting analysis of IC 3476. The star formation activity
of the disc at galactocentric distances $\gtrsim$ 9 kpc, which corresponds to the activity observed in the extended UV discs of unperturbed systems, drastically drops once the loosely
bound gas is removed during the interaction, while it begins at small rates ($SFR$ $\simeq$ 0.01 M$_{\odot}$ yr$^{-1}$)\footnote{In the simulations being the SFR 
measured an interval of 10 Myr, the limiting resolution for the SFR is 
$SFR_\mathrm{lim}$ = 2000 M$_{\odot}$/10$^7$ yr = 2 $\times$ 10$^{-4}$ M$_{\odot}$ yr$^{-1}$
} in the tail (Steyrleithner et al. 2020) once the stripped gas cools and collapses into 
giant molecular clouds after $\sim$ 150 Myr (see Fig. \ref{phases}). For comparison, the observed rate of star formation in the tail of IC 3476 is $SFR$ $\simeq$ 8$\times$ 10$^{-4}$ M$_{\odot}$ yr$^{-1}$,
although this value is probably underestimated because of the stochasticity of the star formation activity at these low levels.

\subsubsection{Kinematics}

\begin{figure*}
   \centering
   \includegraphics[width=1\textwidth]{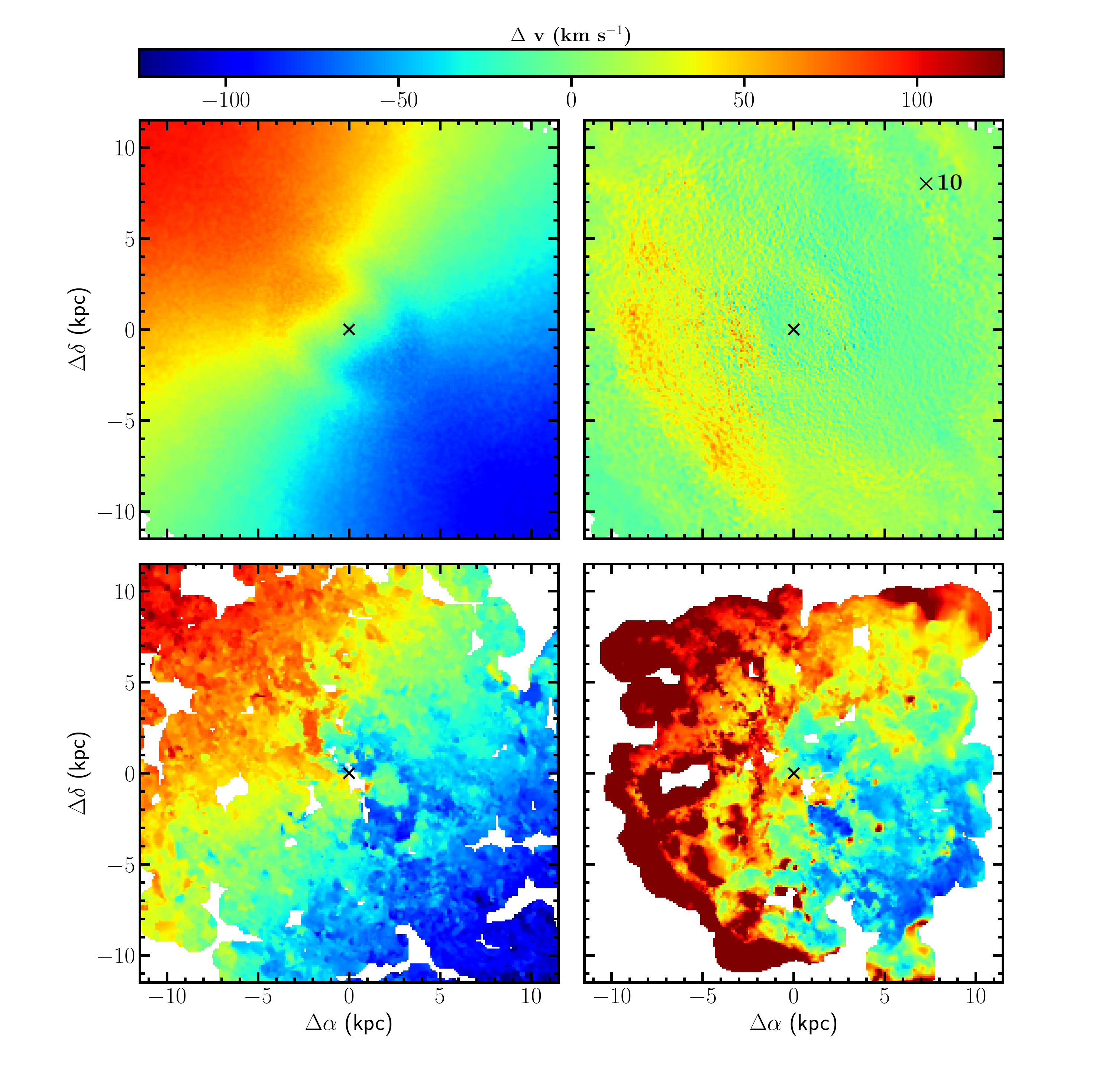}\\
    \caption{Upper left: stellar velocity map for the simulated unperturbed galaxy. Upper right:
    difference of the stellar velocity field of the perturbed vs. unperturbed model galaxy, with velocities multiplied by a factor of 10 
    for visibility. Lower left: velocity map of the molecular gas component for the simulated unperturbed galaxy.
    Lower right: velocity map of the molecular gas component for the simulated perturbed galaxy.
    All plots are for a simulated galaxy 50 Myr 
    after the beginning of the interaction. The black cross shows the position of the photometric centre of the galaxy. 
   }
   \label{vel_sim}%
   \end{figure*}

Figure \ref{vel_sim} shows the velocity field of the gas component and of the stellar component for the unperturbed and 
for the perturbed simulated galaxy 50 Myr after the beginning of the interaction. For a fair comparison with the observations (see Figs. \ref{vel} and \ref{vel_star}),
we choose to plot the velocity field of the molecular gas component measured wherever the star formation density is
$\Sigma_\mathrm{SFR}$ $\geq$ 5 M$_{\odot}$ yr$^{-1}$ kpc$^{-2}$. Indeed, the molecular gas component is the one more tightly associated to the star forming regions detected in the observed spectroscopic data: indeed, a large fraction
of the ionised gas component in the simulations has a very low surface brightness and would be undetectable during the observations. Furthermore, this low column density ionised gas is
mainly produced during the shock at the interface of the galaxy ISM and cluster IGM, and thus might partly be IGM cooled after the interaction. This gas does not necessary follow the rotation of the galaxy.
Table \ref{sim_kin} gives the median of the difference between the velocity of the gas and that of the stars measured on the same region at different epochs after the beginning of the interaction.
The velocity field of the gas is very perturbed and does not allow to estimate its kinematical centre using the same methodology used for the observed data.
Figure \ref{vel_sim} and Table \ref{sim_kin} show that while the stellar velocity field is totally insensitive to the perturbation, that of the gaseous component is strongly perturbed 
even after only 50 Myr, as indeed observed in the IFU spectroscopic data (see Figs. \ref{vel}, \ref{FP}, and \ref{vel_star}), once again confirming that the perturbation of IC 3476 is a recent phenomenon.

\begin{table}
\caption{Median of the difference between the velocity of the gas and that of the stars}
\label{sim_kin}
{
\[
\begin{tabular}{cc}
\hline
\noalign{\smallskip}
\hline
Epoch 	& $<V_\mathrm{gas}-V_\mathrm{stars}>$   \\  
Myr	& km s$^{-1}$			\\    
\hline
0	& 0  \\
25	& 30	\\
50	& 41	\\
75	& 65	\\	
100	& 60	\\
150	& 61	\\
\noalign{\smallskip}
\hline
\end{tabular}
\]
}
\end{table}

\section{Discussion}

\subsection{The galaxy}

The distribution of the star forming regions inside the galaxy and outside the stellar disc clearly suggests an ongoing ram pressure stripping event. The negative redshift of the galaxy indicates
that IC 3476 is crossing the cluster from the backside and moving on the plane of the sky from north-west to south-east. 
The short timescale derived for the quenching episode on the leading-edge of the disc, or for the starburst episode on the boundary regions between the stripped ISM and the
surrounding IGM ($\simeq$ 50 Myr) suggest that the ram pressure stripping event is recent, thus that the galaxy is in a pre-peak or near-peak ram pressure phase.
The continuum-subtracted H$\alpha$ image suggests 
that the interaction with the surrounding intracluster medium is occuring almost edge-on. This interaction formed a star forming structure in the front of the galaxy, where all the most luminous 
HII regions are located, while totally stopping the star formation activity further out in the disc on the leading side. The same dynamical interaction with 
the ICM only mildly affects the star formation activity in the north-western part of the disc, where gas is still present. Overall, while IC 3476 is deficient in HI gas 
($HI-def$ = 0.67, meaning that it has $\sim$ 5 times less HI than similar objects in the field), the galaxy is still very active in star formation. Indeed, because of its total star formation activity 
(0.27 M$_{\odot}$ yr$^{-1}$), the galaxy is $\sim$ 1$\sigma$ above the main sequence drawn by local galaxies in the field, and $\sim$ 2$\sigma$ from the 
main sequence of HI-gas deficient galaxies in the Virgo cluster, being the most active HI-deficient low-mass late-type galaxy among all the \textit{Herschel} Reference Survey (Boselli et al. 2015).
Despite an enhanced activity as observed in the GASP sample of jellyfish galaxies (Vulcani et al. 2018), the ratio of the star formation activity in the tail to that in the disc is only $SFR_\mathrm{tail}/SFR_\mathrm{disc}$ = 2.8
$\times$ 10$^{-3}$, thus significantly lower than that of the jellyfish sample (Gullieuszik et al. 2020).
 
The asymmetric shape of the galaxy and of the tails, with the southern one significantly longer than the northern one, is predicted by ram pressure stripping simulations. Indeed
during an edge-on stripping event, ram pressure is more efficient on the galaxy side rotating as the direction of motion of the galaxy within the ICM than on the other side just because the 
velocity of the rotating spiral arm is summed up to the velocity of the galaxy (Roediger et al. 2014). The velocity field shown in Fig. \ref{vel} indicates that in IC 3476 this 
happens in the southern region (approaching side), where indeed the tail of HII regions formed after the interaction with the surrounding environment is significantly longer than in the northern
tail (receding side). The same figure also shows that the predicted mismatch between the kinematic and the stellar disc centres (Kronberger et al. 2008b), the former located at 
$\sim$ 0.5 kpc north-north-west from the latter, is expected given the direction of motion of the galaxy within the ICM (see Fig. \ref{vel_sim}),
and the perturbed velocity field (Haan \& Braun 2014). The ram pressure stripping event can also explain the observed difference between the systematic 
velocity of the stellar and gaseous component, the latter higher by $\sim$ 20 km~s$^{-1}$ than the former (see Table \ref{sim_kin}), as expected if the gas is decelerated after the interaction of the galaxy with the surrounding ICM
(Kronberger et al. 2008b).

IC 3476 is located at 0.5 Mpc ($r/r_\mathrm{vir}$=0.32, where $r_\mathrm{vir}$ is the virial radius of the Virgo cluster) from the cluster centre (M87), where the most recent estimate of 
the density of the intracluster medium is $n_\mathrm{e}$ $\simeq$ 2 $\times$ 10$^{-4}$ cm$^{-3}$ as derived from \textit{Suzaku} and \textit{Planck} data by Simionescu et al. (2017).
It can thus be compared to the massive SAB(rs)ab galaxy NGC 4569 (M90) which is located at exactly the same distance from the cluster centre and, as IC 3476, is crossing the 
cluster from the back at a similar velocity ($cz_\mathrm{NGC4569}$ = -221 km s$^{-1}$, $cz_\mathrm{IC3476}$ = -170 km s$^{-1}$, see 
Boselli et al. 2008, 2016a, and Vollmer et al. 2004). The effects of the ongoing perturbation on the star formation activity of the two objects are significantly different.
In the more massive NGC4569 ($M_\mathrm{star}$ = 3 $\times$ 10$^9$ M$_{\odot}$) $\sim$ 95\%\ of the total HI gas has been removed during the interaction and the star formation activity drastically reduced
by a factor of $\sim$ 5 (the galaxy is well below the main sequence, Boselli et al. 2016a) mainly in the outer regions, where the potential well is not sufficiently deep to retain the gas necessary
to sustain star formation. Althogh this galaxy is in a post-peak ram pressure stripping phase and thus has already crossed the core of the cluster (Vollmer et al. 2004; Roediger \& Hensler 2005), 
the gas removal and the subsequent decrease of the star formation activity are very recent phenomena ($\sim$ 100-200 Myr, Vollmer et al. 2004, Boselli et al. 2006).
In IC3476 $\sim$ 80\%\ of the gas has been removed mainly from the outer disc decreasing locally the star formation activity on short timescales ($\sim$ 50 Myr). In IC 3476, however,
the interaction boosted the activity of star formation by 
a factor of $\sim$ 2 with respect to similar unperturbed systems consistently with what is observed in some jellyfish galaxies (Vulcani et al. 2018). IC 3476 is in a pre-peak or near-peak 
stripping phase, and might end up totally deprived of its gas content after peak ram pressure, and thus get similar properties than NGC 4569 after a few hundred Myr.

This increase in the star formation activity is probably due to an increase of the molecular gas fraction made possible by the increase of the external
pressure (Blitz \& Rosolowsky 2006), as again observed in some massive jellyfish galaxies (Moretti et al. 2020b) and predicted by hydrodynamic simulations (Henderson \& Bekki 2016; Troncoso-Iribarren et al. 2020).
Despite NGC 4569 being significantly more massive than the dwarf IC 3476, thus able to retain the gas from external perturbations 
because of its deep potential well, the effects induced by the ongoing perturbation on their global star formation activity are opposite.
This significant difference can be explained by the orientation of the motion of the two galaxies within the intracluster medium. NGC 4569 is moving face-on, thus the stripping process removes the gas
pushing it outside the galaxy disc decoupling it from the other regions where star formation is taking place. In IC 3476 the stripping is occuring almost edge-on. In this configuration 
the perturbed gas crosses all the stellar disc and interacts with the rest of the ISM before leaving the galaxy. The pressure induced by this perturbed gas on the giant molecular clouds 
located on the stellar disc can induce gas collapse, an efficient formation of molecular gas, and an increase of star formation as predicted by simulations 
(e.g. Marcolini et al. 2003, Bekki 2014, Roediger et al. 2014, Hendersen \& Bekki 2016). Indeed, the most luminous HII regions are 
located on a star forming structure at the boundary between the galaxy ISM and of the ICM, where the increase of pressure is at its maximum. The analysis presented in Sect. 3.5
has shown an increase of the H$\alpha$ luminosity per unit radius of the HII regions in the front of the galaxy where the front region is located (Fig. \ref{HII}).
This result can be explained by an increase of the gas density, as indeed predicted by hydrodynamic simulations of edge-on ram pressure stripping (Marcolini et al. 2003, 
Kronberger et al. 2008a, Bekki 2014, Roediger et al. 2014, Hendersen \& Bekki 2016). However, it can also be due to
an increase of the hard UV radiation due to the younger age of the stellar populations within the HII regions formed during the interaction on the front of the galaxy.

\subsection{Star formation in the stripped gas}

At the edges of the star forming front region the gas is removed by the tangential friction with the 
ICM. Turbulence and instabilities produced during this interaction induce the formation of giant molecular clouds soon 
decoupled from the potential well of the galaxy (Tonnesen \& Bryan 2012). After being stripped from the galaxy disc, the gas forms long chains of overdensity regions, where star formation can occur. The lack of any stellar component seen in the 
optical bands associated to the star forming complexes in the tail and their H$\alpha$-to-FUV colours strongly suggest that these are very recent 
objects ($\lesssim$ 20 Myr) formed within the gas stripped during the interaction. The typical H$\alpha$ luminosity of the HII regions formed within the tail is $\lesssim$ 10$^{37}$ erg s$^{-1}$,
thus one-to-two orders of magnitude below the mean luminosity of the star forming complexes measured in jellyfish galaxies during the GASP survey (Poggianti et al. 2019),
while consistent with the predictions of the simulations of Tonnesen \& Bryan (2012) and Steyleithner et al. (2020).
This difference is probably related to the limited angular resolution of the GASP data ($FWHM$ $\sim$ 800 pc) for galaxies at a mean redshift of $z$ $\sim$ 0.05 which 
does not enable one to resolve individual HII regions whose typical size is $D_\mathrm{eq}$ $\lesssim$ 400 pc (Rousseau-Nepton et al. 2018).
The striking difference with NGC 4569, where no star formation has been observed in the long tails of stripped gas, questions the importance of the external pressure of the IGM confining the stripped gas
in regulating star formation (Tonnesen \& Bryan 2012) since this parameter is not expected to change significantly with respect to IC 3476, which is located close to NGC 4569 and at the same 
distance from the cluster centre.

\section{Conclusions}

The peculiar morphology of the dwarf ($M_\mathrm{star}$ $\sim$ 10$^9$ M$_{\odot}$) gas-rich galaxy IC 3476 in the Virgo cluster revealed by the very deep NB H$\alpha$ image
gathered during the VESTIGE survey shows a peculiar distribution of the star forming regions principally located along a banana-shaped structure crossing the disc, 
with a few HII regions located along tails well outside the stellar disc. This perturbed morphology indicates that IC 3476 is undergoing an almost edge-on ram pressure stripping
process able to compress the gas from the outer regions of the leading edge towards the stellar disc and removing it along long tails where star formation can occur.
Our incredible set of multifrequency data, combined with the spectacular angular resolution of the VESTIGE images and of the MUSE IFU spectroscopy
and with high-resolution hydrodynamic simulations tuned to reproduce the physical and kinematical properties observed in IC 3476,
allow us to study the process of star formation down to scales of  $r_\mathrm{eq}$ $\simeq$ 40 pc. This analysis indicates that 
the gas of the outer disc compressed on the leading front of the interaction between the hot IGM and the cold ISM increases in density, boosting the 
local and global star formation activity, moving the galaxy above the main sequence relation defined by similar objects in the field. 
The overall increase of the star formation activity of the gas is mainly due to this interface region between the IGM and ISM, where several giant HII regions of luminosity 
$L(H\alpha)$ $\simeq$ 10$^{38}$ erg s$^{-1}$, otherwise rare in unperturbed objects of similar size and mass, are formed. The analysis of the 
SED done using a combined set of photometric and spectroscopic data of the leading edge of the disc, totally quenched because it was deprived of its gas content, and of the 
banana-shaped structure of HII regions crossing the disc, indicates that both the abrupt decrease of the star formation activity and the burst are very recent episodes ($\simeq$ 50 Myr).
These timescales are also consistent with the age of the stellar population within the HII regions observed far from the stellar disc, and with the predictions of the simulations
to reproduce the observed 2D gas and star formation distributions and the observed velocity field of the perturbed galaxy.

This work shows once again the importance of the nearby Virgo cluster as a unique laboratory in the study of the effects of the environment on galaxy evolution:
its proximity and a unique set of multifrequency data of exceptional quality in terms of sensitivity, angular 
and spectral resolution and coverage as those gathered by the VESTIGE, NGVS, GUViCS, and HeViCS surveys, allow the study of the perturbing mechanisms down to the scale of individual HII regions.

\begin{acknowledgements}

We warmly thank the anonymous referee for the
accurate reading of the paper and for the numerous, detailed, 
and constructive comments given in the report which helped improving the quality of the manuscript.
This publication uses the data from the AstroSat mission of the Indian Space Research Organisation (ISRO), archived at the Indian Space Science Data Centre(ISSDC).
We are grateful to the whole CFHT team who assisted us in the preparation and in the execution of the observations and in the calibration and data reduction: 
Todd Burdullis, Daniel Devost, Bill Mahoney, Nadine Manset, Andreea Petric, Simon Prunet, Kanoa Withington.
We acknowledge financial support from "Programme National de Cosmologie and Galaxies" (PNCG) funded by CNRS/INSU-IN2P3-INP, CEA and CNES, France,
and from "Projet International de Coop\'eration Scientifique" (PICS) with Canada funded by the CNRS, France.
This research has made use of the NASA/IPAC Extragalactic Database (NED) 
which is operated by the Jet Propulsion Laboratory, California Institute of 
Technology, under contract with the National Aeronautics and Space Administration
and of the GOLDMine database (http://goldmine.mib.infn.it/) (Gavazzi et al. 2003).
This work used the DiRAC@Durham facility managed by the Institute for Computational Cosmology on behalf of the STFC DiRAC HPC Facility (www.dirac.ac.uk). 
The equipment was funded by BEIS capital funding via STFC capital grants ST/K00042X/1, ST/P002293/1, ST/R002371/1 and ST/S002502/1, Durham University and 
STFC operations grant ST/R000832/1. DiRAC is part of the National e-Infrastructure.
A.L. acknowledges support by the European Research Council Advanced Grant No. 740120 'INTERSTELLAR'.
This work reflects only the authors' view and the european Research Commission is not responsible for information it contains. 
M.B. ackonwledges support from the FONDECYT regular project 1170618.
L.G. was funded by the European Union's Horizon 2020 research and innovation programme under the Marie Sk\l{}odowska-Curie grant agreement No. 839090. This work has been partially supported by the Spanish grant PGC2018-095317-B-C21 within the European Funds for Regional Development (FEDER).
H.K. was funded by the Academy of Finland projects 324504 and 328898.

\end{acknowledgements}

\begin{appendix}

\section{Accuracy on velocity dispersion: spectral resolution matters}

In this study, we made the choice of using Fabry-Perot data rather than MUSE data to measure the velocity dispersion of individual HII regions, 
despite Fabry-Perot data have a coarser spatial resolution and a lower signal-to-noise ratio per pixel ($S/N$). In this Appendix, we justify 
our choice based on spectral resolution arguments and explain the limits in which the Fabry-Perot data can securely be used.

The observed velocity profiles, named $P_{\rm obs}$, have to be corrected for the instrumental spectral response, 
or Line Spread Function (LSF). If we call $P_{\rm corr}$ the profile free from instrumental broadening, the observed profile 
is given by the convolution of the two other functions:

\begin{equation}
P_{\rm obs} = P_{\rm corr}  \otimes {\rm LSF}
 \label{eq:convol}
\end{equation}
The LSF induces a spurious velocity dispersion $\sigma_{\rm LSF}$, thus the velocity dispersion of the observed velocity profile $\sigma_{\rm obs}$ is larger than 
the velocity dispersion free from the instrumental broadening, $\sigma_{\rm corr}$.
The latter can be expressed as:

\begin{equation}
\sigma_{\rm corr} = \sqrt{\sigma_{\rm obs}^2 - \sigma_{\rm LSF}^2}
\label{eq:sig_corr}
\end{equation}
For an infinite signal-to-noise ratio and a sampling going to infinitely small steps, with a LSF perfectly known, corrected velocity profiles 
can be recovered with an infinite accuracy using relation \ref{eq:convol}, whereas the corrected velocity dispersion can be recovered from relation \ref{eq:sig_corr}, 
the various terms being evaluated as the second order moments of both observed and LSF profiles.
However, in practice, none of those assumptions are actually met even when the $S/N$ is very high. This is due to the fact that the LSF varies 
in the field and with the wavelength and the data are often under-sampled while deconvolution processes and the moment method request a strong knowledge 
of the functions over a large spectral range to account for the LSF wings and the high frequencies.
Most specifically, the moment method is quite sensitive to the choice of boundaries defining the velocity profiles and to the evaluation of the continuum level. 
In addition, it is not straightforward to estimate uncertainties with this method.
A common way to overcome these difficulties is to fit 
both the intrinsic line shape and the LSF as Gaussian functions, 
and to estimate the corresponding velocity dispersion terms in Eq. \ref{eq:sig_corr} as the ones of the adjusted Gaussian distributions.
%

Assuming that $\sigma_{\rm obs}$ and $\sigma_{\rm LSF}$ have independent uncertainties ($\delta\sigma_{\rm obs}$ and $\delta\sigma_{\rm LSF}$ respectively), which might be true in practice at first approximation since those two 
quantities are measured independently during different exposures, the uncertainty on the corrected velocity dispersion can be expressed as:
\begin{equation}
 \delta \sigma_{\rm corr} = \sqrt{\delta \sigma_{\rm corr, obs}^2+\delta \sigma_{\rm corr, LSF}^2}
 \label{eq:dsig_corr}
\end{equation}
We further assumed that the approximation of $\sigma_{\rm corr}$ by the first order of its Taylor expansion is close to $\sigma_{\rm corr}$ within $\sigma_{\rm obs} \pm \delta \sigma_{\rm obs}$ and $\sigma_{\rm LSF} \pm \delta \sigma_{\rm LSF}$, in order to use the following simple formulation of uncertainty propagation to derive the two terms:

\begin{equation}
 \delta \sigma_{\rm corr, obs} = \delta \sigma_{\rm obs} \times \frac{\partial \sigma_{\rm corr}}{\partial \sigma_{\rm obs}} = \delta \sigma_{\rm obs} \times \frac{\sigma_{\rm obs}}{\sigma_{\rm corr}}
 \label{eq:dsig_corr_obs1}
\end{equation}
\begin{equation}
 \delta \sigma_{\rm corr,LSF} = \delta \sigma_{\rm LSF} \times \frac{\partial \sigma_{\rm corr}}{\partial \sigma_{\rm LSF}} = \delta \sigma_{\rm LSF} \times \frac{\sigma_{\rm LSF}}{\sigma_{\rm corr}}
 \label{eq:dsig_corr_LSF}
\end{equation}

In practice, during the line fitting process of the MUSE data, the LSF width is fixed and 
is included in the model so that the only free parameter is $\sigma_{\rm corr}$.
This means that the fit does not take into account uncertainties on the LSF and the resulting uncertainty is therefore $\delta \sigma_{\rm corr, obs}$
given by relation \ref{eq:dsig_corr_obs1} (the resulting uncertainty would be $\delta\sigma_{\rm obs}$ if the LSF was not included in the model).
During this process, a normalization factor of the uncertainties is applied to have a reduced $\chi^2$ equal to unity, consistent with the difference 
observed between the variance cube and the observed variation in signal-free regions of the cube.
Excluding pixels with dispersion values at the lower boundary of 1~km~s$^{-1}$ (20\% of the pixels) in our MUSE dataset, 
we measured median values $\delta \sigma_{\rm corr, obs}\sim 6.2$~km~s$^{-1}$ ($\sim 2.5$~km~s$^{-1}$ at the 20th percentile) 
and $\sigma_{\rm corr}\sim 20$~km~s$^{-1}$ ($\sim 13$~km~s$^{-1}$ at the 20th percentile). 
These uncertainties are relatively low, especially where the $S/N$ is the highest, and the expected behavior that the uncertainty gets larger at low corrected velocity dispersion (see Eq. \ref{eq:dsig_corr_obs1}) is not properly recovered on average.
However, this may be due to the fact that, within the assumption on the shape of the line, there is few room for variations for the model 
to be fitted to observations.
In addition, the uncertainties on $\sigma_{\rm corr}$ for almost unresolved lines may be small because the $\chi^2$ minimizing algorithm is 
not able to properly explore the uncertainty on the parameters where the posterior is relatively flat, which might be the case since many 
values of $\sigma_{\rm corr}$ would give a very similar line shape.
Last, since both the line and the LSF might not be pure Gaussian functions, we do expect the $\chi^2$ to be slightly larger than 
unity, which may increase the renormalized uncertainties.
To take those arguments into account, we discarded the statistical fitting uncertainty obtained from the best fit and used instead a term proportional to the LSF width to evaluate the uncertainty on $\sigma_{\rm obs}$:
\begin{equation}
 \delta\sigma_{\rm obs}=\alpha \times \sigma_{\rm LSF}
 \label{eq:dsig_obs}
\end{equation}
where $\alpha$ depends on the $S/N$. This term is independent from the uncertainty on the LSF, leading to rewrite Eq. \ref{eq:dsig_corr_obs1} as:
\begin{equation}
 \delta \sigma_{\rm corr, obs} = (\alpha \times \sigma_{\rm LSF}) \times \frac{\sigma_{\rm obs}}{\sigma_{\rm corr}}
 \label{eq:dsig_corr_obs2}
\end{equation}
Since the width of MUSE LSF is $\sigma_{\rm LSF}\sim 50$~km~s$^{-1}$, we set the factor $\alpha$ to 0.1, so that $\delta\sigma_{\rm obs} = 5$~km~s$^{-1}$, asymptotically matching the range of uncertainties $\delta\sigma_{\rm corr, obs}$ estimated above from the Gaussian fit.

In order to evaluate relation \ref{eq:dsig_corr_LSF},
we have estimated $\delta \sigma_{\rm LSF}$, the uncertainty on the LSF 
dispersion in the MUSE data, using night sky emission lines. To  
reach a large $S/N$ on those lines, we used the deep data of 
the Hubble Ultra Deep Field (HUDF) from Bacon et al. (2017), with exposure times larger than 10h. 
In MUSE data, the LSF is undersampled, which makes its estimation by a simple Gaussian function varying with the sampling, both spatially and spectrally
near the observed H$\alpha$ line.
We have therefore estimated the standard deviation of the line dispersion using a set of eight strong sky lines around $\lambda\sim 6550$~\AA\ 
(as done in Bacon et al. 2017), near the observed H$\alpha$ line, over a box of 100$\times$100 pixels$^2$ at the centre of the HUDF MUSE datacube. 
Assuming the line dispersion to be similar for all considered lines (small spectral range and unresolved lines), we obtained 
$\delta \sigma_{\rm LSF}\sim 0.8$~km~s$^{-1}$ (for a LSF with $\sigma_{\rm LSF} \sim 50$~km~s$^{-1}$). Similarly to
$\delta\sigma_{\rm corr,obs}$ (defined in Eq. \ref{eq:dsig_corr_obs1}), this uncertainty may be underestimated since the LSF might not be Gaussian. We therefore round 
it to $\delta \sigma_{\rm LSF} = 1$~km~s$^{-1}$ to account for possible systematics due to the LSF unknown real shape.

In the case of the Fabry-Perot data, the spectral resolution is $R\sim 10000$, corresponding to $\sigma_{\rm LSF}\sim 13$~km~s$^{-1}$. 
Since the $S/N$ might be a bit lower than in the MUSE data, we expect the parameter $\alpha$ (Eq. \ref{eq:dsig_obs}) to be larger. We therefore made two hypotheses, one pessimistic and the other one more realistic, 
but still conservative.
For the first hypothesis, we assumed similar absolute uncertainties as for MUSE: 
$\delta \sigma_{\rm LSF} = 1$~km~s$^{-1}$, and $\delta \sigma_{\rm obs} = 5$~km~s$^{-1}$, corresponding to $\alpha=0.4$. 
For the second hypothesis, to fairly compare with MUSE, we used a lower value of $\alpha=0.2$, leading to $\delta \sigma_{\rm obs} = 2.5$~km~s$^{-1}$.

\begin{figure}
 \includegraphics[width=0.5\textwidth]{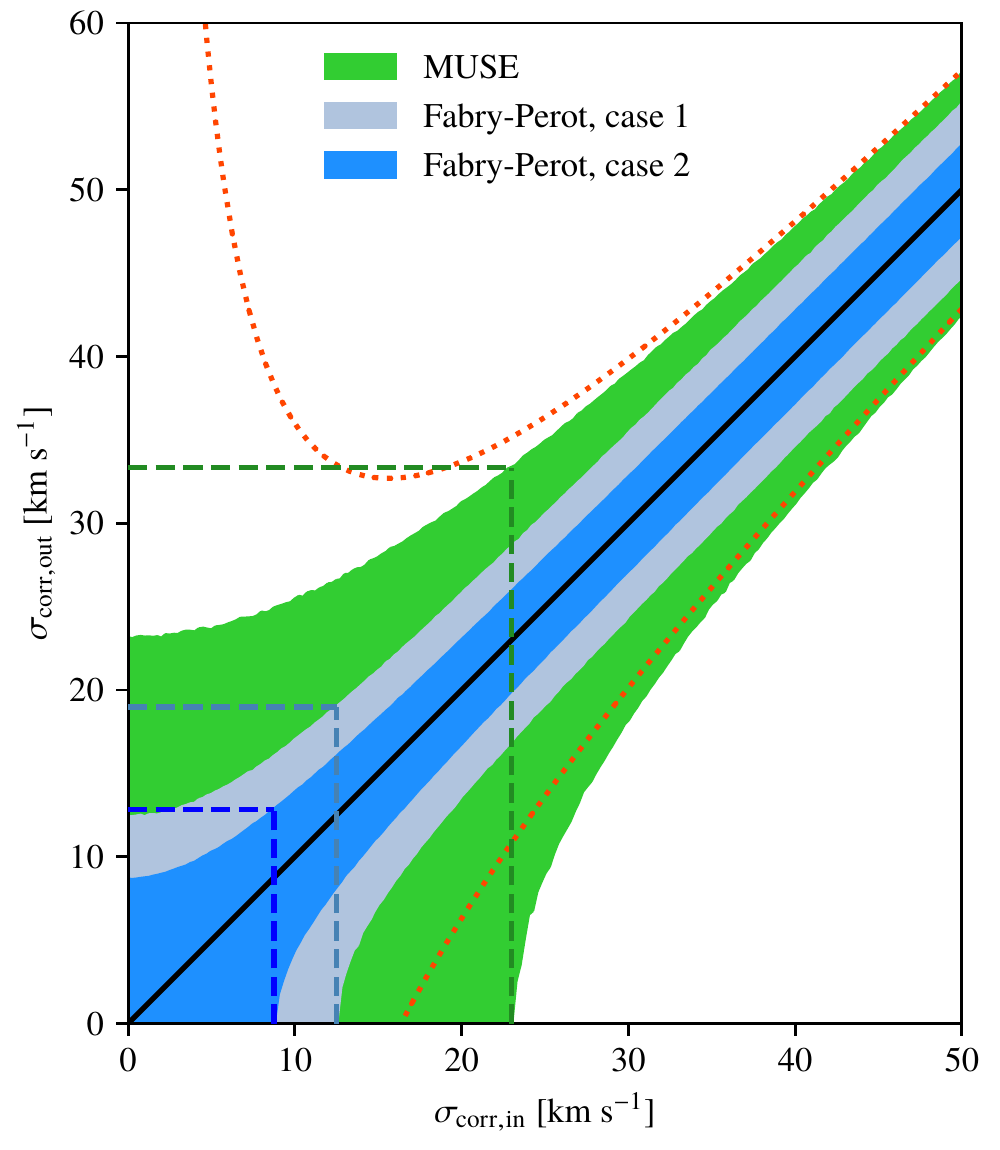}
 \caption{Range of recovered corrected velocity dispersions $\sigma_{\rm corr, out}$ as a function of the input corrected velocity dispersion $\sigma_{\rm corr, in}$
 for several set-ups. The red dotted line corresponds to the analytical propagation of uncertainty of the MUSE set-up, with  $\sigma_{\rm LSF} = 50$~km~s$^{-1}$, $\delta \sigma_{\rm obs} = 5$~km~s$^{-1}$, and $\delta \sigma_{\rm LSF} = 1$~km~s$^{-1}$. 
 The interval determined from the Monte Carlo approach for this set-up is shown as the green area, which is partially hidden by the blue areas for visual convenience.
 The blue areas correspond to the Fabry-Perot set-ups, with $\sigma_{\rm LSF} = 13$~km~s$^{-1}$ and $\delta \sigma_{\rm LSF} = 1$~km~s$^{-1}$.
 The light blue one, which is again partially hidden by the dark blue area, corresponds to the MUSE-like uncertainties (case 1), 
 with $\delta \sigma_{\rm obs} = 5$~km~s$^{-1}$,
 whereas the darker blue one corresponds to the most realistic hypothesis (case 2), with  $\delta \sigma_{\rm obs} = 2.5$~km~s$^{-1}$.
 The black line shows the one to one relation. The three dashed lines (green, light and dark blue) correspond to the three set-ups. 
 The vertical lines show the corrected velocity dispersion where the lower limit gets larger than zero. The horizontal lines show the corresponding 
 upper limit, which defines the range of reliable corrected velocity dispersion.
 }
 \label{fig:delta_sig}
\end{figure}

In Fig. \ref{fig:delta_sig}, we show how the corrected velocity dispersion $\sigma_{\rm corr}$ is expected to be recovered 
within these uncertainties for the MUSE and the Fabry-Perot data. For MUSE, we show the uncertainties obtained from the 
analytical propagation of uncertainty provided in Eq. \ref{eq:dsig_corr} to \ref{eq:dsig_corr_LSF} (red dotted lines). However, for small values of $\sigma_{\rm corr}$, the assumptions made to derive those equations are not valid anymore and the uncertainty propagation therefore diverges. We therefore also derived upper and lower limits using a Monte Carlo approach: 
for any value of $\sigma_{\rm corr, in}$ , and for each instrument hypothesis, (i) we computed the expected mean value of the observed velocity dispersion $\sigma_{\rm obs}$ by inverting Eq. \ref{eq:sig_corr} and using the estimate of the LSF width $\sigma_{\rm LSF}$ of the considered instrument; (ii) we generated two random normal distributions for both $\sigma_{\rm obs}$ and $\sigma_{\rm LSF}$, with corresponding mean values and standard deviations $\delta\sigma_{\rm obs}$ from Eq. \ref{eq:dsig_obs} and $\delta\sigma_{\rm LSF}$, as estimated above (see caption of Fig. \ref{fig:delta_sig}), 
and (iii) analysed the final distribution of the corrected velocity dispersion, $\sigma_{\rm corr, out}$.
We did $10^5$ iterations and 
determined the upper and lower limits from the interval around the median having a probability to occur of 68.27\%, 
which corresponds to the interval within the standard deviation for a Gaussian distribution. This interval is represented 
by the green area for MUSE, and by the light and dark areas for the Fabry-Perot pessimistic and realistic
hypothesis, respectively in Fig. \ref{fig:delta_sig}. The total green area (MUSE) includes the one covered by the blue surfaces (Fabry-Perot),
indicating that the Fabry-Perot data provide more accurate velocity dispersions than the MUSE data. We also notice that 
the analytical uncertainty propagation overestimates the upper limit, whereas it underestimates the lower one.
For MUSE, the lower limit is compatible with zero up to $\sigma_{\rm corr, in} \sim 23$~km~s$^{-1}$, and the corresponding upper limit is 
$\sim 33$~km~s$^{-1}$ (dashed green lines), which indicates that it is not possible to infer with confidence a corrected dispersion lower than $\sim 33$~km~s$^{-1}$.
For the Farby-Perot data, the corrected velocity dispersion can be inferred with confidence only when $\sigma_{\rm corr, in}$ $\geq$ 19~km~s$^{-1}$ 
in the most pessimistic case, and $\sigma_{\rm corr, in}$ $\geq$ 13~km~s$^{-1}$ in the most realistic hypothesis, as shown by the blue horizontal dashed lines.
Reducing by a factor two the parameter $\alpha$, i.e. the uncertainty $\delta \sigma_{\rm corr,2}$, changes the limits on $\sigma_{\rm corr}$ to 
$\sim 24$, $\sim 13$, and $\sim 9$~km~s$^{-1}$ for MUSE, Fabry-Perot cases 1 and 2, respectively.
The distribution in the velocity dispersion shown in Fig. \ref{dispersion} (15 $\lesssim$ $\sigma$ $\lesssim$ 35 km s$^{-1}$)
clearly indicates that the spectral resolution of MUSE is not sufficient to derive the kinematical properties 
of individual HII regions, which require Fabry-Perot data.

\section{Impact of lines mixing to measure velocity dispersion}

In crowded regions, several HII regions can contribute to the spectrum of a spatial resolution element. This is especially true for 
the Fabry-Perot data which have a coarser spatial resolution than the MUSE data. Theoretically, with an infinite $S/N$, lines can be separated 
if their shapes are perfectly known. In practice, when the line shape is not perfectly known (and assumed as Gaussian), it is not possible to asses whether 
their deviation from Gaussianity is due to the line shape or to a double or multiple profile, unless the lines are resolved. This happens when
their separation is larger than $\sim 1.1415$ times the LSF full-width at half maximum (FWHM , Rayleigh criterion adapted to spectroscopic instruments, Namioka 1998), which corresponds to a velocity separation 
of $\sim 133$~km~s$^{-1}$ and $\sim 34$~km~s$^{-1}$ for the MUSE and the Fabry-Perot data, respectively.
According to the velocity field amplitude (Fig. \ref{vel}), adjacent regions can be marginally spectrally resolved with the Fabry-Perot data, but absolutely not with MUSE.
In addition, if the lines have intrinsic velocity dispersions of the order of 10-20~km~s$^{-1}$, 
they are spectrally resolved within the Fabry-Perot data whenever they are separated by at least 38-56~km~s$^{-1}$, which is unlikely.

\begin{figure*}
 \includegraphics[width=0.5\textwidth]{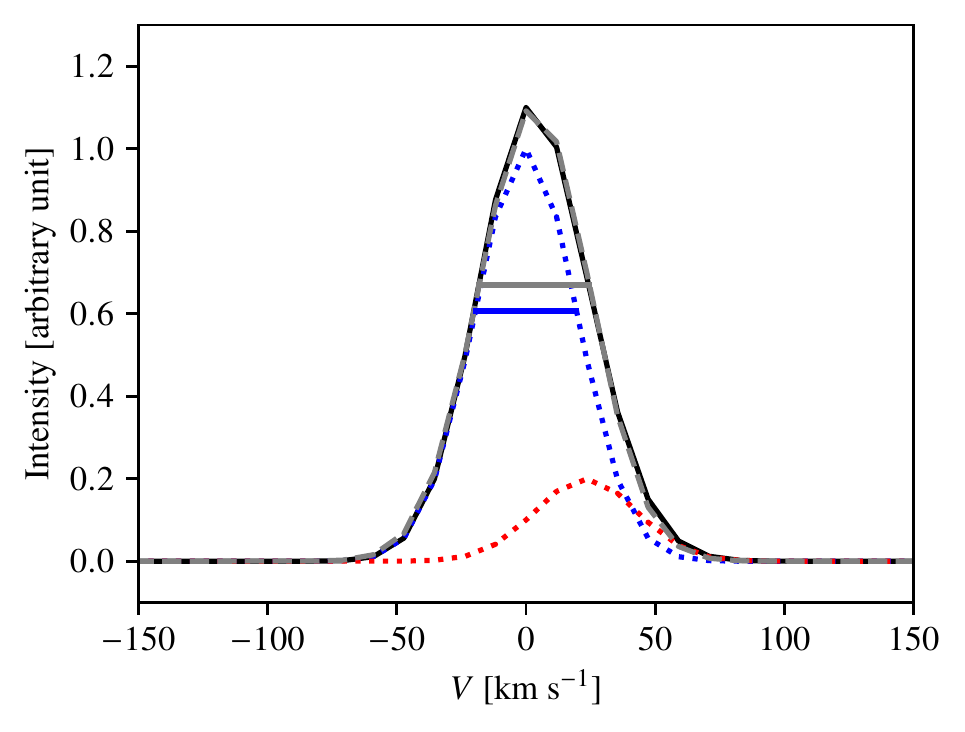}
 \includegraphics[width=0.5\textwidth]{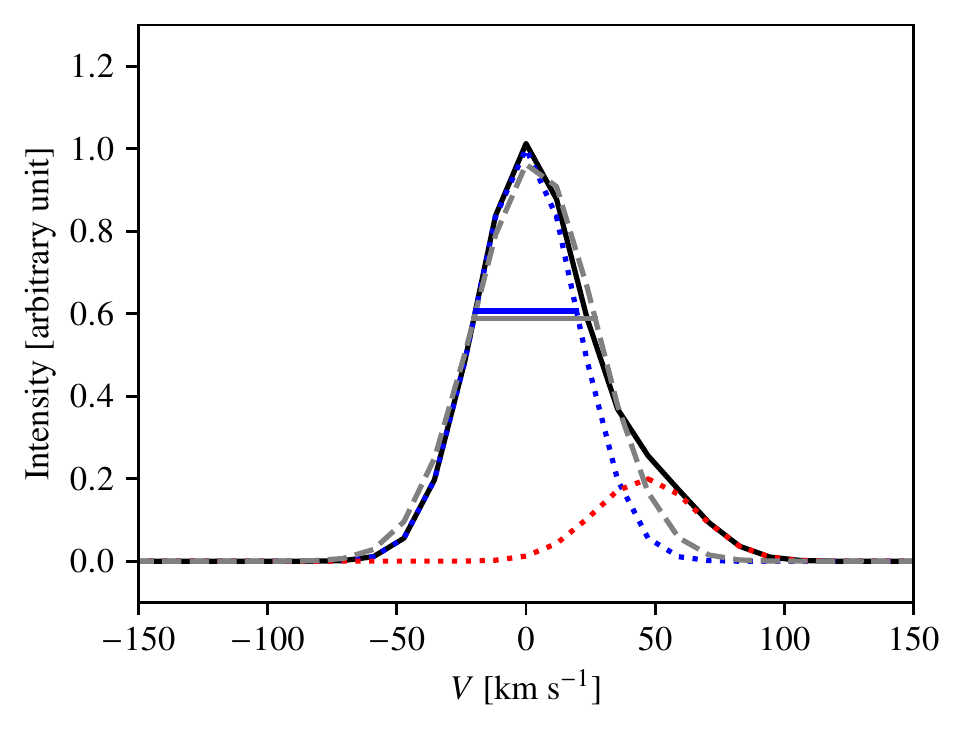}\\
 \includegraphics[width=0.5\textwidth]{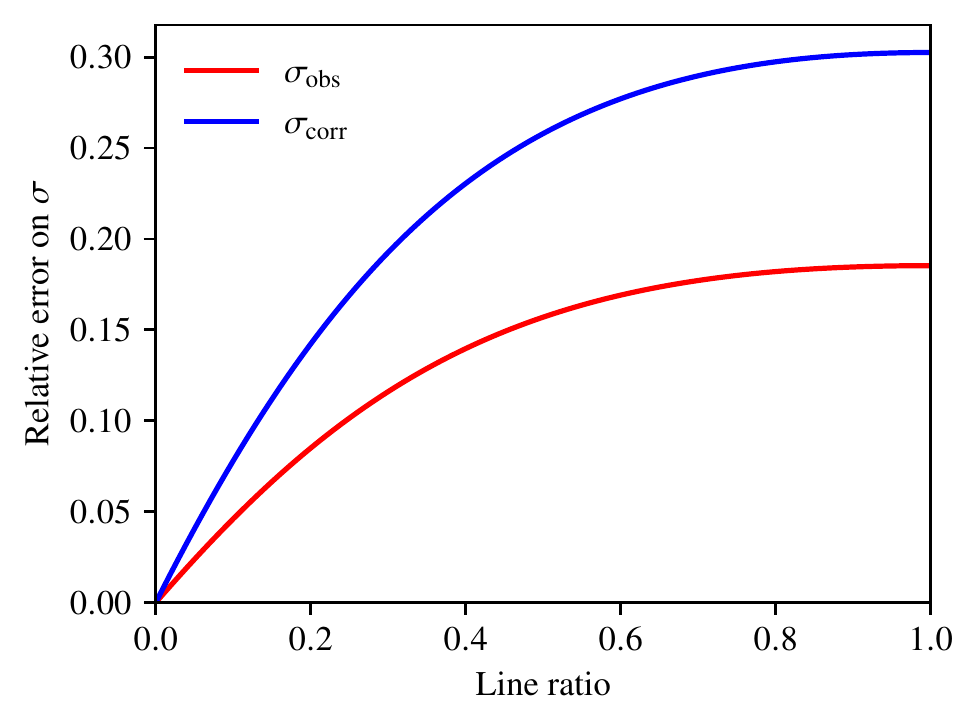}
 \includegraphics[width=0.5\textwidth]{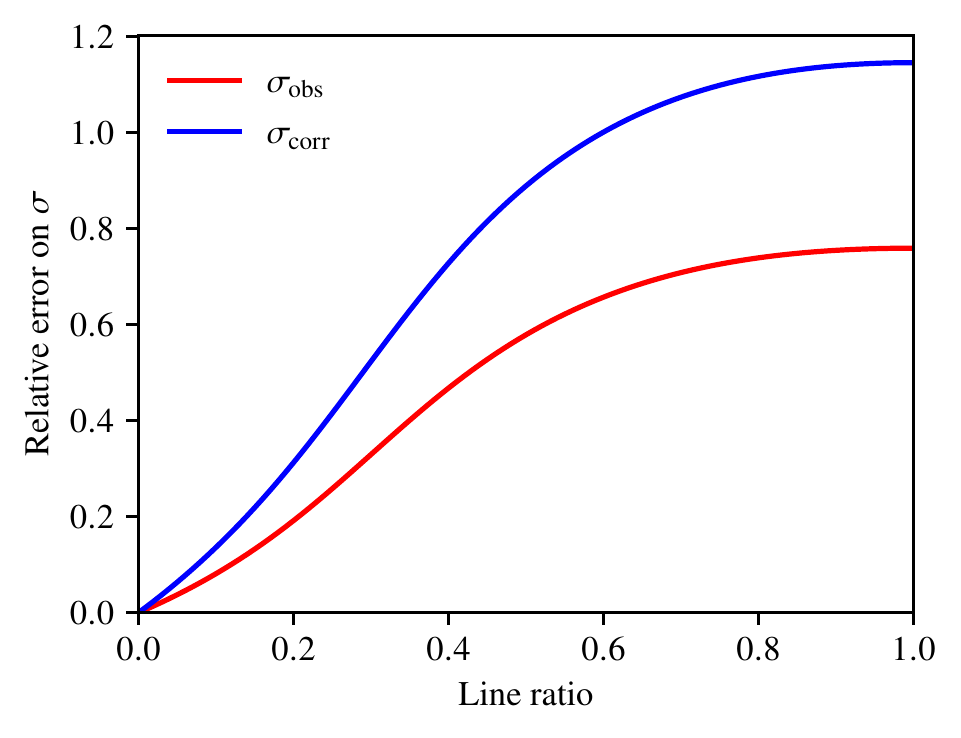}
 \caption{Illustration of the effect of the mix of two lines, having both an intrinsic velocity dispersion $\sigma_{\rm corr}=15$~km~s$^{-1}$, 
 and with a LSF width $\sigma_{\rm LSF}=13$~km~s$^{-1}$, on the extracted velocity dispersion with two line separations: half of a 
 single line FWHM (left) and FWHM (right). The mix of the two lines (red and blue dotted curves) with an intensity ratio of 0.2 is shown 
 on the top panels by the black curve, whereas the single Gaussian fit is shown in dotted gray. The single line and the fitted line velocity dispersions 
 ($2\times \sigma_{\rm obs}$) are shown as blue and gray horizontal segments, respectively. The bottom plots illustrate the variation of the relative 
 uncertainty on both the observed (red) and corrected (blue) velocity dispersion when the line ratio varies between 0 and 1.}
 \label{fig:mix_spectra}
\end{figure*}

In order to measure the impact of mixing lines on the estimation of the velocity dispersion for the Fabry-Perot data, we have simulated two-components 
spectra with a line intensity ratio varying from 0 to 1 and with an intrinsic velocity dispersion of 15~km~s$^{-1}$, which is a lower limit 
to the typical velocity dispersion in the warm component (see e.g. Epinat et al. 2010).
In order to investigate the limit case where lines can be marginally resolved, we first set the separation between the lines to the theoretical observed FWHM of one resulting line, i.e. to 46~km~s$^{-1}$.
We then measured the spectrum properties with a single Gaussian fit and found that the measurement of the observed velocity dispersion of 
the strongest line is biased by less than 10\%\ if the ratio of the line intensities is lower than 12\%.
We then performed the same exercise on a less extreme case, with a separation between the lines twice less, i.e. equals to 23~km~s$^{-1}$, and found that the ratio of 
intensities can reach 24\%\ before affecting the result by more than 10\%\ (see Fig. \ref{fig:mix_spectra}). With this level of contamination, 
the impact on the corrected dispersion is less than 20\%.
As an acceptable compromise, we adopt a limit of 20\%\ in the line intensity ratio in the analysis presented in Sect. 3.5.2. We 
further stress that we did not find clear evidence for multiple peaks in the Fabry-Perot spectra of selected HII regions. If multiple components exist, 
they should be separated by less than a small fraction of the LSF FWHM.

\section{Spatial contamination of adjacent regions}

We aim at estimating the contribution of any neighboring region on the spectrum of a region that we wish to study.
Indeed, given the high angular resolution of the VESTIGE data used to identify the HII regions ($\sim$ 0.74 arcsec), 
some of them are not resolved in the Fabry-Perot data ($\sim$ 2 arcsec).
For any HII region, identified with an index $i$, the \textsc{HIIphot} software computes the flux $F_i$ within the region area $A_i$, and the effective circular 
full width at half-maximum,
$FWHM_{i, \rm eff}$ of a 2D-Gaussian model (see Helmboldt et al. 2005). We note $\sigma_{i, \rm eff}=\frac{FWHM_{i, \rm eff}}{2 \sqrt{2 \log{2}}}$ the dispersion corresponding to the measured region size within the VESTIGE data.
We evaluated the typical radius $r_i$ (in arcsecond) of each region from its measured area $A_i$ in square arcseconds as $r_i = \sqrt{A_i/\pi}$.
Assuming that $F_i$ corresponds to the flux of the 2D symmetric Gaussian within $r_i$, we estimated for each region the total flux $F_{i, T}$ as:
\begin{equation}
 F_{i, T} = \frac{F_i}{1-\exp{\left(-\frac{r_i^2}{2 \sigma_{i, \rm eff}^2}\right)}}
 \label{eq:tot_flux}
\end{equation}
and deduced the central intensity $I_{0, i}$ in the Fabry-Perot data as:
\begin{equation}
 I_{0,i} = \frac{F_{i, T}}{2\pi \sigma_i^2}
 \label{eq:I0}
\end{equation}
where $\sigma_i$ is the observed spatial dispersion of the region within the Fabry-Perot data. 
Assuming that both the VESTIGE and the Fabry-Perot data PSF are well described by Gaussian functions with spatial 
dispersion $\sigma_{\rm V}$ and $\sigma_{\rm FP}$ respectively, we have $\sigma_i^2 = \sigma_{i, \rm eff}^2 + \sigma_{\rm FP}^2 - \sigma_{\rm V}^2$.

For simplicity, we assumed that when we study one region (region 1), we measure the spectrum at its centre. We therefore needed to 
evaluate the intensity of any another region (region $i$) at this location. We estimated a contamination parameter of region $i$ on region 1 
as the ratio of lines intensity at this location:

\begin{equation}
 R_{1i} = \frac{I_i(d)}{I_1(0)} = \frac{F_{i, T}}{F_{1,T}} \times \frac{\sigma_1^2}{\sigma_i^2} \times \exp{\left( - \frac{d_{1i}^2}{2\sigma_i^2}\right)}
\end{equation}
where $d_{1i}$ is the separation between the centres of the two regions.

We have also computed the background contamination in each bin associated to a given region. 
For this purpose we have estimated on the MUSE data the mean diffuse background surface brightness 
of the stellar disc in several extended regions deprived of HII regions: $\Sigma(H\alpha)$ $\sim 2\times 10^{-17}$~erg~s$^{-1}$~cm$^{-2}$~arcsec$^{-2}$. 
We assumed this background to be constant over all regions. 
For each region, we have integrated this background surface brightness over the area of the associated Fabry-Perot bin $A_{i, \rm FP}$. 
We have also integrated the flux of the region over the bin area assuming it corresponds to the flux of a 2D-Gaussian with a 
dispersion $\sigma_i$ within a radius $r_{i,\rm FP}=\sqrt{A_{i, \rm FP}/\pi}$, and deduced a background contamination parameter 
as the ratio between the background flux over the region flux.

In the analysis of the kinematical properties of the HII regions presented in the paper, we have removed all regions 
for which the contamination parameter is larger than $R_{1i}=0.2$ for at least one neighboring region or for the 
background. We further excluded regions for which the barycenter of bins were further away by more than 2 arcseconds from the region centre.
To conclude, with these selection criteria we expect that the velocity dispersion measurements used in the analysis are affected by less than 10\%.

\section{Application to the data}

\begin{figure*}
   \centering
   \includegraphics[width=0.99\textwidth]{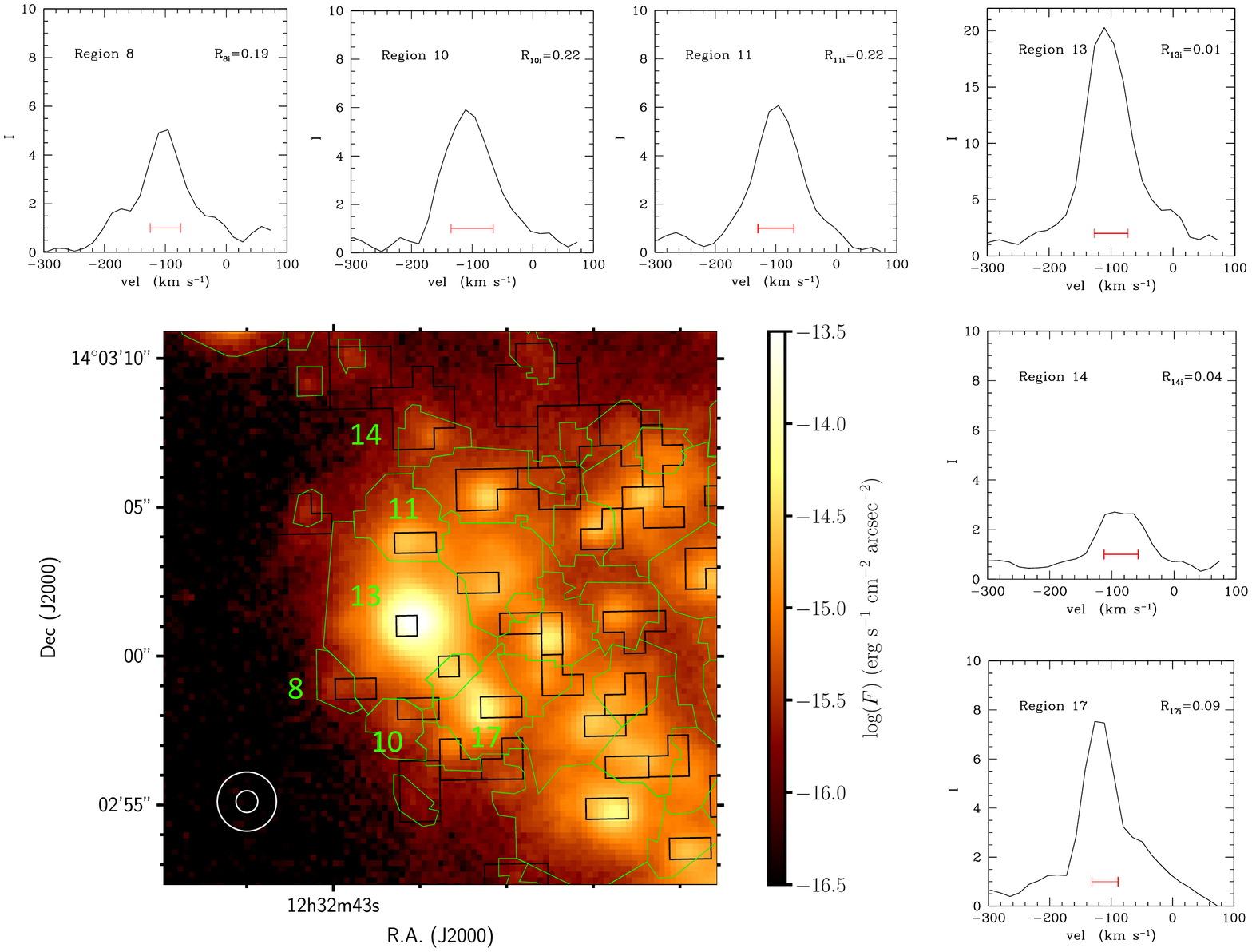}
   \caption{Lower left panel: narrow-band H$\alpha$ image of a crowded region in IC 3476. The boundaries of the individual HII regions 
   identified by the \textsc{HIIphot} pipeline are indicated with the green contours. The corresponding Voronoi bins in the Fabry-Perot 
   data are indicated with black polygons. The limiting seeing of the VESTIGE and of the Fabry-Perot data are marked with two concentric circles
   on the lower left corner of the image. The panels on the top and on the right of the figure show the H$\alpha$ line profiles of the velocity dispersion (in arbitrary units) 
   of the seven selected regions indicated in the lower left panel. Of these, regions 10 and 11 do not satisfy the $R_{1i}$ $\leq$ 0.2 criterion given
   above, and are thus excluded from the analysis of the kinematical properties of HII regions given in the text, while regions 8, 13, 14 amd 17 are included.
   In each panel the red horizontal line gives the measured $\rm FWHM = 2\sqrt{2\ln{2}}\times \sigma$, where $\sigma$ is the velocity dispersion.
 }
   \label{crowded}%
   \end{figure*}

The application of these selection criteria for the identification of the corresponding HII regions 
to the VESTIGE and Fabry-Perot data and the estimate of the velocity dispersion is illustrated in Fig. \ref{crowded}. 
The \textsc{HIIphot} pipeline identifies several HII regions in a crowded region of the galaxy on the deep VESTIGE narrow-band imaging data 
(green contours).
The Voronoi bin with the closest barycenter in the Fabry-Perot data is identified with the black polygons for the brightest HII regions.
Its size increases with the decrease of the H$\alpha$ line intensity of the region, some of which are undetected in the shallower Fabry-Perot data.
For those Voronoi bins satisfying the above criteria (barycenter less than 2 arcsec away from the HII region centre given by \textsc{HIIphot} 
on the VESTIGE image, a contamination $R_{1i}$ $\leq$ 0.2), we extract a velocity dispersion measurement (regions 8, 13, 14 and 17 in Fig. \ref{crowded})
which is used in the analysis of the kinematical properties of the HII regions given along the text. Those not satisfying these criteria 
(regions 10 and 11), are not included in the analysis.

\end{appendix}

\end{document}